\documentclass[aps,prl,preprint,groupedaddress]{revtex4-1}
\usepackage{subfig}
\usepackage{graphicx}
\usepackage{amsmath}
\usepackage{caption}

\begin{document}


\title{Phononic band structure of honeycomb lattice without or with defects, using spectrally formulated finite element method}


\author{Sushovan Mukherjee}
\author{S. Gopalakrishnan}
\affiliation{Indian Institute of Science}


\date{\today}

\begin{abstract}
A spectrally formulated
finite element analysis based methodology has been proposed to
calculate phononic band structure of reticulated
honeycomb lattices having translationally
invariant repetitive elements called unit cells. Bloch
formulation captures dynamics of infinite structure
through that of a unit cell. While conventional FEM
is generally used for analyzing dynamics of such
unit systems, here, constituent structural members are
treated as 1D waveguide and modeled as Timoshenko
beam frame element, enabling application of spectral
FEM, suitable for accurately analyzing the dynamics,
particularly efficient at very high frequencies. Using
exact solutions as shape functions spares dense
meshing. Resulting eigenvalue problem is solved by
Wittrick-Williams method, an iterative scheme.
Subsequently, band structures are obtained for
supercells- units comprising multiple elemental unit
cells; compared and reconciled with those obtained
using elemental cell (termed primitive unit cell
to distinguish from supercell). Primitive cell band
structures are reconstructed from Supercell band
structures. Supercell band structures show some
spurious bands, which are explained in terms of band
folding in the primitive cell band structure.
Supercell allows treatment of defects as a periodic
feature with certain defect density. Of particular
observation in such band structure is the separation
of bands, known as degeneracy breaking.
\end{abstract}


\maketitle
\maketitle

\section{Introduction}

Wave propagation in periodic structures has been studied for a long time from different perspectives in engineering and science. Particularly, the area has been garnering significantly increasing attention over last couple of decades primarily due to the advent of phononic and photonic structures, advances in fabrication technology at micro and nano scales, and discovery of new materials including carbon nanotube and graphene with potentially revolutionizing applications.  Historically, Newton worked on the speed of propagation of sound in air through the use of periodic spring mass system \cite{Brillouin}. Rayleigh, in his treatise, made significant contribution in the field of vibration and acoustics. In the first half of last century Leonard Felix Bloch developed formalism to model electronic wave function in a periodic potential of nucleus which eventually became a cornerstone of solid state physics.  Leon Brillouine\cite{Brillouin} worked on quantum theory of solids and advanced the contemporary development of X-ray scattering based confirmation on crystal lattice theory of solids to propose phononic vibrations. In his seminal work, a treatise  dedicated to the phenomenon of wave propagation in periodic structures, he narrated  band effects and filtering effects through the formation of pass  bands and stop bands. In the later half, there had been work carried out by Mead \cite{Mead-mono, Mead-multi, Mead-research} enumerating novel methods of singly and multiply connected periodic members based on propagation constant. Of late, there have been works by Phani et al., Ruzzenne et al., Gonella et al., Leamy and Farzbod et. al., dealing with wave propagation in mechanical structures having different types of honeycomb lattice. Ruzzene et. al.\cite{Ruzzene} have investigated wave beaming and directional dependency of lattice structures demonstrating application on negetive Poisson ratio lattice .Phani et. al.\cite{Phani} have analysed band structure of periodic lattices using conventional finite element methhod. Farzbod et. al. \cite{Farzbod1} investigated treatment of force in generic lattices in line with Langley \cite{Langley}. They \cite{Farzbod2} have also shown the equivalence of propagation constant method and the Bloch theorem and disproved general applicability of Bloch formulation to nonlinear cases. Gonella et al. \cite{Gonella} investigated the dispersion relationships, wave velocities and band structures of hexagonal and reentrant lattices. Leamy \cite{Leamy} has performed a wave propagation based analysis of lattice structure dynamics accounting for joint inertia. 

Understandably, Bloch theory remains a common constituent to all these works, reducing the computation for the infinite structure to only that of the unit cell, the repeatitive unit of the structure with translational invarience. However, most of them use conventional finite element, rendering the methods computationally prohibitive and inaccurate (particularly at higher frequencies), and whereas Leamy uses an exact method, there remains scope for developing a robust computational 
framework to enable tackle more complicated geometry and to allow for automation. Furthermore, most papers reported in the literature use adhoc method to solve for implicit eigenvalue problem that does not guarantee to provide all eigenvalues- particularly the closely spaced ones. In this paper a method has been proposed which, apart from using Bloch theorem to enable aforesaid reduction of computation to just within a unit cell, uses spectrally formulated dynamic stiffness matrix to capture the dynamic behaviour of the system with accuracy to that of the underlying model for the analysis of beam and rod. For the computation of eigenvalues Wittrick-Williams algorithm \cite{Wittrick} has been used. This algorithm provides a bound on the number of eigenvalues below any particular frequency, thereby allowing for obtaining the eigenvalues iteratively to the desired accuracy without missing any.

An interesting observation in this context can be made while investigating what happens when cerain multiples of unit cells is considered to be a new unit called supercell and the band structure of such a compound cell, is obtained. In spite of underlying structure remaining same, there appears incompatibility between the two band structures arising due to two cases representing different area in wavenumber space, and the existance of some spurious bands, which requires detailed analysis to identify their source. Consideration of supercell makes way to the treatment of defects as a periodic feature with certain incidence density, thereby, making defects to be considered as to be creating a new unit cell with certain deviation from the supercell of the original perfect lattice. A few different types of defects are analysed and their significances are investigated. 

\section{Methods}

\subsection{Bloch Formulation}

One of the most convenient ways to describe the topology of a periodic structure is through the concept of lattice, which is an infinite set of points, the arrangement of which looks the same with reference to any of these points, thus, being endowd with certain symmetry (tranlation and optionally rotational). The dimensionality of the lattice is same as that of the underlying periodic structure and an equal number of translation vectors, in thier integral multiples construct the whole lattice starting from a point. These translation vectors themselves constitute a unit cell containing a unit of the whole structure. Thus, for a two dimensional lattice, for any set of integer values of $m$ and $n$, $\bf{{r}^\prime}$ represents the position of a point equivalent to a point having position $\bf{r}$ in the reference cell, when  
\begin{equation}
{\bf{r^{\prime}}} =  {\bf{r}} + m.{\bf{a_1}} + n.{\bf{a_2}}
\end{equation}

where $\bf{a_1}$, $\bf{a_2}$ are associated unit cell vectors.
The structure, which is presumed to be spread infinitely on both the directions, lends itself to a kinemetic description through the displacement of the points belonging  to a finite size of that of the repeatitive unit cell because of the intrinsic relationship between/amongst similar points, being presentable through their position relative to the reference cell. This can be interpreted in terms of a propagation constant- mainly found in mechanical literature, or in terms of Bloch formulation- primarily occuring in solid state physics context, both being equivalent as shown by Farzbod et al.\cite{Farzbod2}.

The consideration of a lattice in the dual space or reciprocal space is a convenient tool in the analysis related to Bloch formulation. The dual space lattice consists of unit reciprocal cell vectors obtained from the unit cell vector following the relationship $ \bf{a}_i . {\bf{b}}_j={\delta}_{ij}$. 
The reciprocal lattice can be physically interpreted as the space of wavenumbers. 

\begin{figure}[!ht]
\centering
\subfloat[]{
\label{fig_schm_uc1:a}
\begin{minipage}{3.6in}
\includegraphics[width=3.6in]{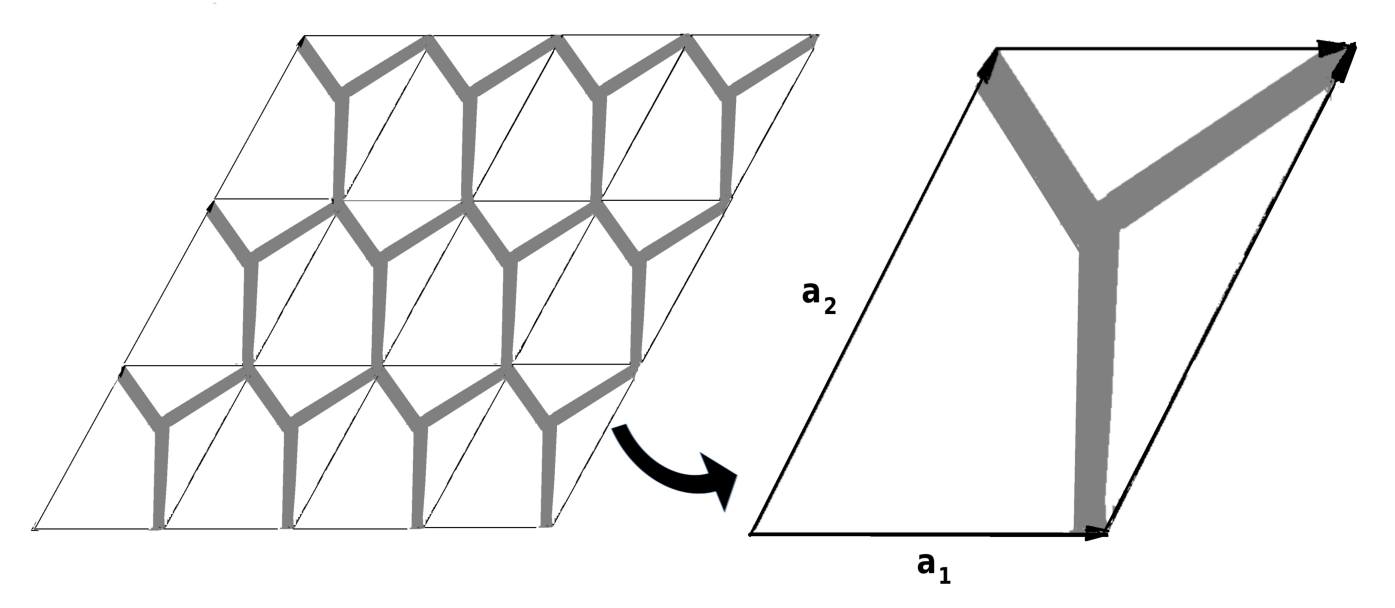}
\end{minipage}}
\subfloat[]{
\label{fig_schm_uc1:b}
\begin{minipage}{1.7in}
\includegraphics[width=1.7in]{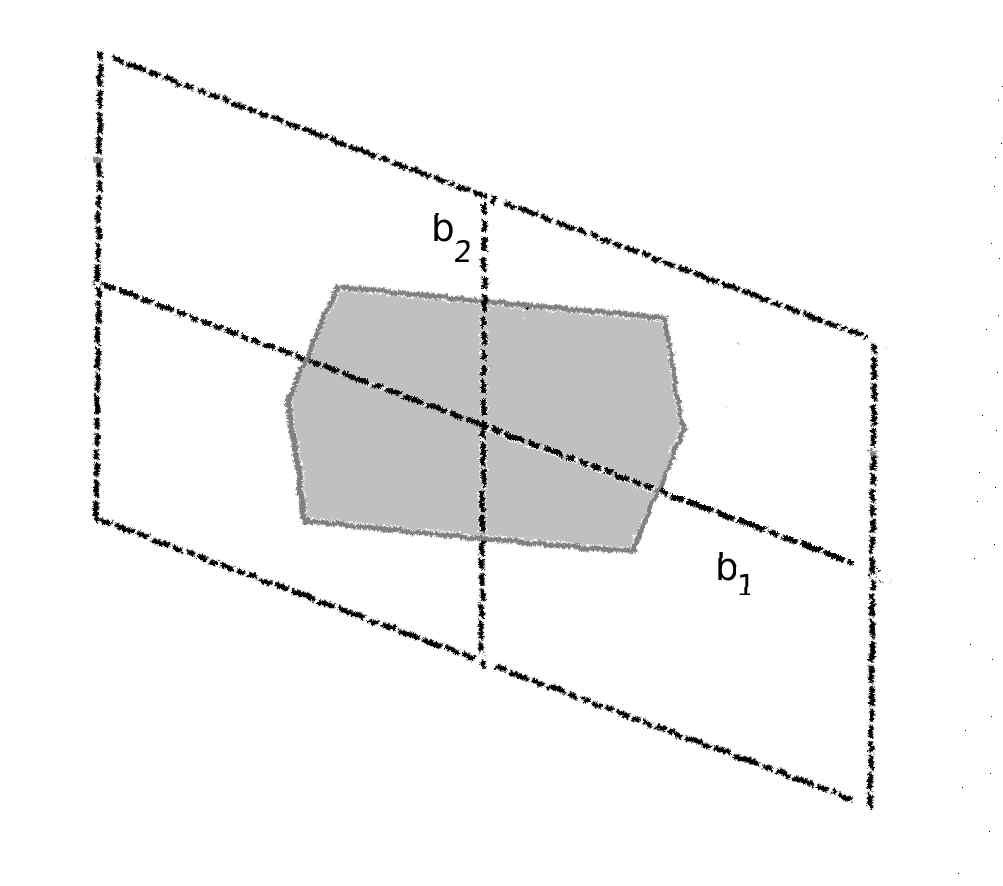}
\end{minipage}}
\caption{(a) Schematic representation of an arbritrary honeycomb structure, seen as formed through the tessalation of a unit cell, with a magnified view of coresponding unit cell having $\bf{a}_1$ and $\bf{a}_2$  as unit cell vectors (b) Unit cell of the reciprocal lattice having $\bf{b}_1$ and $\bf{b}_2$  as unit cell vectors, the Brillouin zone is shown as the grey hexagon} 
\label{fig_schm_uc1}
\end{figure}

Thus, for an arbitrary grillage, as depicted in Fig. \ref{fig_schm_uc1}, the parallogram representing the unit cell which forms the whole structure through tesselation-  $\bf{a}_1$, $\bf{a}_2$ denoting the two vectors for the unit cell, the aforesaid deformation relationship can be expressed as.

\begin{equation}
d_{\bf{k}} ({\bf{r}}^\prime) =  e^{-i({{m.\mu_1+n.\mu_2}})} d_{\bf{k}} ({\bf{r}})
\end{equation}

where ${\mu_1}$ and ${\mu_2}$ are the components of the wavenumber ${\bf{k}}$ in reciprocal space with reference to  unit vectors ${\bf{b}}_1$ and ${\bf{b}}_2$.


Literature in the field of engineering refers to such grillage type of reticulated honeycomb structures as lattice, and is followed here too.  This, however, should not create any ambiguity in understanding as against the formal defination of lattice as a set of points with associated symmetry.

\subsection{Spectral Finite Element Method}
\begin{figure}[!ht]
\centering\includegraphics[width=5.5in]{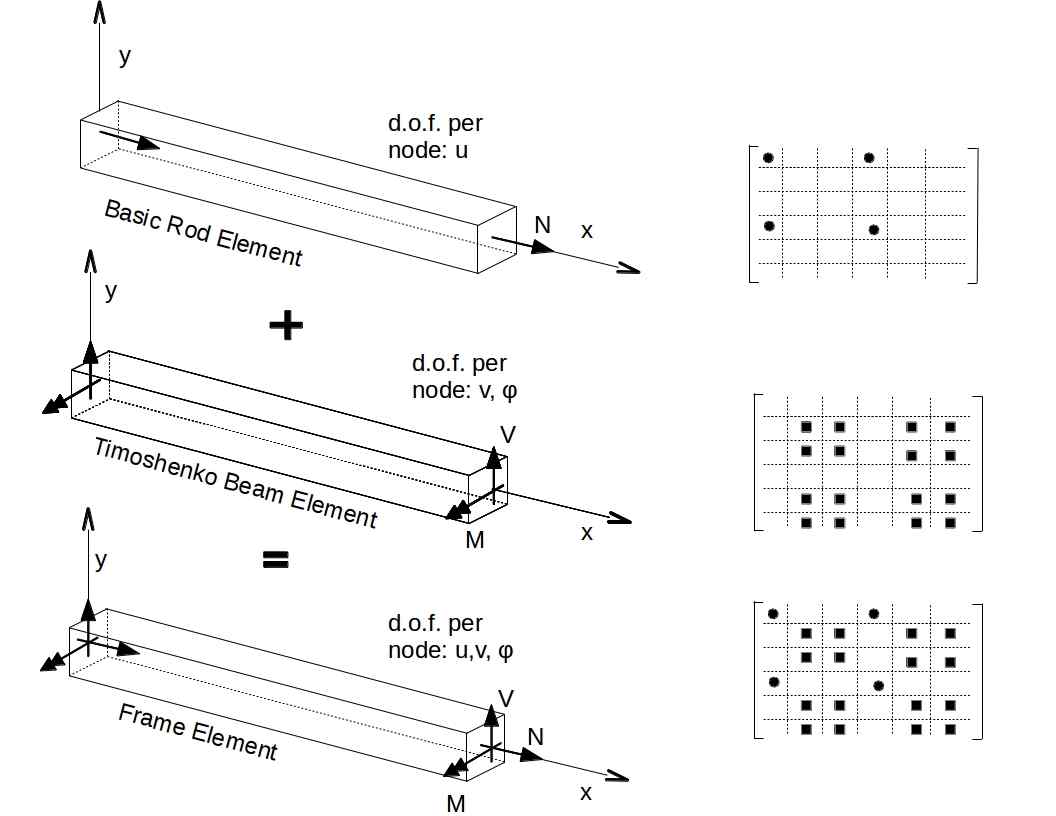}
\caption{Schematic presentation of a frame element as combination of basic rod element and a Timoshenko beam element with associated forces and d.o.f. Matrices on the right depicts contribution of individual element dynamic stiffness matrices into frame element dynamic stiffness matrix }
\label{fig_frame_element}
\end{figure}

The proposed methodology treats the constituent members of the honeycomb as one dimensional waveguide wherein longitudinal, flexural and 
shear waves might propagate. This allows us to model the members as frame element (that can model 
axial, flexural and shear motion together) and in order to aptly account for the aforementioned waves, we use a basic rod model and a Timoshenko beam model combindly as depicted in Fig. \ref{fig_frame_element}. Frequency domain spectral finite element, as propounded by Doyle and co-workers \cite{Doyle2,SG},  is an excellent tool to analyse the wave propagation behaviour through the formulation of dynamic stiffness matrix utilising shape functions formed out of the exact solutions. Due to the use of exact shape functions, it can model the inertial force accurately, thereby rendering the method attractive and particularly suitable for high frequency wave propagation. Spectral formulation, which gives directly the dynamic stiffness matrix for a prismatic member- to be used for subsequent analysis, is provided below. 

\subsubsection{Spectral Rod Element Dynamic Stiffness Matrix using Elementary Rod Theory}
Considering the basic rod model involving axial displacement $u$ as a dependent parameter, the governing equation of motion for a prismatic member made of isotropic, homogeneous material with Young's modulus $E$ and 
Shear modulus $G$ and having geometric parameters as length $L$, cross-sectional area $A$ and area moment of inertia $I$, is
   
\begin{equation}
\label{eq:rod}
\frac {\partial}{\partial x} \Bigg[ E A \frac {\partial u}{\partial x} \Bigg] = \rho A \ddot{u}
\end{equation}

Taking the DFT of the variable $\hat{u}(x,t)$ gives the spectral form \cite{Doyle2}
     
\begin{equation}
\label{eq:dft}
u(x,t)={\sum_n} \hat{u}(x,{\omega}_n) {e}^{i {\omega}_n t}
\end{equation}
Substitution of Eqn (\ref{eq:dft}) into Eqn (\ref{eq:rod}), for a rod of uniform cross section,  results in 

\begin{equation}
\label{eq:rod2}
{\sum_n} \bigg[E A  \frac {\partial^2 {\hat{u}}}{\partial x^2} +\rho A {{\omega}_n}^2 {\hat{u}} \bigg] {e}^{i {\omega}_n t}=0
\end{equation}

All ${e}^{i {\omega}_n t}$ being independent,  Eqn. (\ref{eq:rod2}) implies term within the square bracket is zero for every value of integer index n. 	  
 
and the complete solution in this case takes the form

\begin{equation}
\hat{u}(x,\omega_n)= M e^{-i k_1 x}+ N e^{i k_1 x}
\end{equation}

where  

\begin{equation}
 k_1=\omega_n \sqrt{\frac{\rho A}{E A}}
\end{equation}

and $M, N$ are frequency dependent amplitude.  When M and N are presented in terms of the nodal displacement $\hat{u}(0)$ $\hat{u}(L)$, general longitudinal displacement at any arbitrary point can be rewritten as

\begin{equation}
\hat{u}(x)= \hat{g}_1 (x) \hat{u}(0) + \hat{g}_2 (x) \hat{u}(L)
\end{equation}

where

\begin{equation}
\begin{split}
\hat{g}_1 (x)= \big{[} e^{-i k_1 x}-  e^{-i k_1 (2 L-x)} \big{]}/\big{[}1-e^{-i 2 k_1 L}\big{]}   \\
\hat{g}_2 (x)= \big{[}-e^{-i k_1 (L+x)}-  e^{-i k_1 (L-x)}\big{]}/\big{[}1-e^{-i 2 k_1 L}\big{]}
\end{split}
\end{equation}

The internal force can be obtained as 

\begin{equation}
\hat{F}(x) =E A {\frac{\partial{\hat{u}}}{\partial x}}=E A [\hat{g}^{'}_1 (x) \hat{u}(0) + {{\hat{g}}_2}^{'} (x) \hat{u}(L)]
\end{equation}
where  $\hat{g}^{'}_1$ and $\hat{g}^{'}_2$ are first spatial derivitive of $\hat{g}_1$ and $\hat{g}_1$ respectively.

This results in the dynamic stiffness matrix relating the nodal forces $\hat{F}(0)$, $\hat{F}(0)$ to the nodal diplacement $\hat{u}(0)$ $\hat{u}(L)$  as

\begin{equation}
 \begin{Bmatrix}
\hat{F}_1 \\\hat{F}_2
 \end{Bmatrix}  =\frac{E A}{L} \frac{i k_1 L}{(1 - e^{2 i k_1 L} )}
 \begin{bmatrix}
 1 - e^{2 i k_1 L} & -2 e^{2 i k_1 L} \\
 -2 e^{2 i k_1 L} &  1 - e^{2 i k_1 L}
 \end{bmatrix} 
  \begin{Bmatrix}
\hat{u}_1 \\\hat{u}_2
 \end{Bmatrix}
 \end{equation}
 
\subsubsection{Spectral Beam Element Dynamic Stiffness Matrix using Timoshenko Beam Theory}
In similar line, for the same structure, considering Timoshenko Beam model involving two dependent variables $v$ and $\phi$ representing transverse displacement and rotation 
respectively, the governing equations of motion can be written as

\begin{equation} \label{eq:timo1}
\begin{aligned}
G A K_1 \frac {\partial}{\partial x} \Bigg[ \frac {\partial v}{\partial x}  -\phi \Bigg]= \rho A \ddot{v} \\
E I \frac{\partial^2 \phi} {\partial^2 x}+ G A K_1 \Bigg[ \frac {\partial v}{\partial x}  -\phi \Bigg]=\rho I K_2 \ddot{\phi}
\end{aligned}
\end{equation}
 
 and the associated (generalised) forces are 
 \begin{equation} \label{eq:timo_force}
 \begin{split}
V=G A K_1 \Bigg[ \frac {\partial v}{\partial x}  -\phi \Bigg], \\
 M=E I \frac {\partial \phi}{\partial x}
\end{split} 
 \end{equation}

Substitution of an assumed solution for the dependent variables of the form $v = v_0 e^{-i(k x - \omega t )}$ and  $\phi = \phi_0 e^{-i(k x - \omega t )}$ leads to \cite{Doyle2}

\begin{equation} \label{eq:timo_mat0}
\begin{split}
G A K_1 k^2 v_0 e^{-i(k x - \omega t )} -i k G A K_1 \phi_0 e^{-i(k x - \omega t )}= \rho A \omega^2 v_0 e^{-i(k x - \omega t )}  \\
i k G A K_1 v_0 e^{-i(k x - \omega t )}+ (E I k^2 +G A K_1) \phi_0 e^{-i(k x - \omega t )} =\rho I K_2 \omega^2 \phi_0 e^{-i(k x - \omega t )}
\end{split}
\end{equation} 

upon cancellation of exponential term, which is nonzero in general, Eqn. (\ref{eq:timo_mat0}) can be written in a more compact matrix form as 

\begin{equation} \label{eq:timo_mat}
\begin{bmatrix} 
G A K_1 k^2 -\rho A \omega^2 & -i k G A K_1 \\[0.3em]
i k G A K_1 & E I k^2 +G A K_1 -\rho I K_2 \omega^2
\end{bmatrix}
\begin{Bmatrix}
v_0 \\[0.3em]
\phi_0
\end{Bmatrix} =0
\end{equation} 

which gives the characteristic equation
\begin{equation} \label{eq:timo_char}
\begin{split}
[ G A K_1 E I ] k^4 -[G A K_1 \rho I K_2 \omega^2 +E I \rho A \omega^2] k^2\\
+ [\rho I  K_2 \omega^2 - G A K_1] \rho A \omega^2=0
\end{split}
\end{equation} 
which is a standard quadretic equation in $k^2$.
Thus, the solutions of Eqn (\ref{eq:timo_char}) are, $ \pm {k}_1$ and $\pm {k}_2$, $k_1$ and $k_2$ being the wavenumbers given by,  

\begin{equation}
\begin{split}
k_1 = \sqrt{\frac{-Y+\sqrt{Y^2-4XZ}}{2a}} \hspace{0.5in} k_2 = \sqrt{\frac{-Y-\sqrt{Y^2-4XZ}}{2a}} \\
where \\
X=[ G A K_1 E I ] \\
Y=-[G A K_1 \rho I K_2 \omega^2 +E I \rho A \omega^2] \\
and \\ 
Z=[\rho I  K_2 \omega^2 - G A K_1] \rho A \omega^2
\end{split}
\end{equation}

thus, giving spectral relationship for the beam, and the complete solution for the dependent variables takes the form

\begin{equation} \label{eq:timo_sol1}
\begin{split}
{\hat{v}(x)}=  R_1 M e^{-i k_1 x}+ R_2 N  e^{-i k_2 x} + R_1 P e^{i k_1 x} + R_2 Q e^{ik_2 x}\\
{\hat{\phi}(x)}= \hspace{0.4cm} M e^{-i k_1 x}+ \hspace{0.4cm} N e^{-i k_2 x} + \hspace{0.4cm} P e^{i k_1 x} + \hspace{0.4cm} Q e^{ik_2 x}
\end{split}
\end{equation}

where $M,N,P$ and $Q$ are frequency dependent amplitudes and  $R_i, i=1,2$ are the amplitude ratios defined as
  
\begin{equation}
R_i=\frac{i k_i G A K_1}{G A K_1 k_i^2 -\rho A \omega^2}
\end{equation}

Thus displacements at two nodes can be written as

\begin{equation} \label{eq:timo_mat1}
\left\{ \begin{array}{l}
 \hat v_1  \\ 
 \hat \phi _1  \\ 
 \hat v_2  \\ 
 \hat \phi _2  \\ 
 \end{array} \right\} = \left\{ \begin{array}{l}
 \hat v(0) \\ 
 \hat \phi (0) \\ 
 \hat v(L) \\ 
 \hat \phi (L) \\ 
 \end{array} \right\} = [T_1 ]\left\{ \begin{array}{l}
 A \\ 
 B \\ 
 C \\ 
 D \\ 
 \end{array} \right\}
\end{equation}

The aforementioned solution, as in Eqn (\ref{eq:timo_mat1}), enables us to write the constants $A$, $B$, $C$, $D$ in terms of the nodal 
values of the associated degrees of freedom, i.e. $\hat{v}(0)$,  $\hat{\phi}(0)$, $\hat{v}(L)$ $\hat{\phi}(L)$. On the other hand, the generalised 
forces at nodal points, i.e. $V(0)$,  $M(0)$, $V(L)$, $M(L)$ can be obtained upon substituting Eqn (\ref{eq:timo_sol1}) into Eqn 
(\ref{eq:timo_force}).
\begin{equation} \label{eq:timo_mat2}
\left\{ \begin{array}{l}
 \hat V_1  \\ 
 \hat M_1  \\ 
 \hat V_2  \\ 
 \hat M_2  \\ 
 \end{array} \right\} = \left\{ \begin{array}{l}
  - \hat V(0) \\ 
  - \hat M(0) \\ 
 \hat V(L) \\ 
 \hat M(L) \\ 
 \end{array} \right\} = [T_2 ]\left\{ \begin{array}{l}
 A \\ 
 B \\ 
 C \\ 
 D \\ 
 \end{array} \right\}
\end{equation}

This allows us to write generalised forces in terms of associated degrees of freedom  as below and the resulting 4x4 coefficient matrix relating the two
is termed as the dynamic stiffness matrix which is frequency dependent.
\begin{equation} \label{eq:timo_mat3}
\left\{ \begin{array}{l}
 \hat V_1  \\ 
 \hat M_1  \\ 
 \hat V_2  \\ 
 \hat M_2  \\ 
 \end{array} \right\} = [T_2 ][T_1 ]^{ - 1} \left\{ \begin{array}{l}
 \hat v_1  \\ 
 \hat \phi _1  \\ 
 \hat v_2  \\ 
 \hat \phi _2  \\ 
 \end{array} \right\} = [\hat K]\left\{ \begin{array}{l}
 \hat v_1  \\ 
 \hat \phi _1  \\ 
 \hat v_2  \\ 
 \hat \phi _2  \\ 
 \end{array} \right\},\hspace{0.5cm} [\hat K] = [T_2 ][T_1 ]^{ - 1} 
\end{equation}

 The elements of the $[\hat{\bf{K}}]$ are as follows \cite{Doyle2, Gopalakrishnan}
 \begin{equation}
 \begin{split}
 {\hat{k}}_{11}= \left( {k_2}^2 - {k_1}^2 \right) \left( {r_2}{z_{22}} - {r_1}{z_{21}} \right)/ {\bf{\Delta}} \\
 {\hat{k}}_{13}= \left( {k_1}^2 - {k_2}^2 \right) \left( {r_1}{z_{21}} - {r_2}{z_{22}} \right)/ \bf{\Delta} \\ 
 {\hat{k}}_{12}=[-i k_2 \left( {r_1}{z_{11}} + {r_2}{z_{12}} \right) +i k_1\left( {r_1}{z_{11}} - {r_2}{z_{12}} \right)]/ \bf{\Delta} \\ 
  {\hat{k}}_{14}=[-i k_1 \left( {r_1}{z_{11}} - {r_2}{z_{11}} \right) +i k_1\left( {r_1}{z_{11}} - {r_2}{z_{12}} \right)]/ \bf{\Delta} \\ 
  {\hat{k}}_{22}=\left(-i k_1 R_2 + i K_2 R_1 \right) \left( {r_1}{z_{22}} - {r_2}{z_{21}} \right) / \bf{\Delta} \\   
  {\hat{k}}_{24}=\left(i k_1 R_2 - i K_2 R_1 \right) \left( {r_1}{z_{21}} - {r_2}{z_{22}} \right) / \bf{\Delta} \\ 
   {\hat{k}}_{23}= -{\hat{k}}_{14}, \hspace{0.2in} {\hat{k}}_{33}= {\hat{k}}_{11}, \hspace{0.2in}  {\hat{k}}_{34}= -{\hat{k}}_{12},  \hspace{0.2in} {\hat{k}}_{44}= {\hat{k}}_{22}\\
   where \\  
   r_1=\left( R_1 -R_2 \right) z_{11},  \hspace{0.2in} r_2=\left( R_1 + R_2 \right) z_{12}, \hspace{0.2in}  \Delta=\left({r_1}^2-{r_2}^2 \right) /EI \\
   z_{11}=1-e^{-i k_1 L} e^{-i k_2 L} , \hspace{0.2in}  z_{12}=e^{-i k_1 L}- e^{-i k_2 L} \\
   z_{22}=1+e^{-i k_1 L} e^{-i k_2 L} , \hspace{0.2in}  z_{12}=e^{-i k_1 L}+ e^{-i k_2 L}
 \end{split}
 \end{equation}

\subsection{Free Vibration Analysis} 

\begin{figure}[!ht]
\includegraphics[width=3.0in]{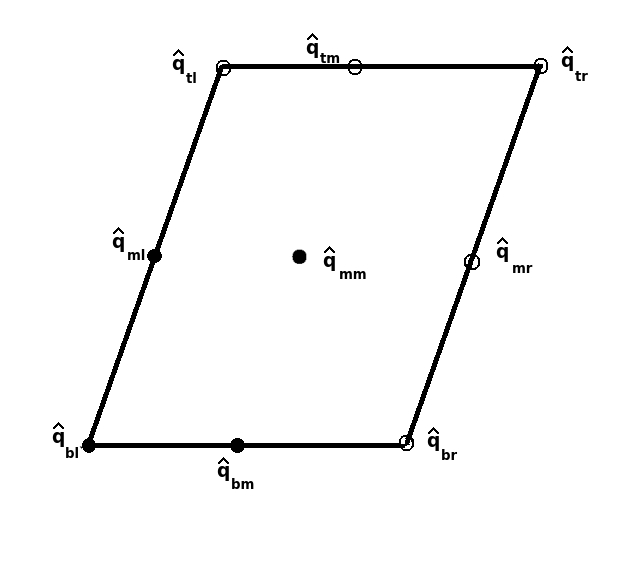}
\caption{Schematic representation of a generic unit cell with associated deformation vectors shown as beads, the solid beads may be chosen to be independent degrees of freedom and the hollow ones to be dependent on them}
\label{fig_schm_uc}
\end{figure}

Once the dynamic stiffness matrices have been obtained for all the individual members of the unit cell, those can be assembled to generate the global dynamic stiffness 
matrix for the whole unit cell. Thus, in spectral domain, if the nodal degrees of freedoms for a particular node is represented as ${\bf{\hat{q}}}_j = [\hspace{0.05cm} \hat{u}_j \hspace{0.2cm} \hat{v}_j\hspace{0.2cm} {\hat{\theta}}_j \hspace{0.05cm} ]^T$, 
corresponding equation of motion in the spectral domain can be written as 

\begin{equation}
 \hat{\bf{K}}{\hat{\bf{q}}}={\hat{\bf{f}}}
\end{equation}
 where $\hat{\bf{K}} $ is the aforesaid global  dynamic stiffness matrix for the unit cell $\hat{\bf{q}}$ is the displacement vector combining all ${\bf{\hat{q}}}_j$ for the unit cell and 
$ {\hat{\bf{f}}}$ is the corresponsing spectral force vector. In the equation above and hereafer, boldface small letters have been used to indicate vector and boldface capitals to indicate matrix. In line with the conventional schematic representation of a generic unit cell, nodal spectral displacements are
sub-suffixed with two indices based on the position of the nodes, first one presenting top (t), middle (m) or bottom (b) and second one presenting left (l), middle (m) or right (r)
, as  depicted in Fig. \ref{fig_schm_uc}. Thus, $\hat{\bf{q}}_{br}$ would indicate the vector for spectral displacement of the bottom right juction of a unit cell. Bloch theory provides their interrelationship as below,

\begin{equation}
 \begin{split}
  \hat{\bf{q}}_{mr}=e^{i\mu_1} \hat{\bf{q}}_{ml}, \hspace{0.5cm} \hat{\bf{q}}_{tm}=e^{i\mu_2} \hat{\bf{q}}_{bm},\\
  \space  \hat{\bf{q}}_{br}=e^{i\mu_1} \hat{\bf{q}}_{bl},\hspace{0.5cm}  \hat{\bf{q}}_{tr}=e^{i({\mu_1+\mu_2})} \hat{\bf{q}}_{bl} ,\hspace{0.5cm}  \hat{\bf{q}}_{tl}=e^{i\mu_2} \hat{\bf{q}}_{bl} 
 \end{split}
\end{equation}

In similar fashion, spectrally decomposed forces are interrelated as below

\begin{equation}
 \begin{split}
  \hat{\bf{f}}_{mr}=e^{i\mu_1} \hat{\bf{f}}_{ml}, \hspace{0.5cm} \hat{\bf{f}}_{tm}=e^{i\mu_2} \hat{\bf{f}}_{bm},\\
  \space  \hat{\bf{f}}_{br}=e^{i\mu_1} \hat{\bf{f}}_{bl},\hspace{0.5cm}  \hat{\bf{f}}_{tr}=e^{i({\mu_1+\mu_2})} \hat{\bf{F}}_{bl} ,\hspace{0.5cm}  \hat{\bf{f}}_{tl}=e^{i\mu_2} \hat{\bf{f}}_{bl} 
 \end{split}
\end{equation}

Due to force equilabrium, the forces satisfy the equation

\begin{equation}
 \hat{\bf{f}}_{tr}+e^{i\mu_1} \hat{\bf{f}}_{tl}+ e^{i\mu_2} \hat{\bf{f}}_{br}+ e^{i({\mu_1+\mu_2})} \hat{\bf{f}}_{bl}=0  
\end{equation}

It follows, after enforcing Bloch condition, that

 \begin{equation}
 \begin{split}
 {\hat{\bf{q}}}={\bf{T}}{\hat{\bf{q}}_{r}}  \hspace{1.5in} \\
where \hspace{1.67in} \\
{\bf{T}}=\begin{bmatrix}
   I & 0 & 0 & 0 \\
  I e^{i{\mu}_1} & 0 & 0 & 0 \\
  0 & I & 0 & 0 \\
  0 & I e^{i{\mu}_2} & 0 & 0 \\
   0 & 0 & I & 0 \\
  0 & 0 & I e^{i{\mu}_1} & 0 \\
  0 & 0 & I e^{i{\mu}_2} & 0 \\
  0 & 0 & I e^{i({\mu}_1+{\mu}_2)} & 0 \\
  0 & 0 & 0 & I 
  \end{bmatrix}
  and \hspace{0.5cm}
  {\hat{\bf{q}}_{r}} =
  \begin{Bmatrix}
   {\hat{\bf{q}}_{ml}} \\ 
  {\hat{\bf{q}}_{bm}} \\
  {\hat{\bf{q}}_{bl}} \\
  {\hat{\bf{q}}_{mm}} \\
  \end{Bmatrix}
  \end{split}
 \end{equation}

 resulting in,

\begin{equation}
 \hat{\bf{K}}_r{\hat{\bf{q}}_r}={\hat{\bf{f}}_r} \hspace{0.5cm} where \hspace{0.5cm} \hat{\bf{K}}_r={{\bf{T}}^{H}}\hat{\bf{K}}{\bf{T}} \hspace{0.5cm} and \hspace{0.5cm}    {\hat{\bf{f}}_r}={{\bf{T}}^{H}}{\hat{\bf{f}}}
\end{equation}

For the free vibration analysis, this leads to,
\begin{equation}
 \hat{\bf{K}}_r{\hat{\bf{q}}_r}=0
\end{equation}

\subsection{Wittrick Williams Algorithm}
Unlike dynamic stiffness matrix of the form $\bf{K} -\omega^2 \bf{M}$, obtained from conventional finite element method which has got an explicit expression for frequency, the dynamic stiffness method obtained with the help of spectral finite element method involves frequency implicitly. Thus, once Bloch formulation is applied to bring in the two wavenumbers corresponding to the unit vector directions of the reciprocal lattice, the resulting eigenvalue problem involves the explicit wavenumber variables and the implicit frequency. This leads to a difficulty in systematic solution of the equation and generally ad-hoc methods have been used in such a situation, however, apart from being conspicuously inefficient, the ad-hoc methods bear the risk of missing certain frequencies particularly so for the closedly spaced ones.

To get rid of such complications we have used Wittrick-Williams algorithm  which not only allows to obtain frequencies iteratively to the desired accuracy in a systematic fashion, but also ensures that all frequencies are accounted for, including closed spaced or repeated frequencies.

As per Wittrick-Williams method, the number of natural frequencies of the system lying below a certain frequency $\omega^*$ can be written as  
\begin{equation}
J(\omega^*)=J_0(\omega^*)+S(K(\omega^*))
\end{equation}

\noindent where, 

$J (\omega^* )$ is the number of natural frequency below    $\omega^*$

$J_0 (\omega^* )$ is the number of natural frequency of the constrained system below $\omega_*$

$S ( K (\omega^* ))$  is the sign count, i.e. the number of sign changes of principal minors
$K _1 (\omega^* ), K _2 (\omega^* ), K _3 (\omega^* ).... K _r (\omega ^*)$

 The principle minor of order $m$ is the determinant of the matrix comprising of first $m$ rows and columns.  The sign change is also equal to the number of negative diagonal entries once the matrix is brought to upper diagonal form. 

 Number of natural frequency of the constrained system refers to the number of natural frequency of the system when all the associated nodal degrees of freedom have been constrained. Thus, it would be equal to the total number of fixed-fixed natural frequency of all the elements when each element is treated as frame element.  For Timoshenko beam, these can be obtained numerically, by solving for $b$ , the non-dimensionalise frequencies and hence for $\omega$, the intended natural frequencies, in the equation below \cite{Huang} 
 \begin{equation}
 \begin{split}
 2-2 cosh(b \alpha) cos(b \beta)+ \frac{b [b^2 s^2 (r^2-s^2)^2+(3 s^2-r^2)]}{(1-b^2 r^2 s^2)^\frac{1}{2}}  sinh(b \alpha) sin(b \beta)      =0  \\
  where  \hspace{0.1in}                                          
 \phi=K_1 A G  ,        \hspace{0.15in}
  b^2=\rho A L^4 \omega^2/(E I) ,        \hspace{0.15in}  
  r^2=I/(A L^2)   ,        \hspace{0.15in}
  s^2=E I/(\phi L^2)   \\
 \alpha=\frac{1}{\sqrt{2}} \bigg{[}-(r^2 + s^2)+\big{[}(r^2 - s^2)^2+(4/b^2)\big{]}^{0.5}\bigg{]}^{0.5}  \\ 
  when \hspace{0.2in}
  \big{[}(r^2 - s^2)^2+(4/b^2)\big{]}^{0.5} >(r^2 + s^2) \\ 
  \alpha=\frac{1}{\sqrt{2}} \bigg{[}(r^2 + s^2)+\big{[}(r^2 - s^2)^2+(4/b^2)\big{]}^{0.5}\bigg{]}^{0.5} \\
 when \hspace{0.2in}
  \big{[}(r^2 - s^2)^2+(4/b^2)\big{]}^{0.5} <=(r^2 + s^2) \\
 \beta=\frac{1}{\sqrt{2}} \bigg{[}(r^2 + s^2)+\big{[}(r^2 - s^2)^2+(4/b^2)\big{]}^{0.5}\bigg{]}^{0.5}     \\
 \end{split}
 \end{equation}
 
 For the rod, i.e. for the longitudinal vibration, this comes as
 \begin{equation}
  \omega_n= n \pi \sqrt{(E/ \rho)}
 \end{equation}

 Aforesaid methods put together, allow for the calculation of band structure, more efficiently- as for the same desired level of accuracy, than the conventional method of calculating of band structure through the use of conventional finite element. 
 
 The band structure, as customary, is represented along the boundary of the irreducible Brillouin zone (IBZ). It might be alluded here that it has been shown in literature that this presentation might not vouch for correct calculation of Band gaps, wherein the minimum band gaps have been shown to be occuring somewhere within the IBZ, the proposed is no way tied to this particular presentation and may very well be used to calculate band structure for the whole area within IBZ, if desired. 
 
 \subsection{Supercell Method and Defects}
\begin{figure}[!ht]
\centering
\subfloat[]{
\label{fig_supercell1:a}
\begin{minipage}{5.5in}
\centering\includegraphics[width=5.5in]{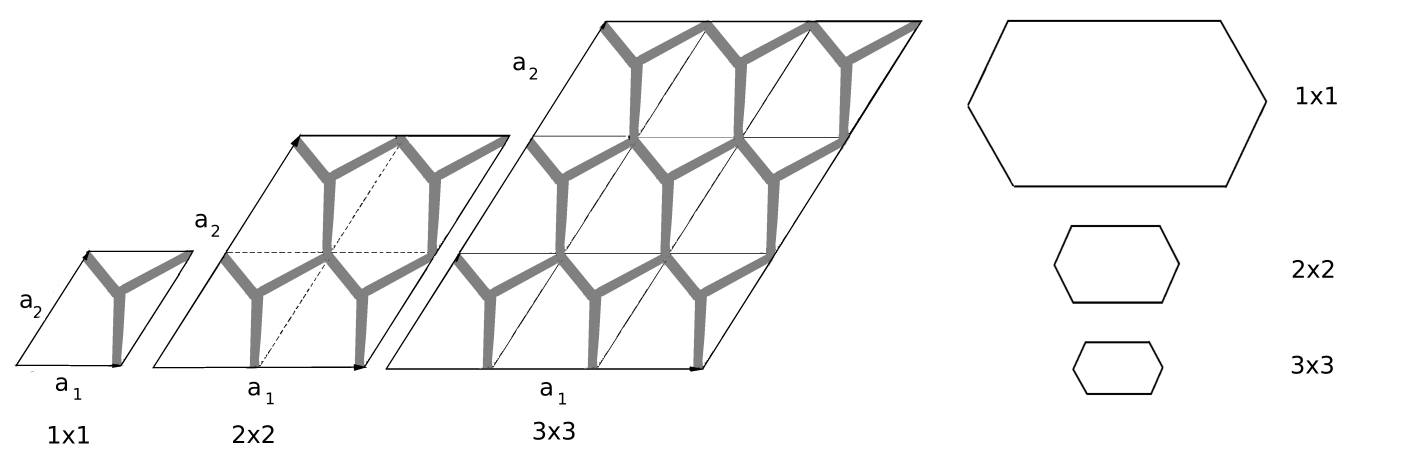}
\end{minipage}}
\vspace{0.1in}
\subfloat[]{
\label{fig_supercell1:b}
\begin{minipage}{3.0in}
\centering\includegraphics[width=3.0in]{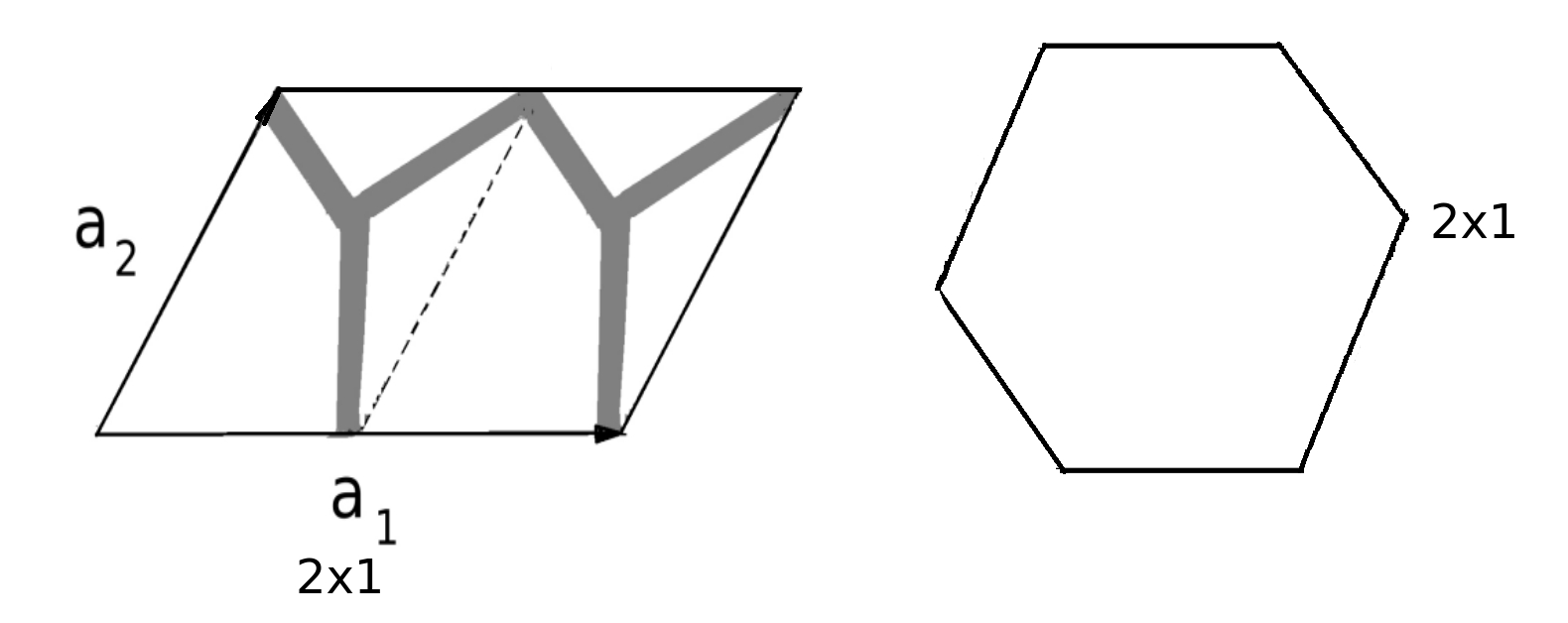}
\end{minipage}}
\caption{(a) Primitive unit cell, 2x2 supercell and 3X3 supercell for the arbritrary hexagonal honeycomb shown in Fig. \ref{fig_schm_uc1:a} and the corresponding Brillouine zones all having same shape.
(b) A 2x1 supercell for the same lattice with associated Brillouin zone, the Brilloiun zone shape is different from those in Fig. \ref{fig_supercell1:a} }
\label{fig_supercell1}
\end{figure}

 It is possible, instead of considering a primitive unit cell- comprising the smallest possible repeatitive unit, to consider a nonprimitive unit cell formed out of multiple primitive unit cells, as shown in Fig. \ref{fig_supercell1}, as the constituent unit for the analysis of periodic structure. Though this artifact, in spite of the underying physical structure having remained the same, changes the band structure of the model due to the enforcement of periodicity related constraints at a higher level , it brings out certain intersting characteristic of the vibration phenomenon of such strutures and sets in the premises for the comparision of the band structure at multiple levels, thereby enabling us with a powerful mechanism for analytical treatment of defects. 
 
 Supercell method has been in vougue in solid state physics but has rarely been found to have applications in mechanical vibration. Supercell of a two dimensional lattice is made upon, first choosing a primitive unit cell and then taking certain muliples of unit cell vectors on both the direction as the new unit cell vectors for the newly formed unit cell, thus forming a new Brillouin zone. Though it is possible to take unequal number of multiples along two primitive unit cell vector direction, taking equal value for the two has got the particular advantage of retaining the shape of the Brillouin zone, thus making room for a comparison for the band structure at different levels. Hereafter in this paper, only $n$x$n$ supercell is considered.  
 
Once the supercell method is established, along with the mechanism to compare them at different levels, the stage is set for the treatment of defects. Supercell method, as aforesaid,  has been used in solid state physics to handle defects in  materials. In the context of engineering structures, literature on the treatment of defects in periodic structure is not vast.  There have been works \cite{Bendiksen}  on the defects in structure having rotational periodicity, e.g. turbines and gears, which allow complete description of defects because of finiteness of the structure, to demonstrate mode localization, first proposed by Anderson \cite{Anderson} in his seminal contribution dealing with transport properties of disordered solids.. Hodges \cite{Hodges} analysed mode localization for a system of coupled pendula and for a vibrating string with mass and spring constraint. There are papers on the mathematical treatment of mode localizations. Supercell allows treatment of defects in an infinite structure with periodicity with the assumption of a periodic nature of defects. It is understood that a finite number of defects in an otherwise infinite periodic structure would not bring any effect except in the viscinity of the defects. Therefore a periodic distribution of defects become imperative for the treatment of periodic structure with infinite spread.

 \section{Results}
 
The proposed method has been demonstrated with three different lattice types, namely hexagonal honeycomb, kagome and square. The first two have got a reciprocal lattice unit cell of the shape of regular hexagon and the same for a square lattice is a square. The details of the geometric and material properties of the constituent members of all the lattices used for the examples have been given in Table \ref{table_example}.

\begin{table}[!h]
\centering
\caption{Material and Geometric properties of the constituent members of the honeycombs}
\label{table_example}
\begin{tabular}{llllll}
\hline
\centering
Density & Young'	s modulus & Poisson's Ratio & Length & Width & Depth \\
\centering
(kg/m) &    (GPA) & & (m) & (m) & (m) \\
\hline \\
25000 & $210*10^9$ &0.25 &0.125 & 0.0125 & 0.0125 \\\\
\hline
\end{tabular}
\end{table}

\subsection{Primitive Unit Cell}
\begin{figure}[!hp]
\centering
\subfloat[]{
\label{fig_hex:a}
\begin{minipage}{3.5in}
\centering\includegraphics[width=3.5in]{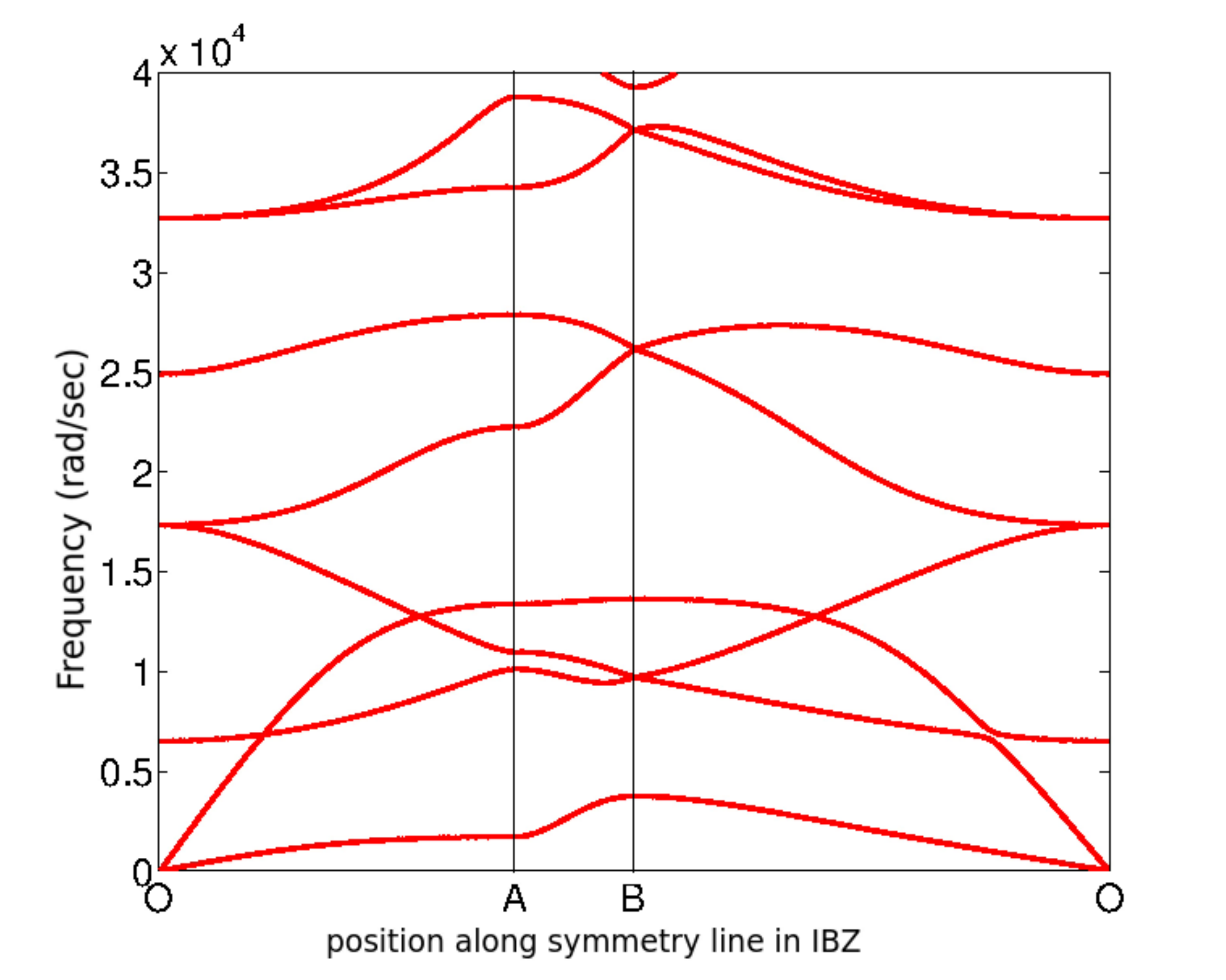}
\end{minipage}}
\subfloat[]{
\label{fig_hex:b}
\begin{minipage}{1.5in}
\centering
\vspace{0.4in}
\includegraphics[width=1.0in]{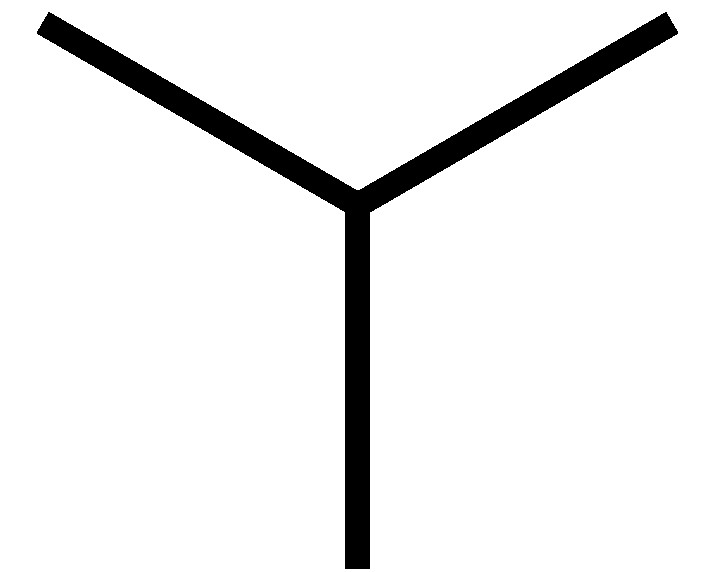}
\vspace{0.1in}
\includegraphics[width=1.5in]{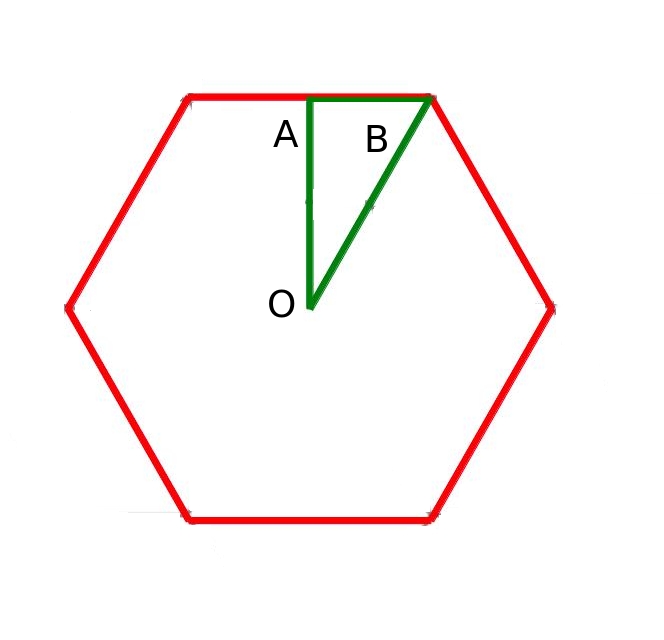}
\end{minipage}}
\caption{(a) Band structure of the hexagonal honeycomb using primitive unit cell. (b) the primive unit cell of the hexagonal honeycomb used for calculation (top) and the first Brillouin zone with the tiangular irreducible Brilloun zone marked as OAB (bottom)  }
\label{fig_hex}
\end{figure}

First, the band structure for the hexagonal honeycomb with the primitive unit cell has been been generated. Fig. \ref{fig_hex}(a) shows the band structure for the hexagonal lattice computed along the IBZ boundary, as shown in the bottom subfigure in Fig. \ref{fig_hex}(b) by the triangle ABC marked within the hexagonal first Brillouin zone. The primitive cell used is shown in the top subfigure in  Fig. \ref{fig_hex}(b). For the purpose of calculation, a single spectral frame element has been used between two joints totaling only three elements for the whole unit cell. This is, in comparison with around 30 elements \cite{Leamy} needed for the same unit cell through conventional finite element based calculation, as in \cite{Phani}, requiring even finer elements for higher frequencies. This gives significant computational advantage in reducing the problem size drastically. So, though spectral based formulation involves calculation of dynamic stiffness matrix for each frequency, the reduction in matrix size convincingly outweighs the additional cost. 

\begin{figure}[!hp]
\centering
\subfloat[]{
\label{fig_kag:a}
\begin{minipage}{3.5in}
\centering\includegraphics[width=3.5in]{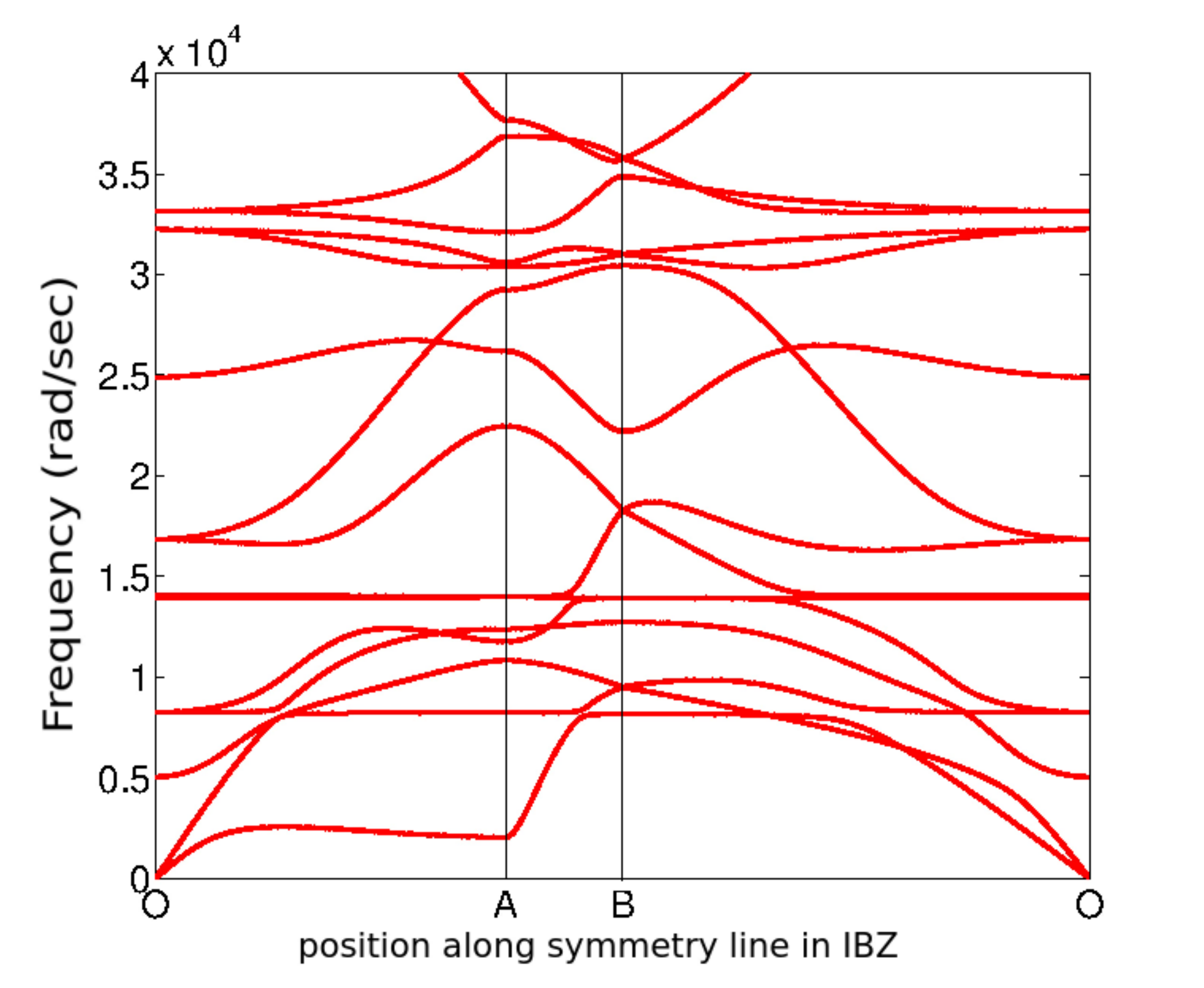}
\end{minipage}}
\subfloat[]{
\label{fig_kag:b}
\begin{minipage}{1.5in}
\centering
\includegraphics[width=0.95in]{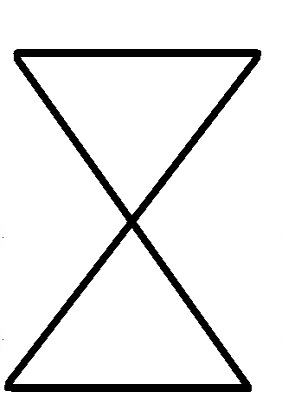}
\includegraphics[width=1.5in]{f1_1x1_hexagon_brillouin.jpg}
\end{minipage}}
\caption{Band structure computed using primitive unit cell of a kagome lattice (b) the primive unit cell of the kagome lattice used for calculation (top) and the first Brillouin zone with the triangular irreducible Brilloun zone marked as OAB (bottom)  }
\label{fig_kag}
\end{figure}

Susequently, the band structure for the primitive cell for the kagome lattice has been generated as shown in Fig. \ref{fig_kag}(a). The unit cell comprices six frame elements, as in Fig. \ref{fig_kag}(b) top subfigure. The Brillouin zone and the IBZ for this lattice, shown in Fig. \ref{fig_kag}(b) bottom subfigure, is same as that of hexagonal honeycomb.

\begin{figure}[!htp]
\centering
\subfloat[]{
\label{fig_square:a}
\begin{minipage}{3.5in}
\centering\includegraphics[width=3.5in]{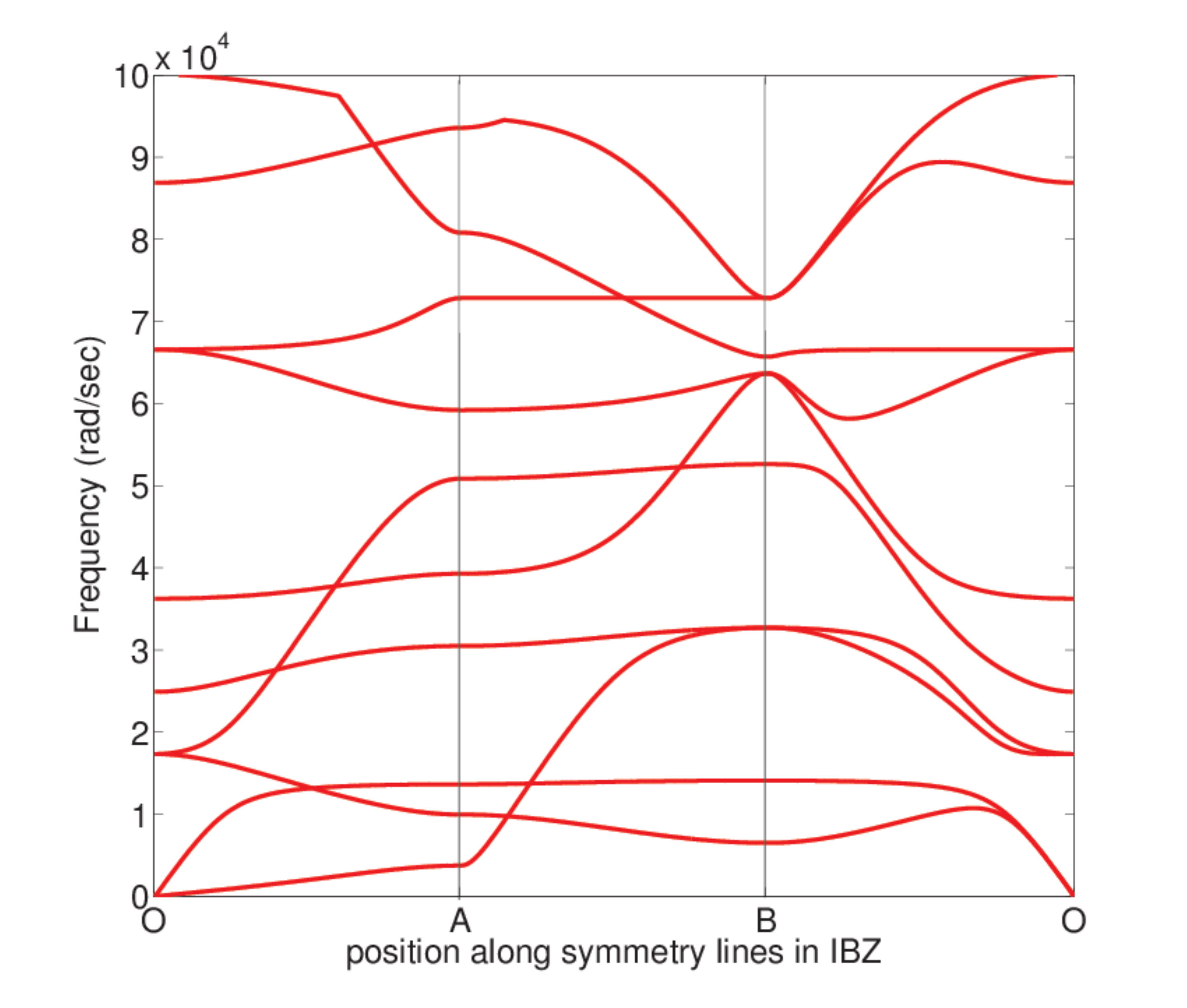}
\end{minipage}}
\subfloat[]{
\label{fig_square:b}
\begin{minipage}{1.5in}
\centering
\includegraphics[width=0.95in]{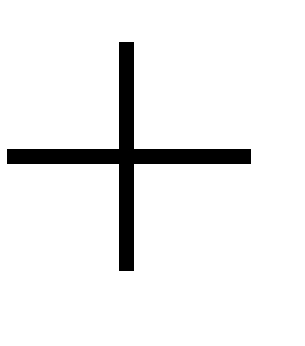}
\includegraphics[width=1.5in]{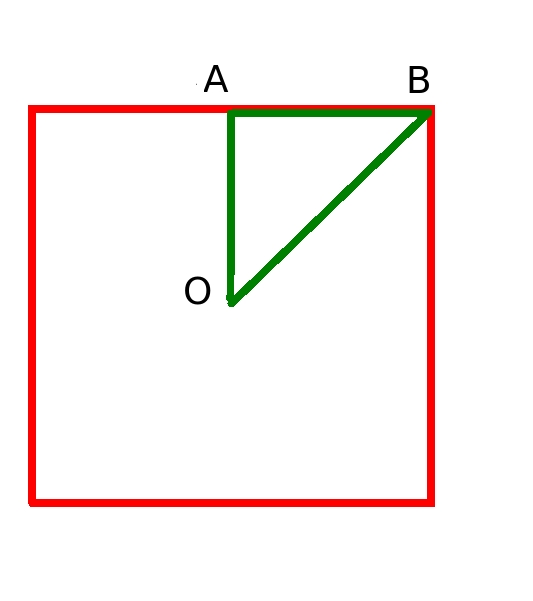}
\end{minipage}}
\caption{Band structure computed using primitive unit cell of a square lattice (b) the primive unit cell of the square lattice used for calculation (top) and the first Brillouin zone with the triangular irreducible Brilloun zone marked as OAB (bottom) }
\label{fig_square}
\end{figure}

The band structure of square lattice- the other lattice considered here, is obtained next and presented in Fig. \ref{fig_square}(a) along with the primitive lattice used- Fig. \ref{fig_square}(b) top subfigure, and the square Billouine zone and the triangular IBZ ABC marked therein- as shown in Fig. \ref{fig_square}(b) bottom subfigure. 

These results will be used as a basis for our analysis of supercell with or without defects to identify the spurious bands. 

 \pagebreak

\subsection{Supercell}
Rather than being related to the physical lattice itself, the analysis and results pertaining to the supercell band structure expounded here relate more congruently to the shape of the Brillouin zone, and  hence subgrouped acordingly.

\subsubsection{Hexagonal Brillouin Zone}
 
The band structures of the hexagonal honeycomb supercell, with 2x2 primitive unit cells, and the kagome lattice supercell with 3x3 primitive unit cells  have been generated for the purpose of demonstration of the concepts and  subsequently the $n$x$n$ case has been analysed. 

\begin{figure}[h]
\centering
\subfloat[]{
\label{fig_hex2:a}
\begin{minipage}{2.7in}
\includegraphics[width=2.7in]{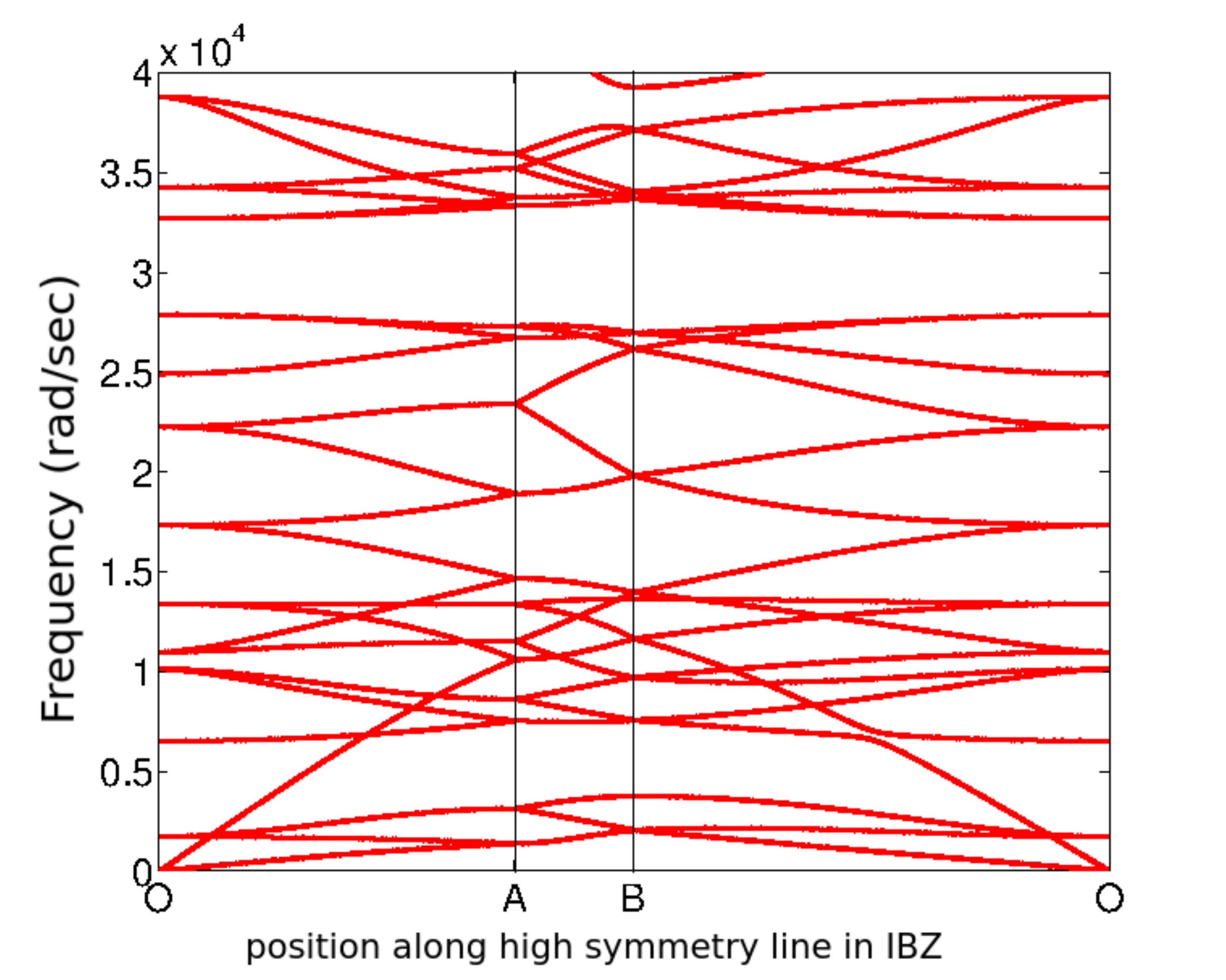}
\end{minipage}}
\subfloat[]{
\label{fig_hex2:b}
\begin{minipage}{2.7in}
\includegraphics[width=2.7in]{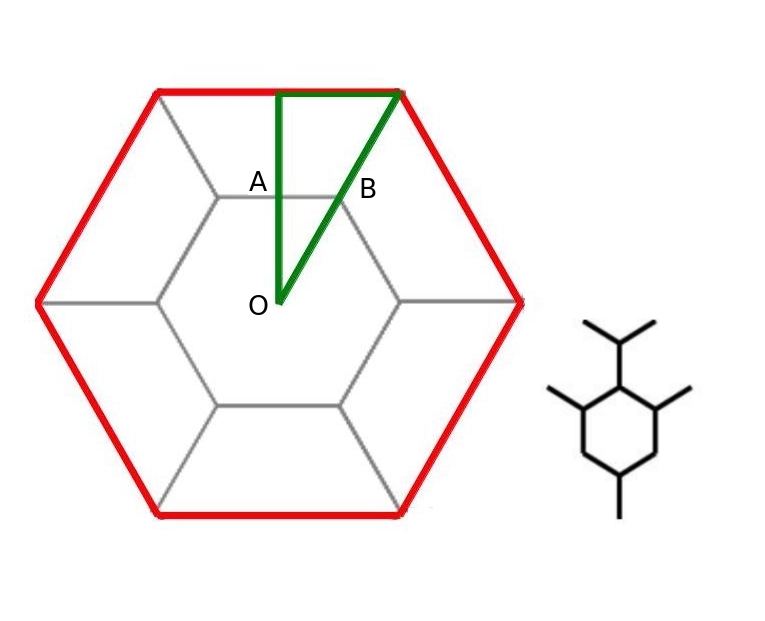}
\end{minipage}}
\vspace{0.01in}
\subfloat[]{
\begin{minipage}{2.7in}
\label{fig_hex2:c}
\includegraphics[width=2.7in]{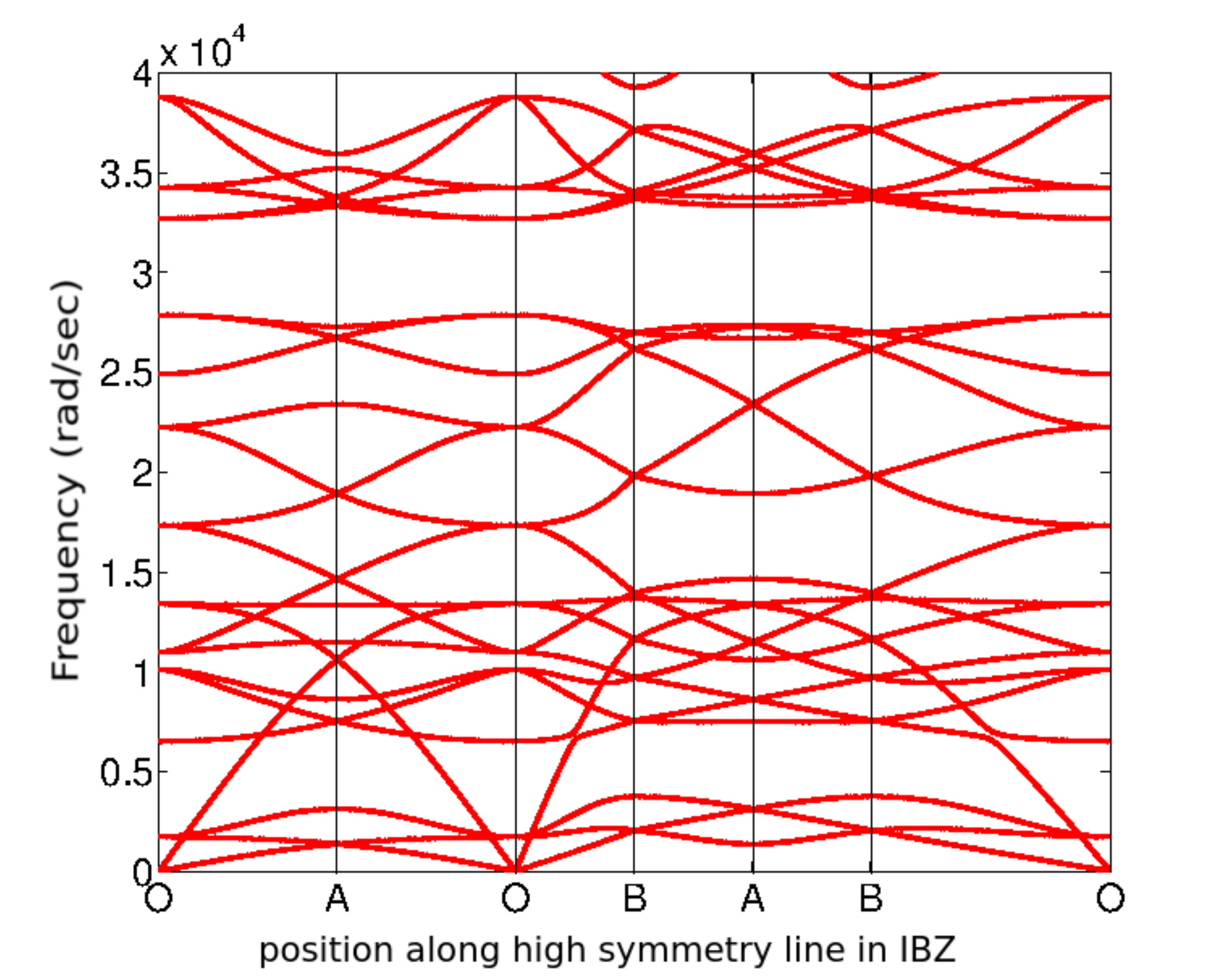}
\end{minipage}}
\subfloat[]{
\begin{minipage}{2.7in}
\label{fig_hex2:d}
\includegraphics[width=2.7in]{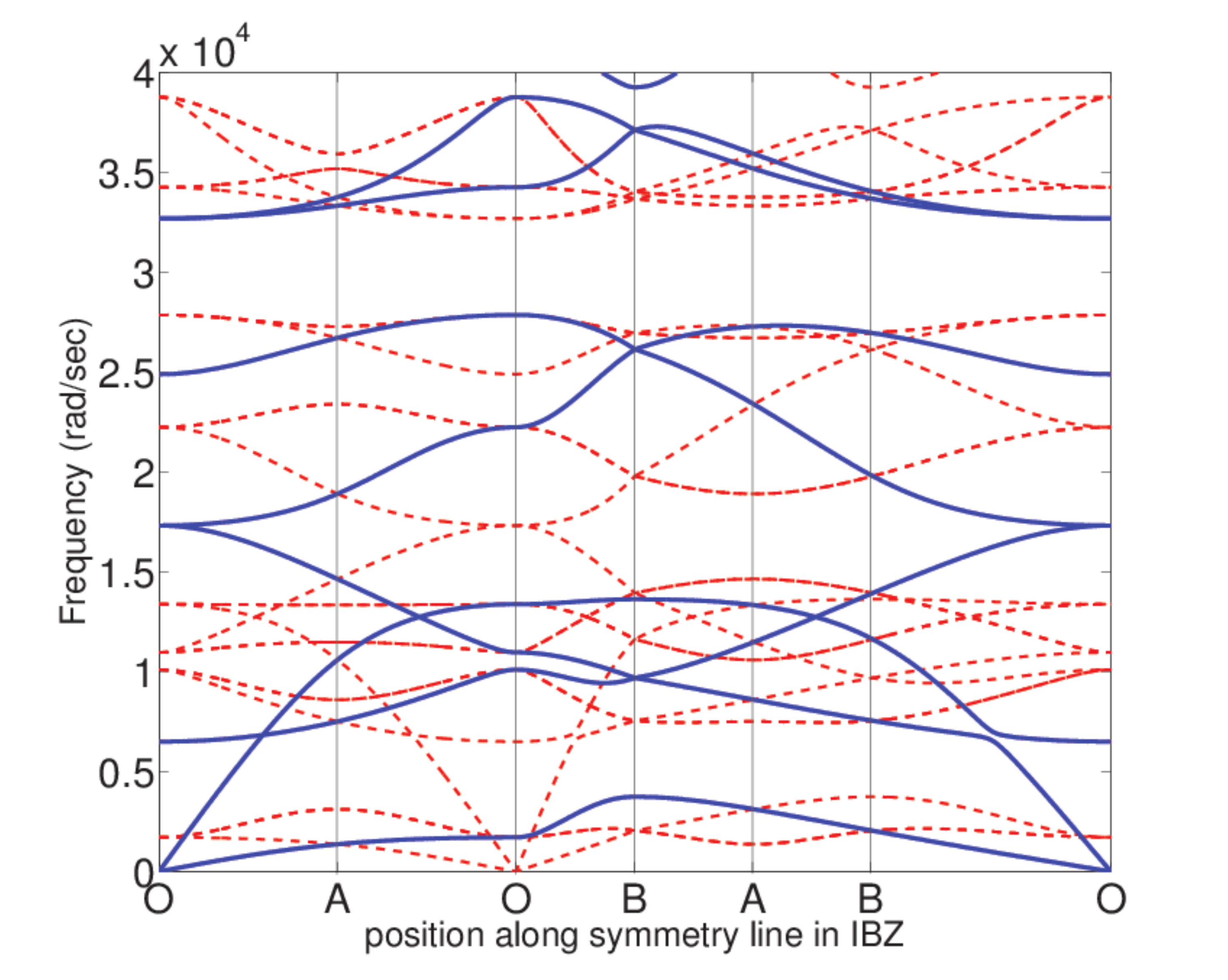}
\end{minipage}}
\caption{(a) Band structure of the hexagonal honeycomb using 2x2 supercell. (b) The Brillouin zone for the supercell and the primitive cell. (left) and the supercell used for calculation (right) (c) The juxtaposed band structure to compare with the primitive cell band structure, (d) The same figure as in Fig. \ref{fig_hex2:c} with overlaping primitive unit cell band structure }
\label{fig_hex2}
\end{figure}

Fig. \ref{fig_hex2}(a) shows the band structure for the 2x2 supercell of a hexagonal honeycomb lattice and the respective unit cell is shown in Fig. \ref{fig_hex2}(b) right subfigure; this band structure is different from the one in  Fig. \ref{fig_hex}(a). The left subfigure in Fig. \ref{fig_hex2}(b)  shows the supercell Brillouin zone- the inner hexagon,  as against the primitive cell Brillouin zone- the outer hexagon. Toward addressing the anomaly in band structure at different levels of supercell, invariably arising out of the fact of different supercell representing different area in k-space as the first Brillouine zone- the Brillouine zone shape remaining unchanged notwithstanding, it would be legitimate to ask for the possibility to reconstruct the band structure of the primitive cell from the band structure of the supercell. The answer would be in affermative, and to this end, it can be seen that a particular way of juxtaposition of the zones along the boundary line of the Irriducible Brillouin Zone (IBZ), can generate the bands as in the primitive cell.  The first Brillouin zone of the supercell along with proximal halvs of each of the six neighbouring Brillouin zones of the same constitute the first Brillouin zone of the primitive cell. Consiquently, the IBZ of the primitive cell, the bigger triangle marked, is composed of the smaller triangle OAB indicating the supercell IBZ and three more pieces of the same (accounting for two reflected ones). Hence the primitive cell IBZ boundary is composed of the elements in supercell IBZ boundary. This makes the bands in primitive cell band structure, drawn along it's IBZ boundary, have a coincident band in a composite band structure made of a particular ordered juxtaposition of different segments in supercell band structure presented along its IBZ boundary. As seen in Fig. \ref{fig_hex2}(b) left subfigure, based on the previously used nomenclature (i.e. the centre of IBZ as O, the mid point of side of BZ as A and the vertex as B), the juxtaposition for this case is OA-AO-OB-BA-AB-BO. Fig. \ref{fig_hex2}(c) shows the juxtaposed band structure to compare with the primitive cell band structure, and Fig. \ref{fig_hex2}(d) shows the same with overlaping primitive unit cell band structure.

\begin{figure}[]
\centering
\subfloat[]{
\label{fig_kag3:a}
\begin{minipage}{2.7in}
\includegraphics[width=2.7in]{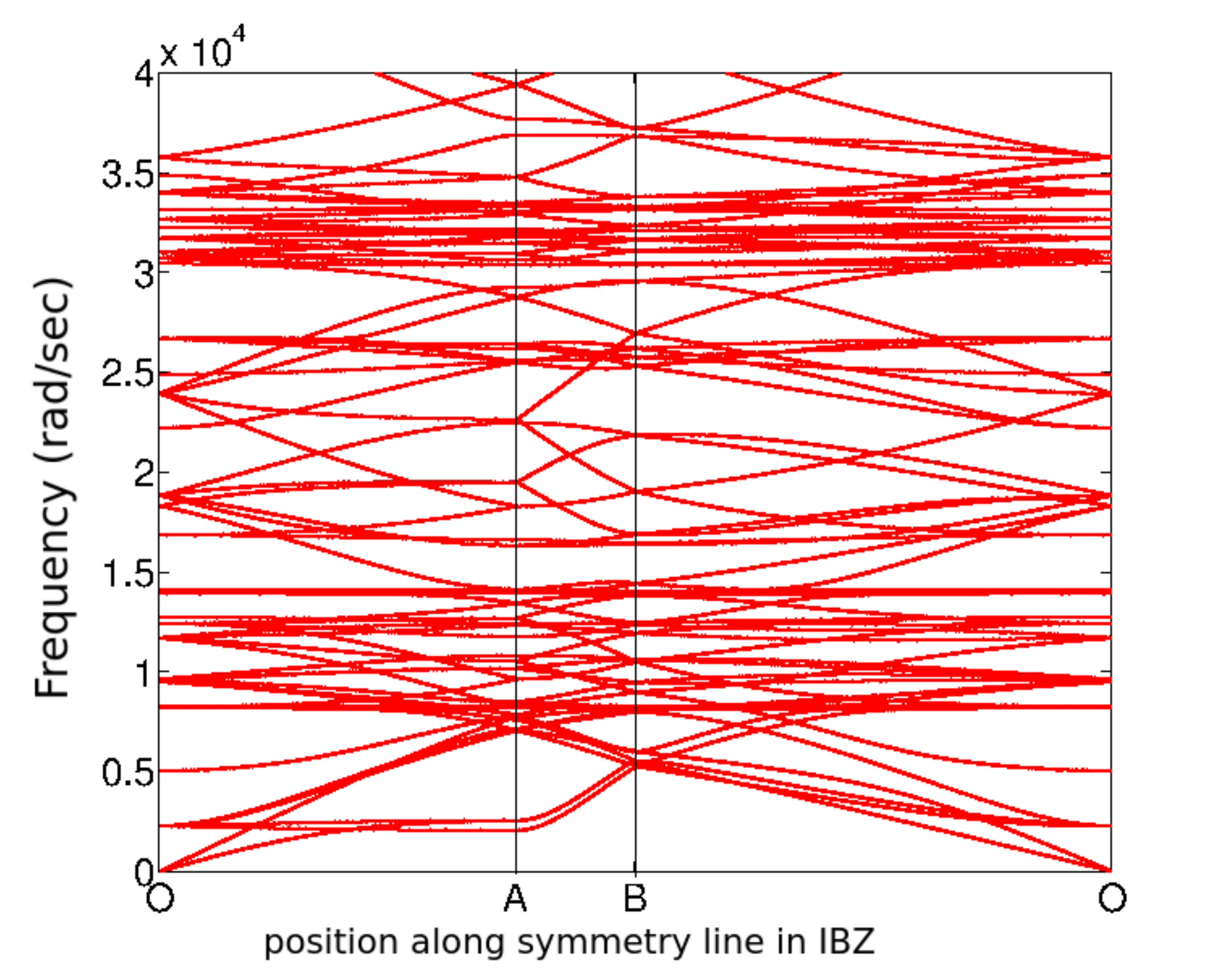}
\end{minipage}}
\subfloat[]{
\label{fig_kag3:b}
\begin{minipage}{2.7in}
\includegraphics[width=2.7in]{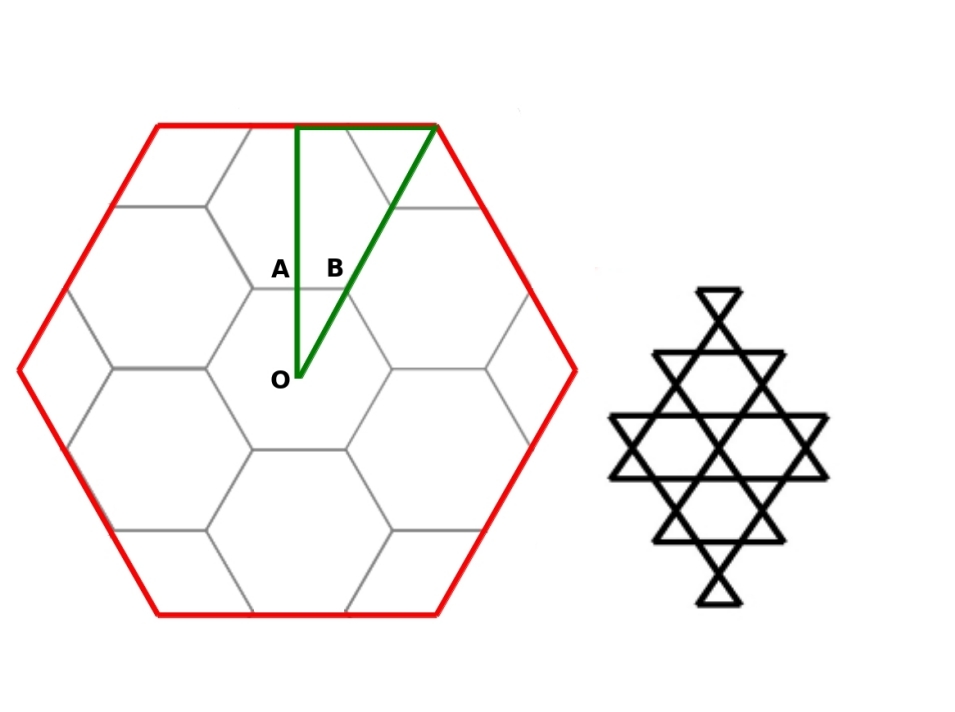}
\end{minipage}}
\vspace{0.01in}
\subfloat[]{
\begin{minipage}{2.7in}
\label{fig_kag3:c}
\includegraphics[width=2.7in]{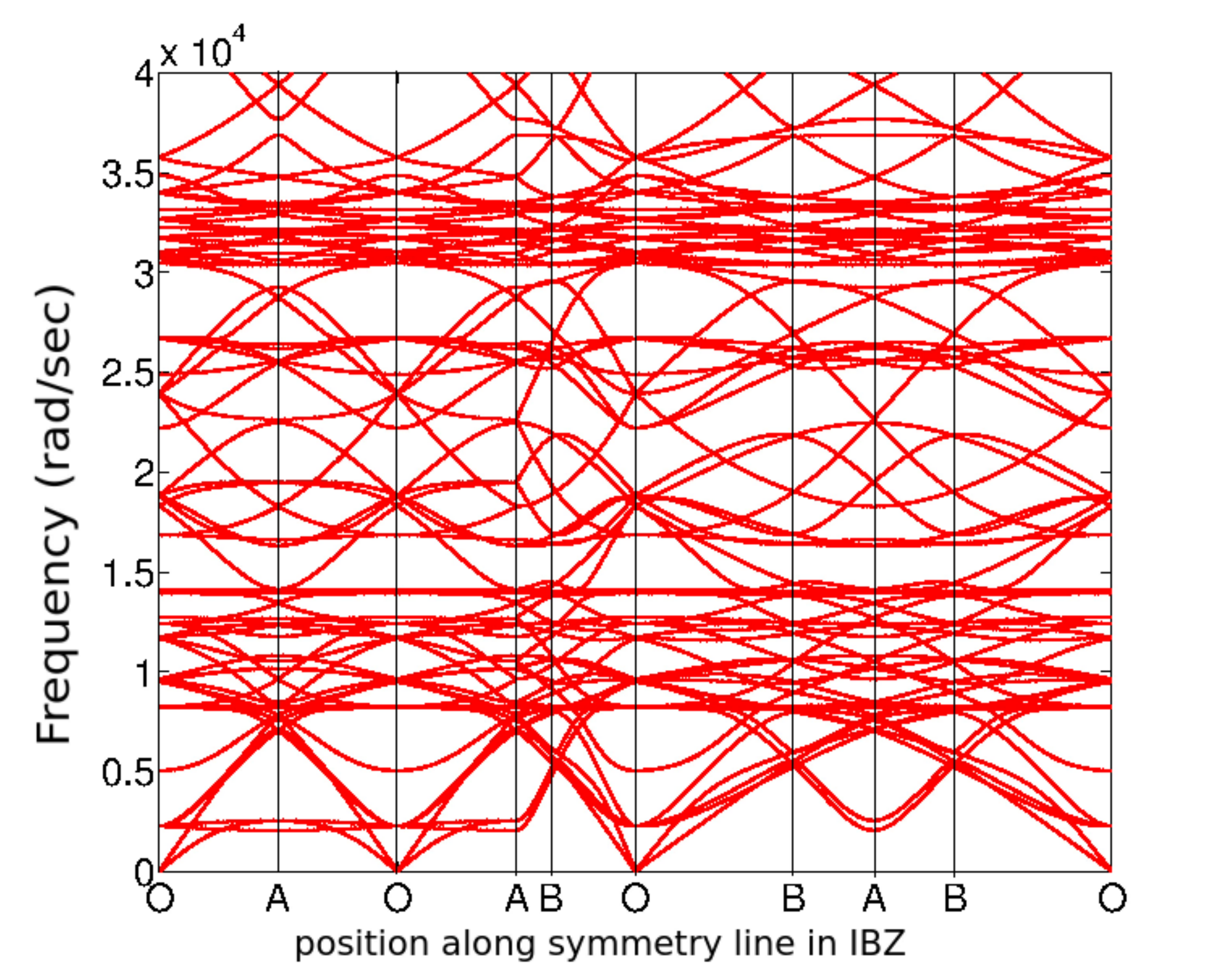}
\end{minipage}}
\subfloat[]{
\begin{minipage}{2.7in}
\label{fig_kag3:d}
\includegraphics[width=2.7in]{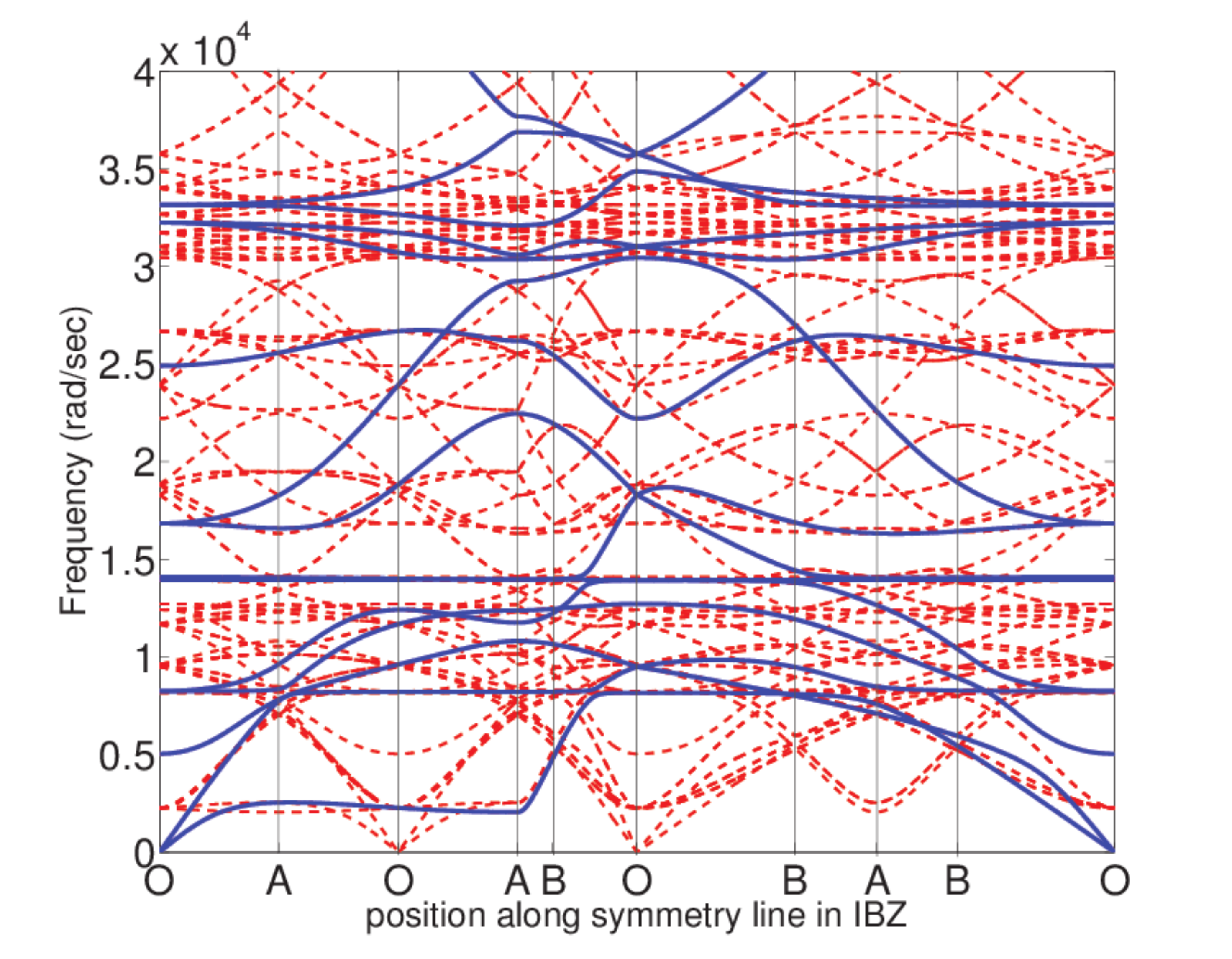}
\end{minipage}}
\caption{(a) Band structure of the kagome lattice using 3x3 supercell. (b) The Brillouin zone for the supercell and the primitive cell. (left) and the supercell used for calculation (right) (c) The juxtaposed band structure to compare with the primitive cell band structure, (d) The same figure as in Fig. \ref{fig_kag3:c} with overlaping primitive unit cell band structure }
\label{fig_kag3}
\end{figure}

 Next, Fig. \ref{fig_kag3}(a) shows the band structure for the 3x3 supercell of a kagome lattice, the unit cell being shown in Fig. \ref{fig_kag3}(b) right subfigure. The differnce can, again, be seen from the primitive cell band structure  as in Fig. \ref{fig_kag}(a). The phenomenon remains the same except the primitive cell Brillouin zone being composed of, in addition to the supercell Brillouin zone, the whole of the six neighbouring Brillouin zone of the latter as well as proximal one-third of six Brillouin zone of the same-one at each of the vertices, as in Fig. \ref{fig_kag}(b) left subfigure, and resulting implications to IBZ. Consiquently, the juxtaposition here is OA-AO-OA-AB-BO-OB-BA-AB-BO. Fig. \ref{fig_kag3}(c) shows the juxtaposed band structure, and Fig. \ref{fig_hex2}(d) shows the same with primitive unit cell band structure superposed to it.
    
For an $n$x$n$ supercell, the case falls in line with either 2x2 or 3x3 depending on n being even or odd and can be generalised as below.
    
    \vspace{0.2in}
    $n$ even: \hspace{0.2in}    $\frac{n}{2}$(OA-AO)-$\frac{n}{2}$(OB-BA-AB-BO)
    
    \vspace{0.1in}
    
    $n$  odd:  \hspace{0.23in} $(\frac{n-1}{2})$(OA-AO)-OA-AB-BO-$(\frac{n-1}{2})$(OB-BA-AB-BO) 
    \vspace{0.2in}
    
    Or, more generally,
    
    \vspace{0.2in} 
        $n$:   \hspace{0.23in} $ floor (\frac{n}{2})$(OA-AO)-$remainder(\frac{n}{2})$(OA-AB-BO)-$floor (\frac{n}{2})$(OB-BA-AB-BO) 
    \vspace{0.2in}  
        
\begin{figure}[]
\includegraphics[width=5in]{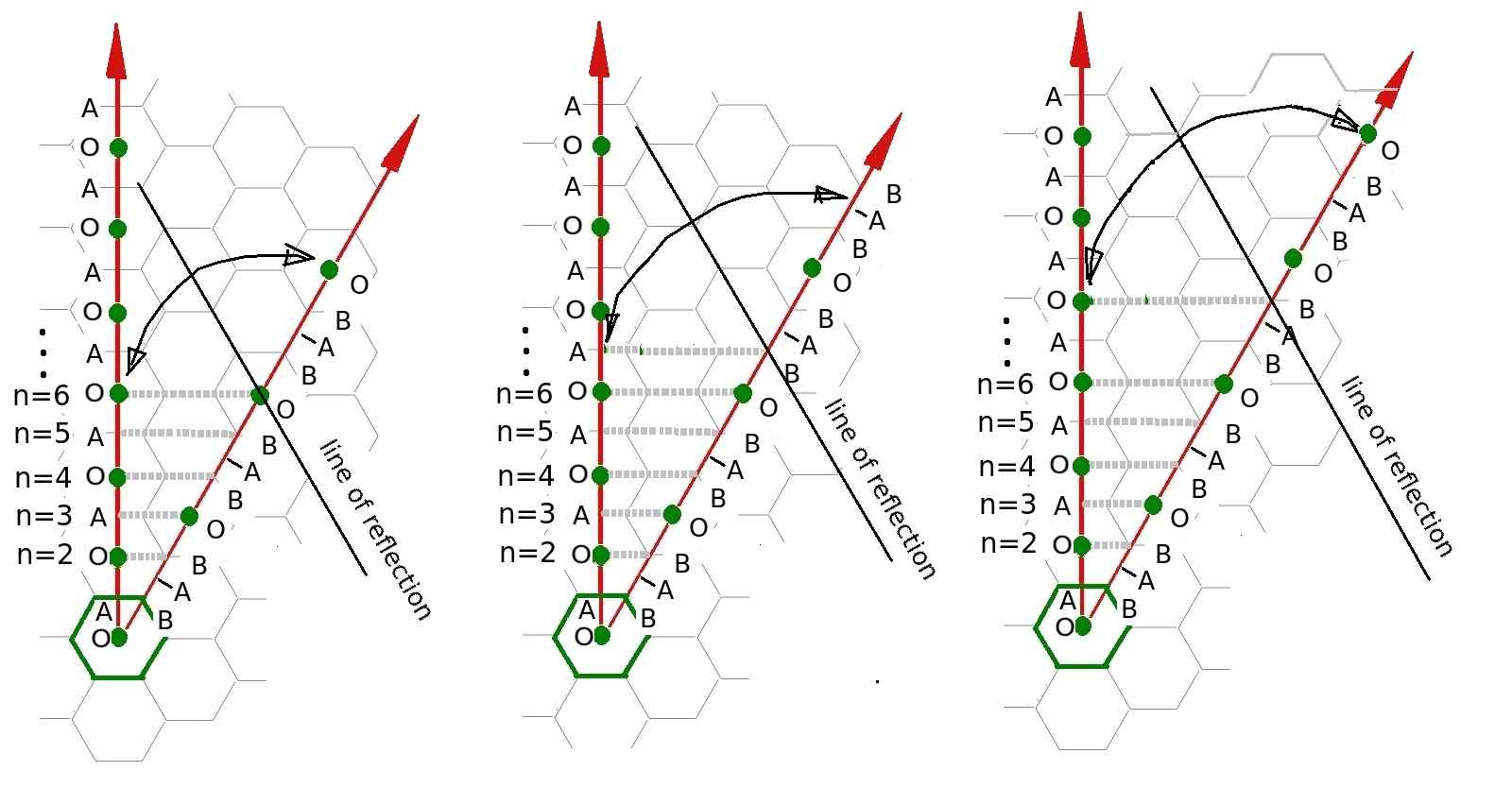}
\caption{explanation of juxtaposition for the lattice with regular hexagonal Brillouin zone. The vertical part consists of alternate OA-AO, For the rest, there can be total six categories three each for even and odd value of $n$, accounting for the three possible catagories of reflection}
\label{fig_hex_band_juxt}
\end{figure}

As shown in Fig.\ref{fig_hex_band_juxt} the vertical line consits of OA and AO continuing alternaticely till the count reaches $n$. Whereas, the other lines, if seen from the back (i.e. from the point O in the concluding BO segment), consist of the section OB-BA-AB-AB, and as shown in the figure, can be explain by means of an imaginary line of reflection, which might coincide with either point O, or B (two cases) depending on the value of $n$. So, when read from the back it continues with the pattern OB-BA-AB-BO till it finds the meetiing point with the verical section (point A or O depending on $n$ being odd or even). 

\begin{figure}[]
\centering
\subfloat[]{
\label{fig_hex_kag:a}
\begin{minipage}{2.7in}
\includegraphics[width=2.7in]{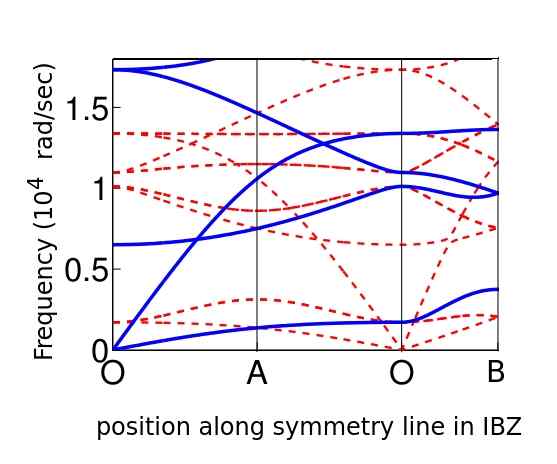}
\end{minipage}}
\subfloat[]{
\label{fig_hex_kag:b}
\begin{minipage}{2.7in}
\includegraphics[width=2.7in]{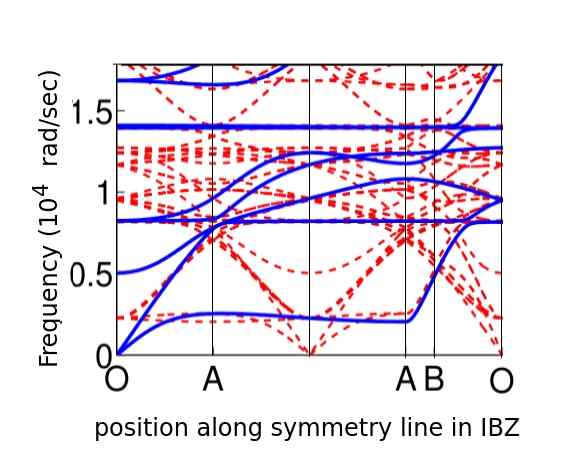}
\end{minipage}}
\caption{A magnified view of part of  (a) Fig. \ref{fig_hex2}(d) and (b) Fig. \ref{fig_kag3}(d) showing primitive cell band structure superinpsed on supercell band structure, dotted line supercell band structure, solid line primmitive cell band structure }
\label{fig_hex_kag}
\end{figure}

A magninified version of part of Fig. \ref{fig_hex2}(d) and Fig. \ref{fig_kag3}(d) are shown in Figs. \ref{fig_hex_kag}(a) and \ref{fig_hex_kag}(b) with dashed line indicating supercell band structure and solid line indicating primitive cell band structure. It may be seen that when the primitive cell band structure is superposed on the duly juxtaposed band structure of the supercell, though the bands in the primitive cell band structure perfectly overlap some of the bands in the supercell, there remain the spurious bands which come through the phenomenon known as band folding in the reduced zone \cite{Ashcroft}. This occurs because of continuation of the bands from the neighbouring and other parts of the Brillouine zone belonging to the Brillouin zone of the primitive unit cell arising out of the decreased periodicity of the supercell. So, in order to retain the whole set of information that were there in the primitive cell Brillouin zone in the reduced space of supercell Brillouin zone, the bands manifests as folded bands. Thus, when the band structure of the supercell is attempted to be arrived at from the band structure of the primitive cell, bands along the equivalent lines of the symmetry as well as nonsysmmtry zones in the primitive Brillouin zone get projected onto the high symmetry lines of the IBZ of the supercell and shown in the results section.

\begin{figure}[]
\centering
\subfloat[]{
\label{fig_band_folding_hex:a}
\begin{minipage}{1.8in}
\includegraphics[width=1.8in]{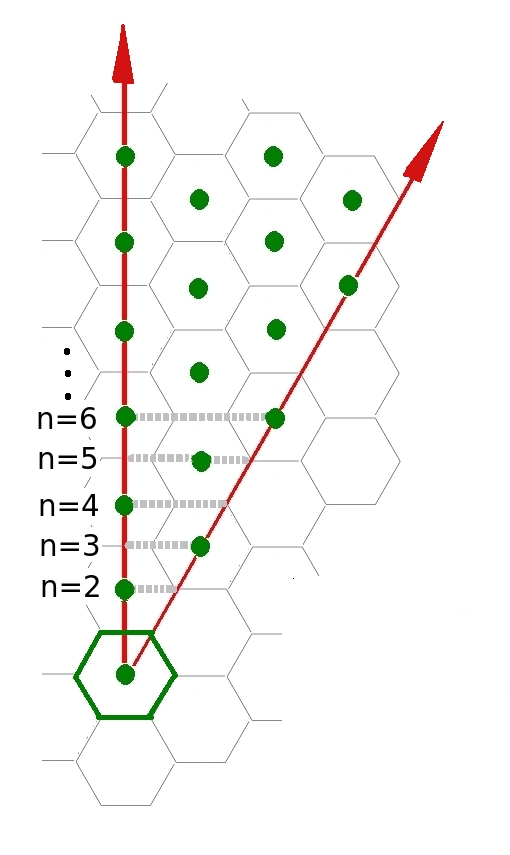}
\end{minipage}}
\subfloat[]{
\label{fig_band_folding_hex:b}
\begin{minipage}{2.7in}
\includegraphics[width=2.7in]{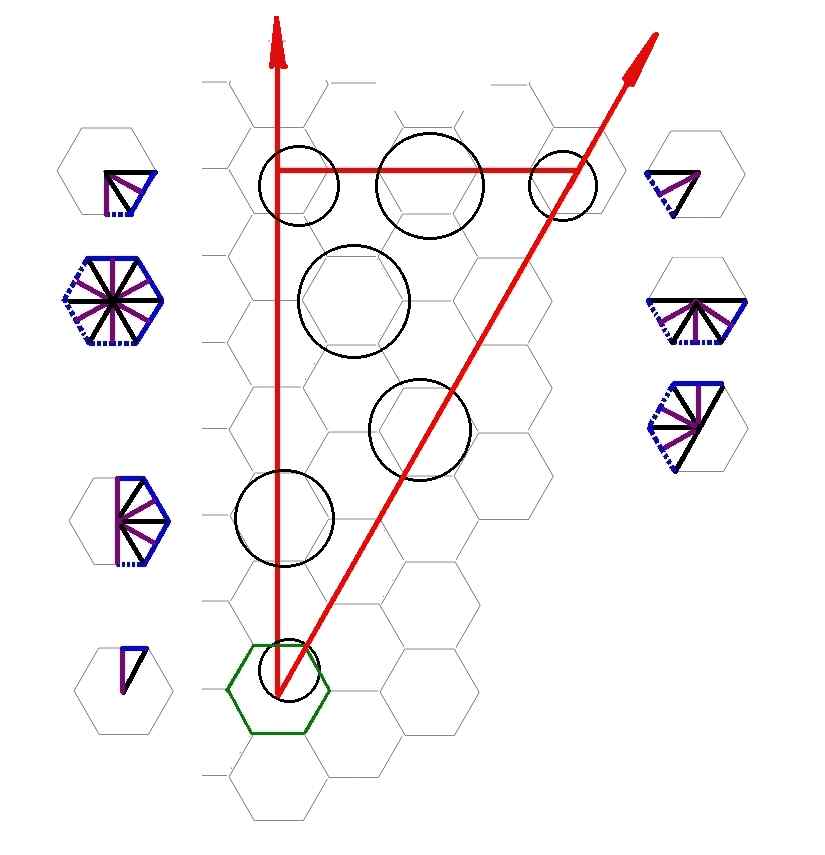}
\end{minipage}}
\vspace{0.01in}
\subfloat[]{
\begin{minipage}{3.4in}
\label{fig_band_folding_hex:c}
\includegraphics[width=3.2in]{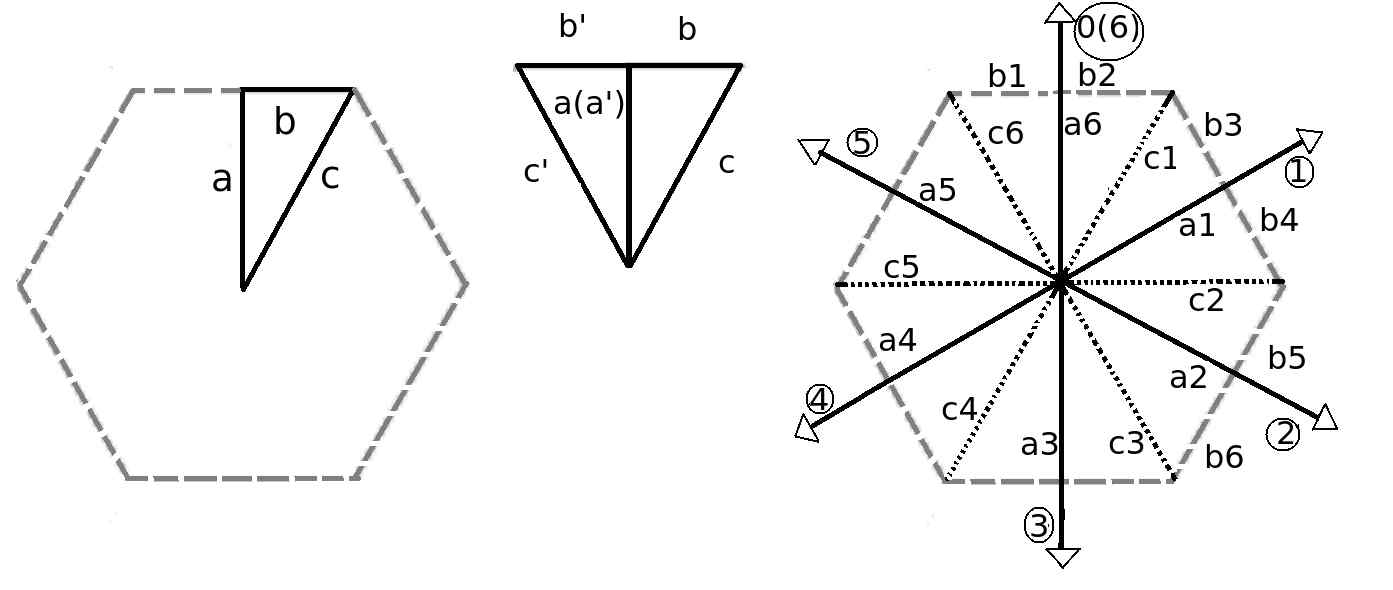}
\end{minipage}}
\subfloat[]{
\begin{minipage}{2.0in}
\label{fig_band_folding_hex:d}
\includegraphics[width=2.0in]{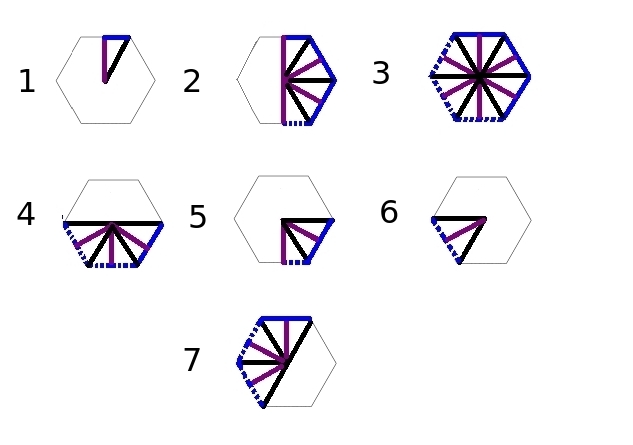}
\end{minipage}}
\caption{Schematic explanation of band folding for the lattices having regular hexagonal Brillouin zone. (a) primitive cell Brillouin zones contributing to the band structure of supercell of size $n$ (b) their involvement in supercell band structure according to their position and associated components of symmetry lines (c) IBZ boundariy lines filling Brillouin zone symmetry lines with one reflection and six rotation symmetries (d) primitive cell Brillouin zone with all seven possible arrangements of contributing components}
\label{fig_band_folding_hex}
\end{figure}

As shown in Fig. \ref{fig_band_folding_hex}(a), the hexagon with solid (green) boundary line depicting the first Brillouin zone of the supercell and the portion of it between the arrowed (red) straight lines indicating the IBZ of the same. The zone contained within the arrowed (red) lines and the dashed grey line would indicate the IBZ of premitive cell with reference to the given supercell, having a particalar value of $n$, as depicted. The solid round (green) beads within this zone, including those on the boundaries, indicate, for any particalar value of $n$, the centres of those Brillouin zones of the supercell which have got at least some part of it housed within the IBZ of the primitive cell. 

At this juncture, bearing in mind the fact that the IBZ with a reflective symmetry and six-fold rotational symmetry, fills the whole of the Brillouin zone, it may be noted that the IBZ boundary lines too, through such reflection and six-fold rotations represnt the whole set of symmetry lines in the Brillouin zone. This has been depicted in Fig. \ref{fig_band_folding_hex}(c) showing first the Brillouin zone and the IBZ with boundaries marked, and subsequently the IBZ boundaries with reflection symmetry and the sixfold rotational symmetry, duly marked. The arrowed lines show the rotations with encircled counter 1 to 6. rotation 6 comes back to the starting point 0. 
 The bands in the primitive cell IBZ, for each of the lines in the supercell Brillouin zone, for all the supercell Brillouin zones positioned winthin the primitive cell IBZ, having centres marked with solid circular (green) beads, get folded to the respective line in supercell IBZ. Fig. \ref{fig_band_folding_hex}(c) is a variation of Fig. \ref{fig_band_folding_hex}(a) with the aforesaid contributing symmetry lines marked corresponding to the position of primitive cell brillouin zone in the supercell Brillouin zone. There could be upto eighteen such contributions, due to six lines each for the three lines in IBZ boundary (dashed lines are duplicates  and hence discounted, due to their overlapping with other supercell Brillouin zones and having been considered there) .  There are seven such configureations possible in which symmetry lines in the interior and boundary supercell Brillouin zone contribute to the band folding, as shown by encircling their respective position in Fig. \ref{fig_band_folding_hex}(c). Fig. \ref{fig_band_folding_hex}(d) shows all the seven possible combinations from the symmetry lines in Brillouin zone and detailed in Table \ref{table_hex}.   
 
\begin{table}[!]
\centering
\caption{Possible arrangement of combination of the Brilluoin zone symmetry lines in their contribution to band fonding in supercell band structure of lattices with regular hexagonal Brillouin zone }
\label{table_hex}
\begin{tabular}{lll}
\hline
Serial & Number of  components  & Components present\\ 
\hline \\
1 & 3 & a6, b2, c1 \\ 
\hline
2 & 12 & a6,a1,a2,a3,b2,b3,b4,b5,b6,c1,c2,c3 \\
\hline
3 & 18 & a6,a1,a2,a3,a4,a5,b1,b2,b3, b4,b5,b6,c1,c2,c3,c4,c5,c6 \\
\hline
4 & 9  & a2,a3,a4,b5,b6,c2,c3,c4,c5 \\
\hline
5 & 5  & a2,a3,c2,c3,b5,b6 \\
\hline
6 & 3  & a4,c4,c5 \\
\hline
7 & 9  & a6,a4,a5,b1,b2,c1,c4,c5,c6 \\ 
\hline
\end{tabular}
\end{table}

\begin{figure}[!ht]
\subfloat[]{
\label{fig_nonsym_hex2:a}
\begin{minipage}{1.8in}
\includegraphics[width=1.4in]{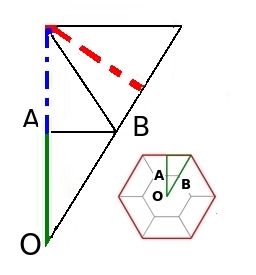}
\end{minipage}}
\subfloat[]{
\label{fig_nonsym_hex2:b}
\begin{minipage}{2.0in}
\includegraphics[width=1.4in]{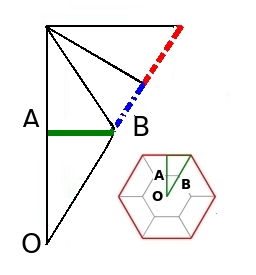}
\end{minipage}}
\subfloat[]{
\label{fig_nonsym_hex2:c}
\begin{minipage}{1.8in}
\includegraphics[width=1.4in]{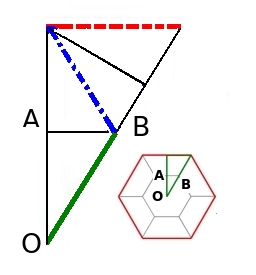}
\end{minipage}}
\par\medskip
\subfloat[]{
\label{fig_nonsym_hex2:d}
\begin{minipage}{1.8in}
\includegraphics[width=1.4in]{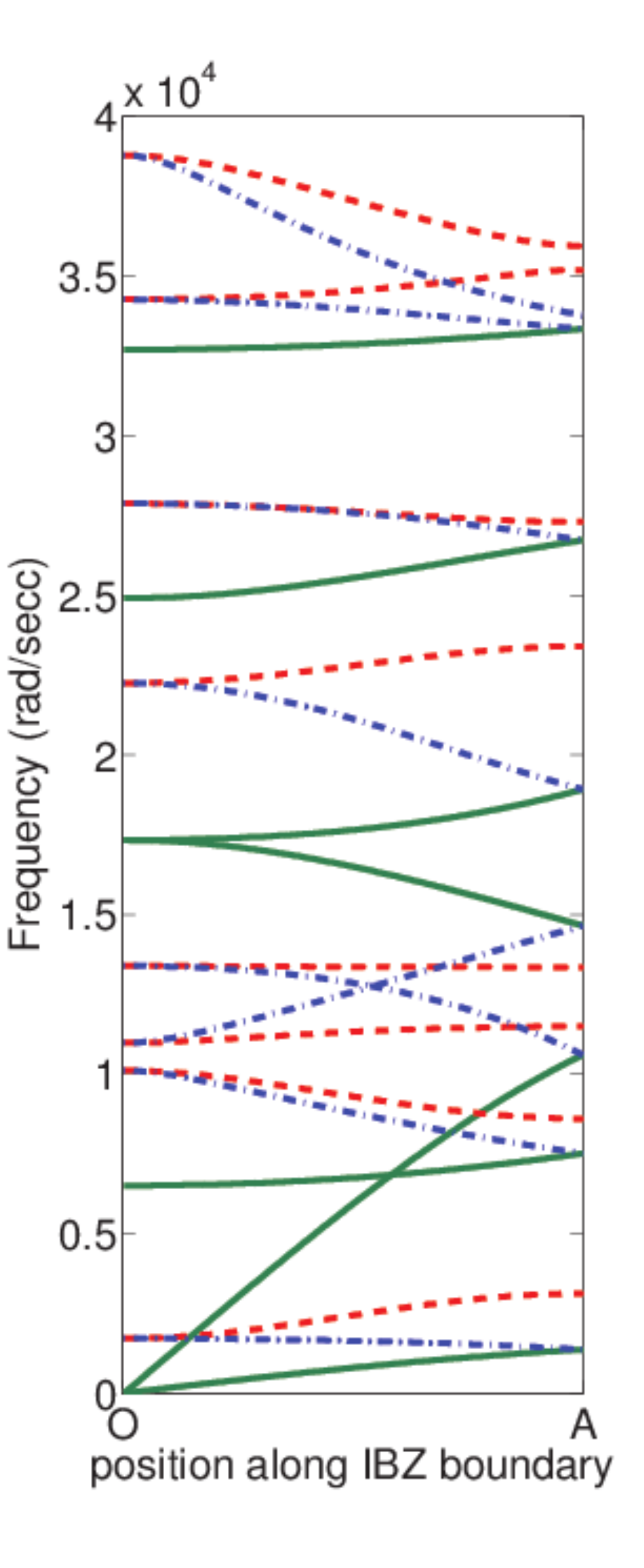}
\end{minipage}}
\subfloat[]{
\label{fig_nonsym_hex2:e}
\begin{minipage}{1.2in}
\includegraphics[width=1.4in]{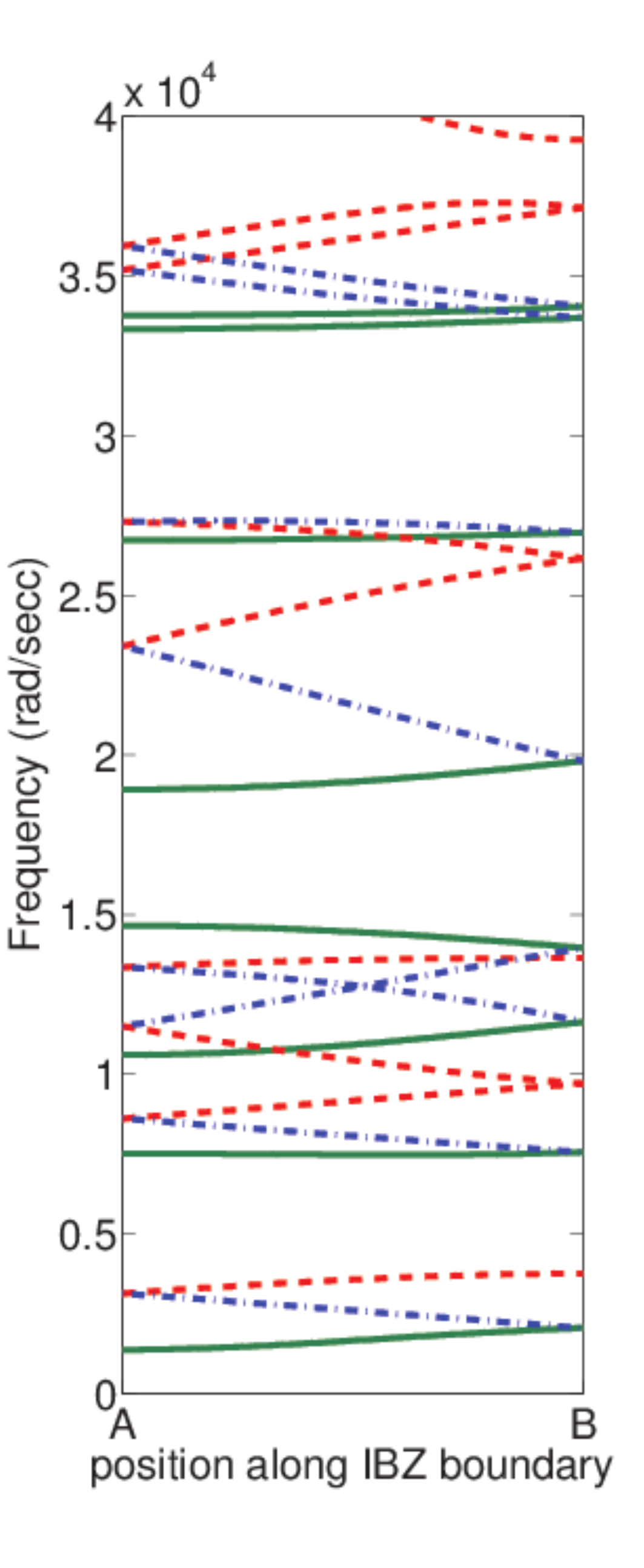}
\end{minipage}}
\hspace{0.4in}
\subfloat[]{
\label{fig_nonsym_hex2:f}
\begin{minipage}{1.4in}
\includegraphics[width=1.4in]{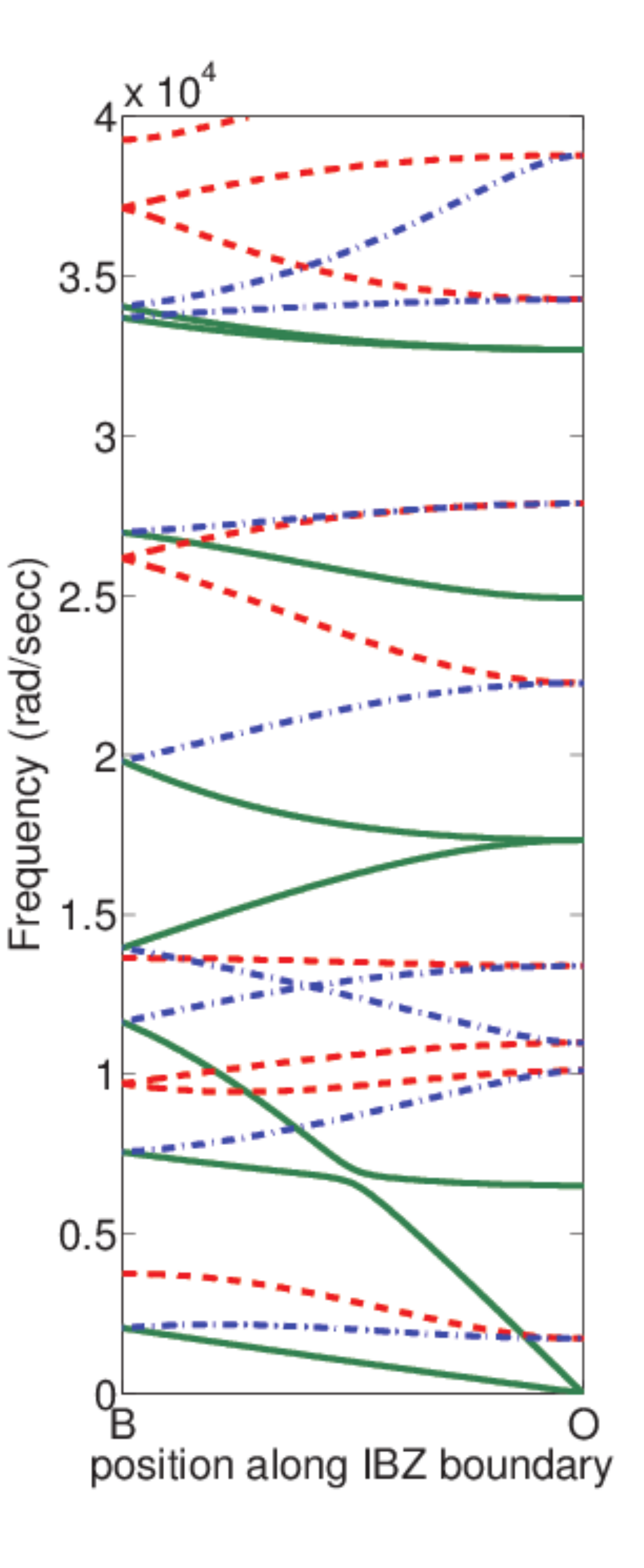}
\end{minipage}}
\hspace{0.4in}
\caption{Construction of band structure of 2x2 supercell from that of primitive cell for a hexagonal honeycomb latiice having a regular hexagonal Brillouin zone using equivalent symmetry and non-symmetry lines in the Brillouin zone. Explains origin of spurious modes. There are 6 such lines as  shown with different line style (and colors) on the IBZ at the top (a), (b), (c) and the corresponding lines in the band structure at the bottom (d),(e),(f) for the segments OA,AB and BO respectively}
\label{fig_nonsym_hex2}
\end{figure}

\begin{figure}[!ht]
\centering
\subfloat[]{
\label{fig_nonsym_kag3:a}
\begin{minipage}{1.8in}
\includegraphics[width=1.4in]{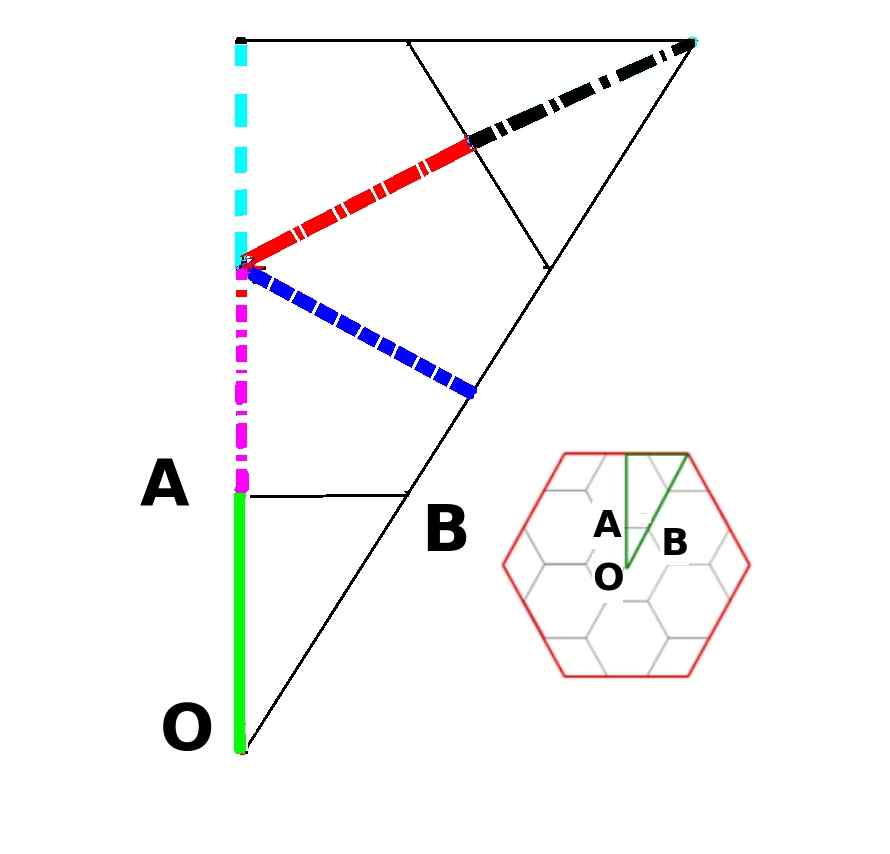}
\end{minipage}}
\subfloat[]{
\label{fig_nonsym_kag3:b}
\begin{minipage}{1.8in}
\includegraphics[width=1.4in]{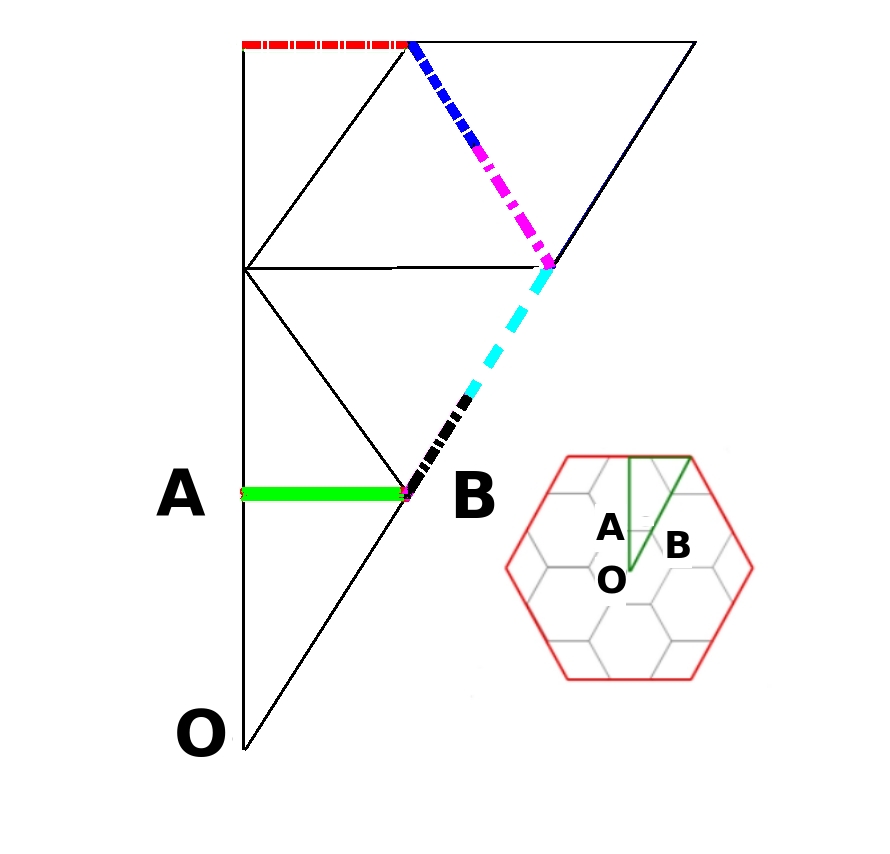}
\end{minipage}}
\subfloat[]{
\label{fig_nonsym_kag3:c}
\begin{minipage}{1.8in}
\includegraphics[width=1.4in]{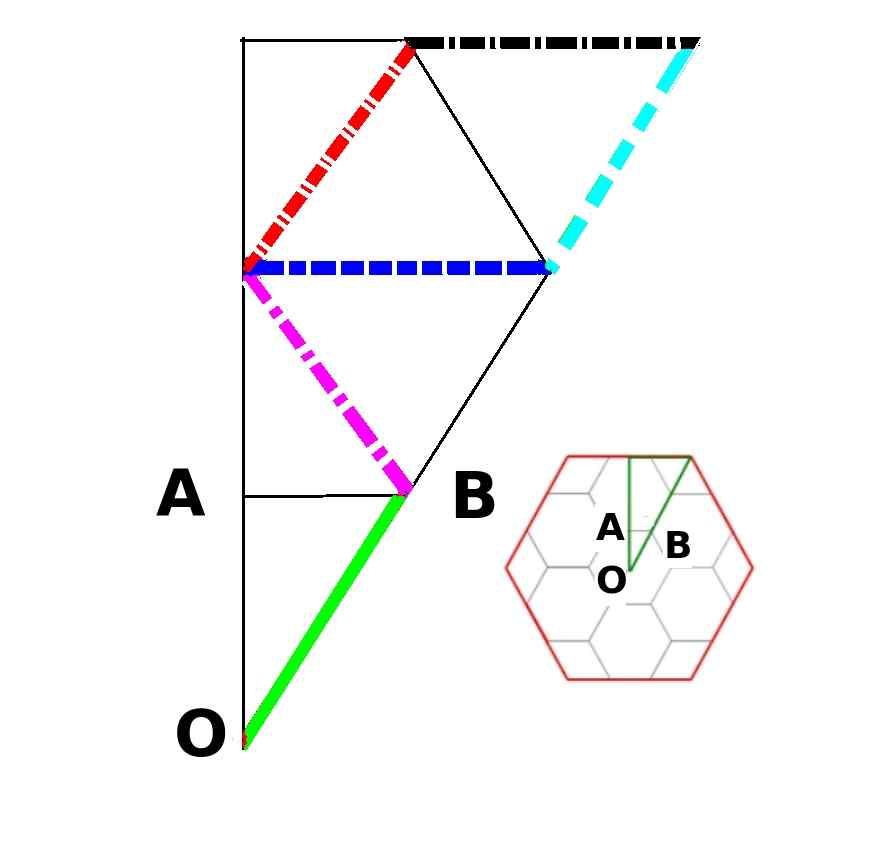}
\end{minipage}}
\par\medskip
\subfloat[]{
\label{fig_nonsym_kag3:d}
\begin{minipage}{1.8in}
\includegraphics[width=1.4in]{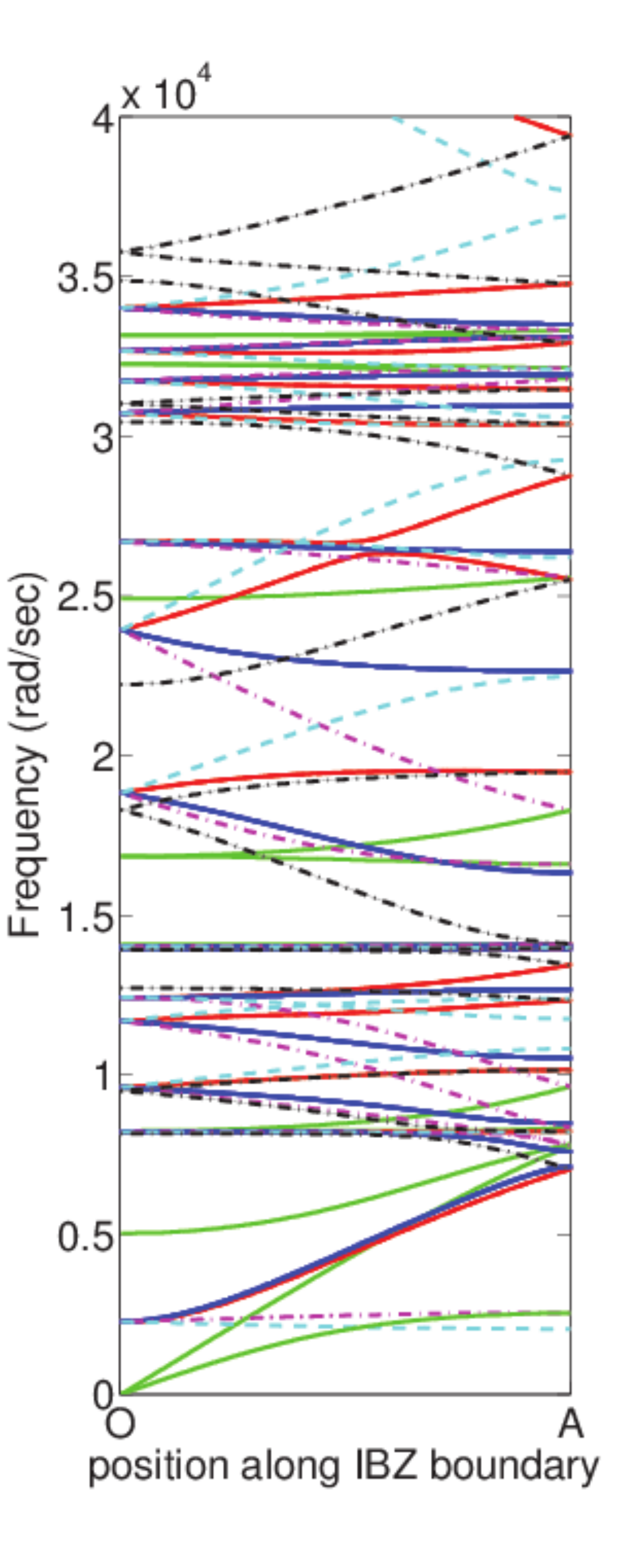}
\end{minipage}}
\subfloat[]{
\label{fig_nonsym_kag3:e}
\begin{minipage}{1.8in}
\includegraphics[width=1.4in]{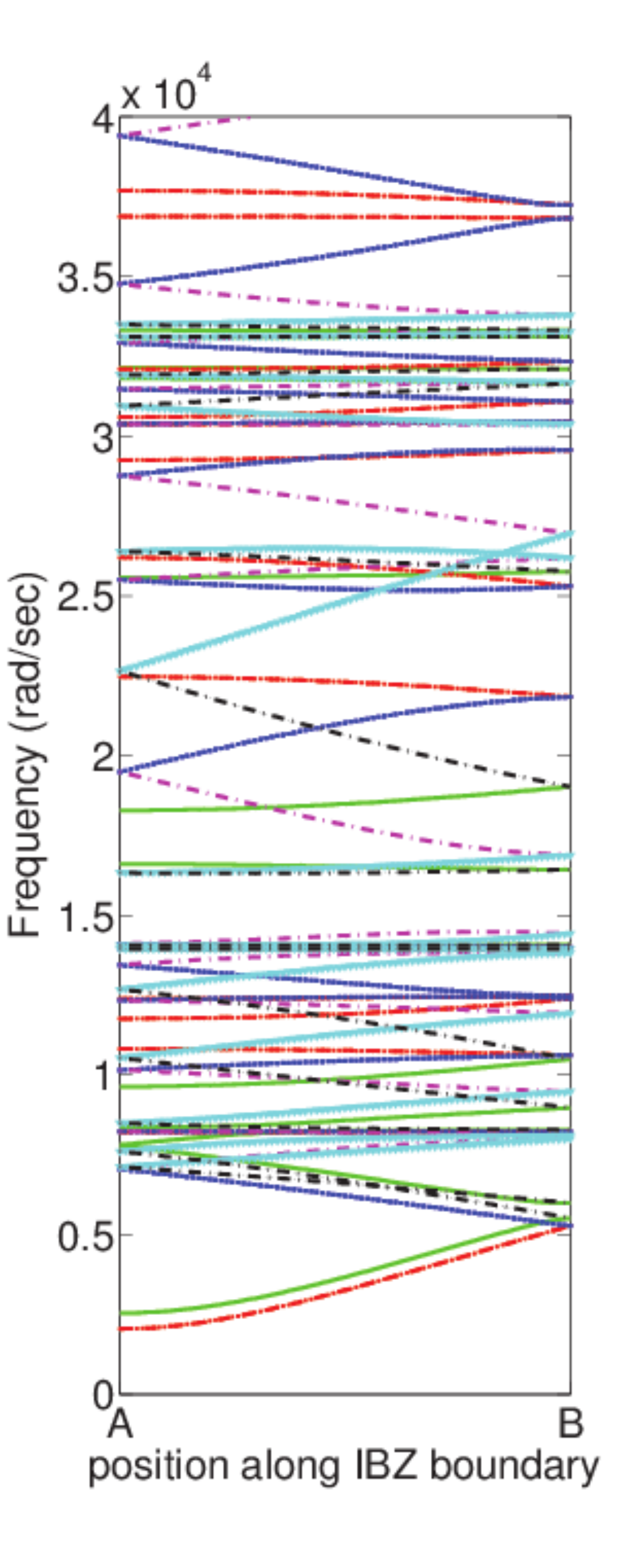}
\end{minipage}}
\subfloat[]{
\label{fig_nonsym_kag3:f}
\begin{minipage}{1.8in}
\includegraphics[width=1.4in]{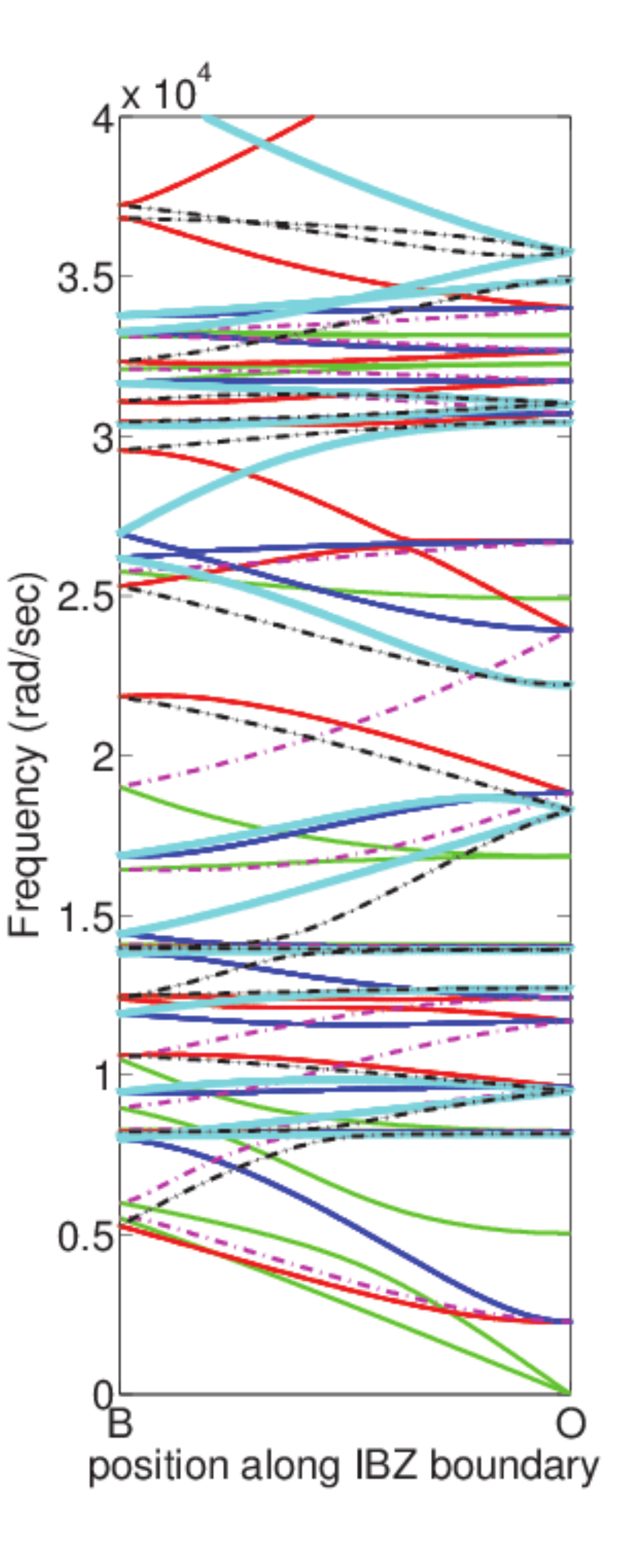}
\end{minipage}}
\caption{Construction of band structure of 3x3 supercell from that of primitive cell for a kagome latiice having a regular hexagonal Brillouin zone using equivalent symmetry and non-symmetry lines in the Brillouin zone. Explains origin of spurious modes. There are 6 such lines as  shown with different line style (and colors) on the IBZ at the top (a), (b), (c) and the corresponding lines in the band structure at the bottom (d),(e),(f) for the segments OA,AB and BO respectively}
\label{fig_nonsym_kag3}
\end{figure}

Fig. \ref{fig_nonsym_hex2} and Fig. \ref{fig_nonsym_kag3} exemplifies the phenomenon of the band folding. The bottom subfigures in both these figures, i.e.  Figs. \ref{fig_nonsym_hex2}(d), \ref{fig_nonsym_hex2}(e), \ref{fig_nonsym_hex2}(f) and Figs. \ref{fig_nonsym_kag3}(d), \ref{fig_nonsym_kag3}(e), \ref{fig_nonsym_kag3}(f)  separately represent the zones in the band structure corresponding to the three segments in IBZ boundary, OA, AB and BO. The line of incidence winthin primitive cell IBZ, marked with a particular line style (and particular color), as shown in the top subfigures Figs. \ref{fig_nonsym_hex2}(a), \ref{fig_nonsym_hex2}(b), \ref{fig_nonsym_hex2}(c) and Figs. \ref{fig_nonsym_kag3}(a), \ref{fig_nonsym_kag3}(b), \ref{fig_nonsym_kag3}(c), points to, in the supercell, the origin of all the bands with that particular line style (and color). For example, Figs. \ref{fig_nonsym_hex2}(d) shows the band folding arising out of the OA and its equivalent segments in primitive cellas shown in Figs. \ref{fig_nonsym_hex2}(a). It may be noted that the number of source lines of incidence are same for the three zones. It may also be nnoted that some of the contributing segmets are not lying on the symmetry lines of the primitive cell. 

It can be seen that for the case of 2x2 supercell, as in Fig. \ref{fig_nonsym_hex2}, within primitive cell IBZ, for each of the three zones, there are three different source lines. Though, generated from the primitive cell, having put the three zones together, the band structure formed this way is identical to the supercell band structure as in Fig. \ref{fig_hex2}. Thus, Figs. \ref{fig_nonsym_hex2}(d), \ref{fig_nonsym_hex2}(e), \ref{fig_nonsym_hex2}(f) put together make the supercell band structure. 

For the 3x3 supercell, as in Fig. \ref{fig_nonsym_kag3} for each of the three zones, the number of source lines of incidence witin the primitive cell IBZ goes to six. As before, having put the three zones together, the band structures formed this way is identical to the supercell band structure as in Fig. \ref{fig_kag3}.

\subsubsection{Square Brillouin Zone}

\begin{figure}[]
\centering
\subfloat[]{
\label{fig_square3:a}
\begin{minipage}{2.7in}
\includegraphics[width=2.7in]{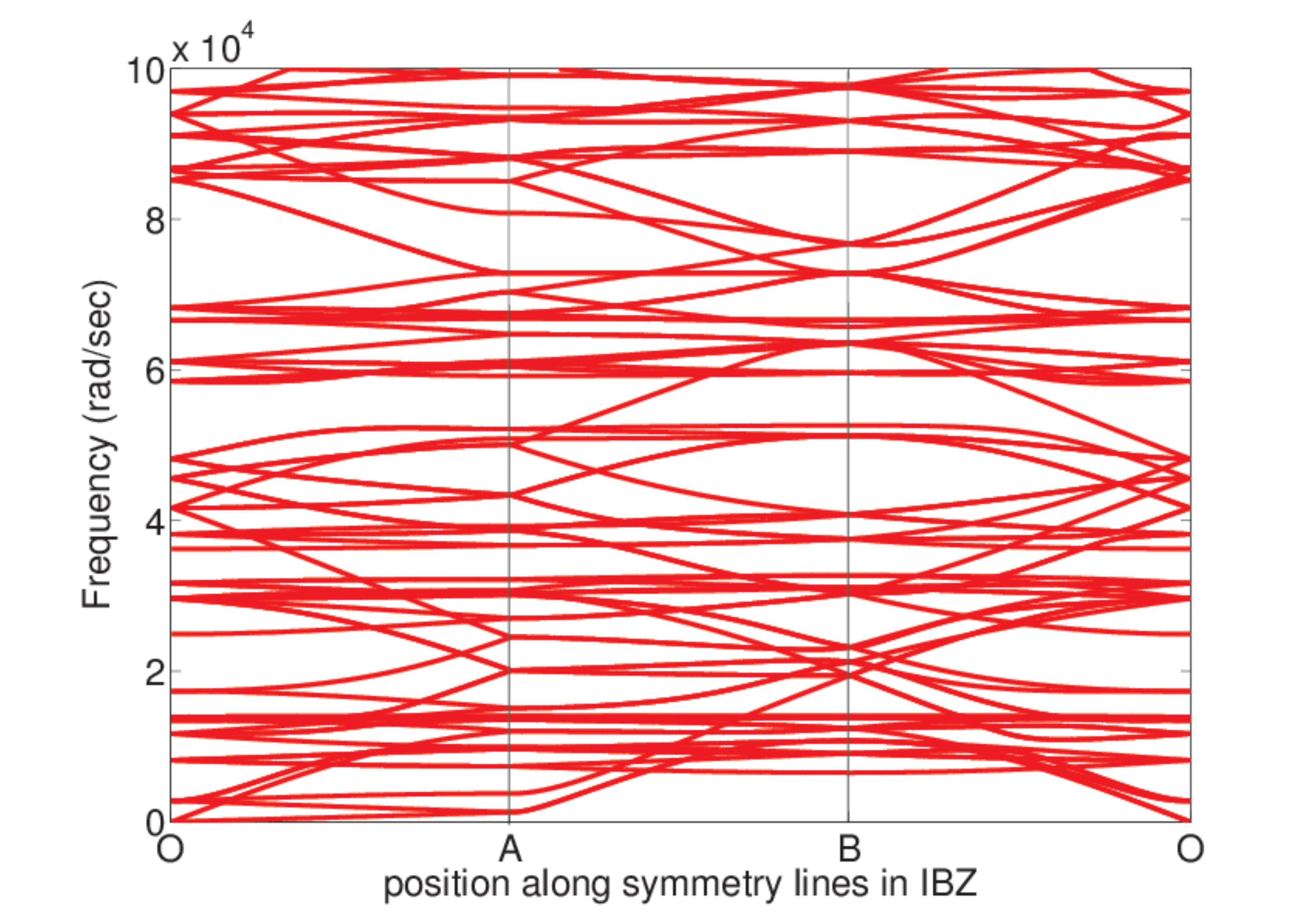}
\end{minipage}}
\subfloat[]{
\label{fig_square3:b}
\begin{minipage}{2.7in}
\includegraphics[width=2.7in]{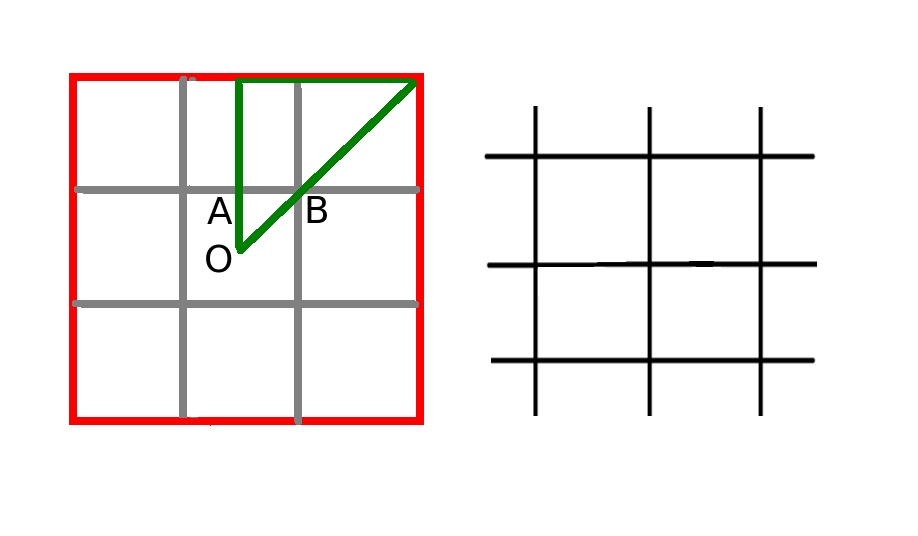}
\end{minipage}}
\vspace{0.01in}
\subfloat[]{
\begin{minipage}{2.7in}
\label{fig_square3:c}
\includegraphics[width=2.7in]{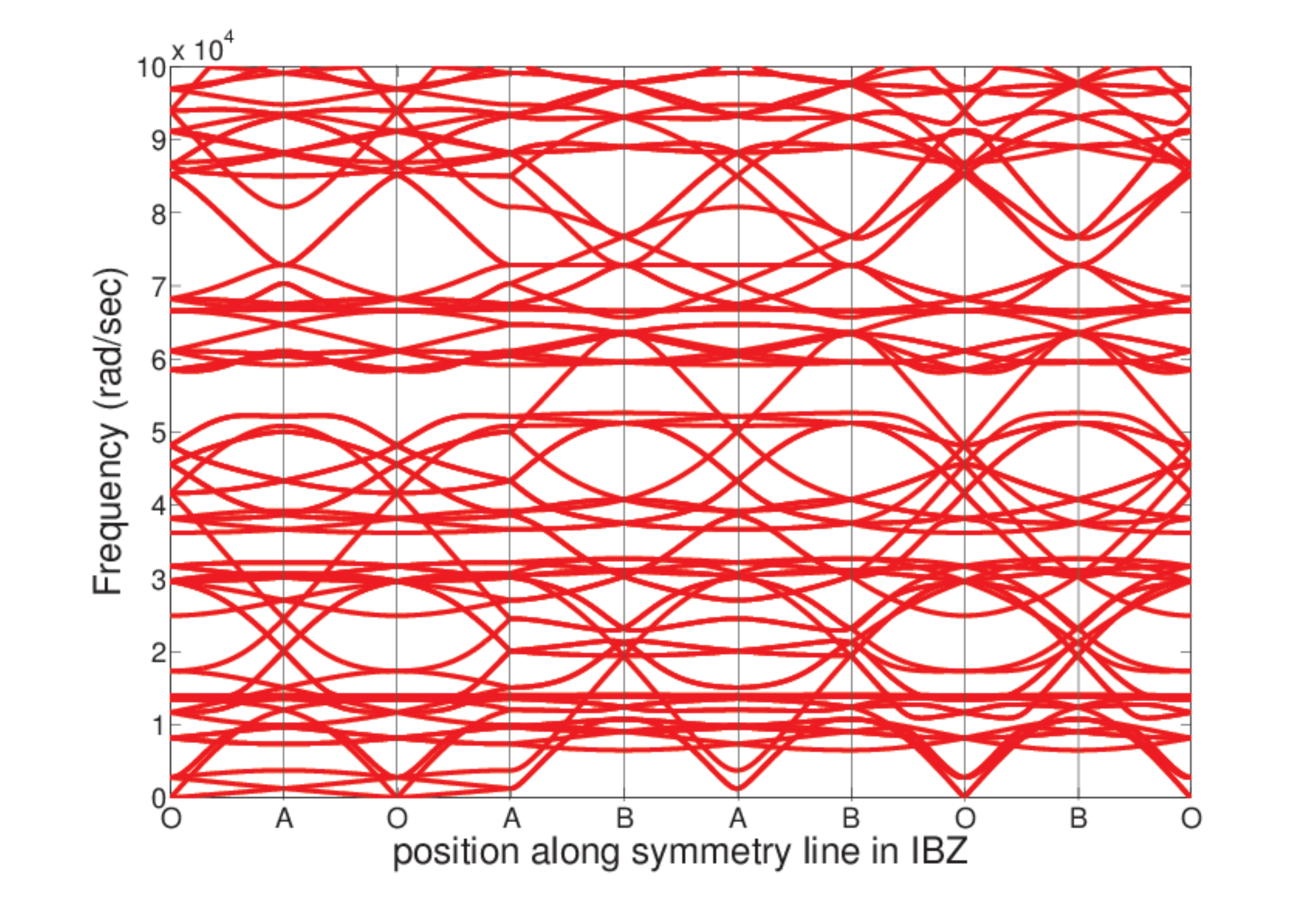}
\end{minipage}}
\subfloat[]{
\begin{minipage}{2.7in}
\label{fig_square3:d}
\includegraphics[width=2.7in]{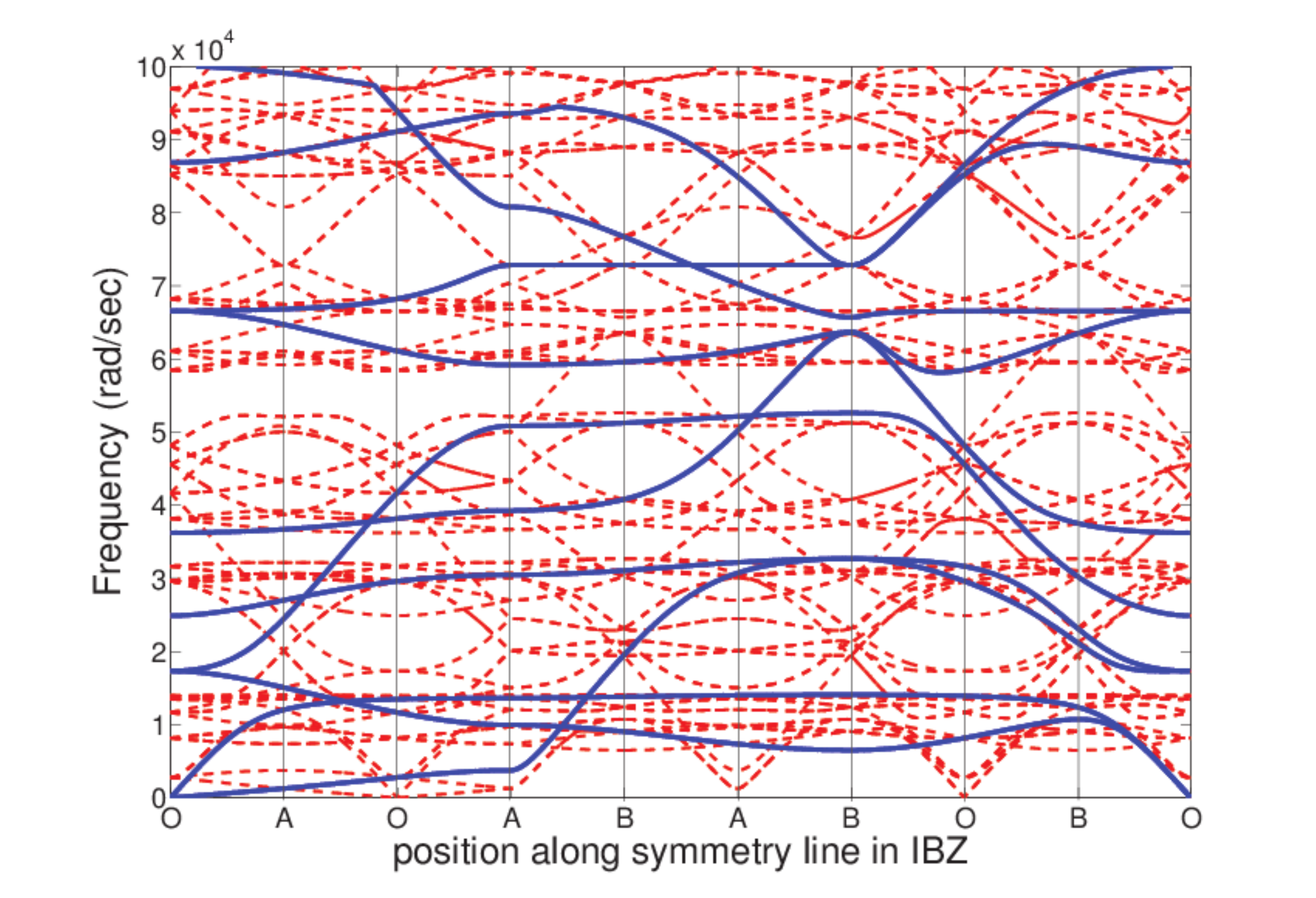}
\end{minipage}}
\caption{(a) Band structure of the square lattice using 3x3 supercell. (b) The Brillouin zone for the supercell and the primitive cell. (left) and the supercell used for calculation (right) (c) The juxtaposed band structure to compare with the primitive cell band structure, (d) The same figure as in Fig. \ref{fig_square3:c} with overlaping primitive unit cell band structure }
\label{fig_square3}
\end{figure}

The next example taken in realation to supercell is a square lattice with a 3x3 supercell. Its band structure is shown in \ref{fig_square3}(a). The lattice taken and the associated supercell Brillouin zone- the inner grey square, along with IBZ marked as triangle ABC is shown in  Fig. \ref{fig_square3}(b) right subfigure and left subfigure respectively. In line similar to that of previous subsection, the difference can be noted between Fig. \ref{fig_square3}(a) and Fig. \ref{fig_square}(a). The juxtaposed supercell band structure is shown in Fig. \ref{fig_square3}(c) and the Fig. \ref{fig_square3}(d) shows Fig. \ref{fig_square3}(c) with the primitive cell band structure superposed to it. In this case also, the bands in the primitive cell band structure maches perfectly with some of the bands in the juxtaposed supercell band structure. 
 The juxtaposition for any $n$, arguing in similar line to that of a hexagonal Brillouin zone, may be written as

\vspace{0.2in} 
  $n$ even: \hspace{0.2in} $n$(OA-AO)-($\frac{n}{2}$)(OB-BO)
  
\vspace{0.2in} 
 $n$ odd:  \hspace{0.23in} ($\frac{n-1}{2}$)(OA-AO)-OA-AB-($\frac{n-1}{2}$)(BA-AB)-BO-($\frac{n-1}{2}$)(OB-BO)
\vspace{0.2in}

\begin{figure}[]
\centering
\subfloat[]{
\label{fig_band_folding_square:a}
\begin{minipage}{2.4in}
\includegraphics[width=2.4in]{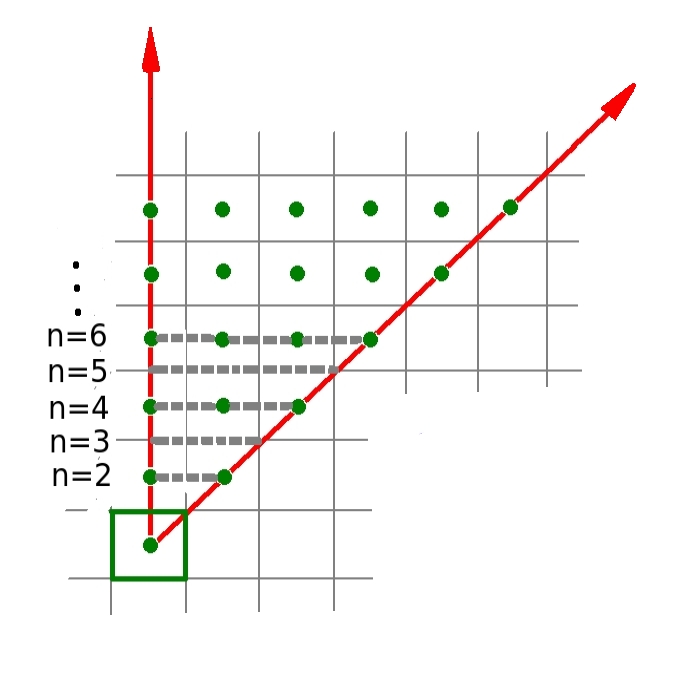}
\end{minipage}}
\subfloat[]{
\label{fig_band_folding_square:b}
\begin{minipage}{2.7in}
\includegraphics[width=2.7in]{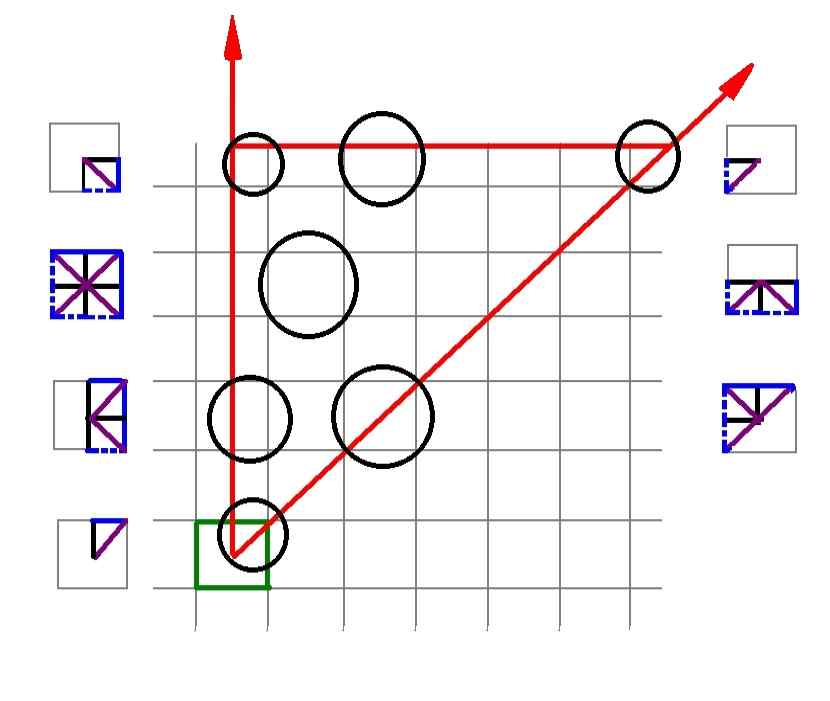}
\end{minipage}}
\vspace{0.01in}
\subfloat[]{
\begin{minipage}{3.0in}
\label{fig_band_folding_square:c}
\includegraphics[width=3.0in]{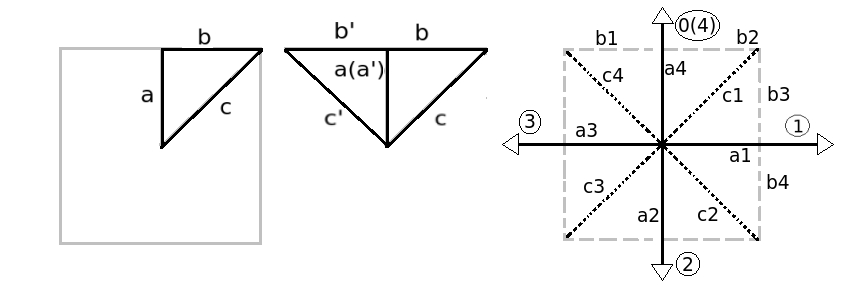}
\end{minipage}}
\subfloat[]{
\begin{minipage}{2.2in}
\label{fig_band_folding_square:d}
\includegraphics[width=2.2in]{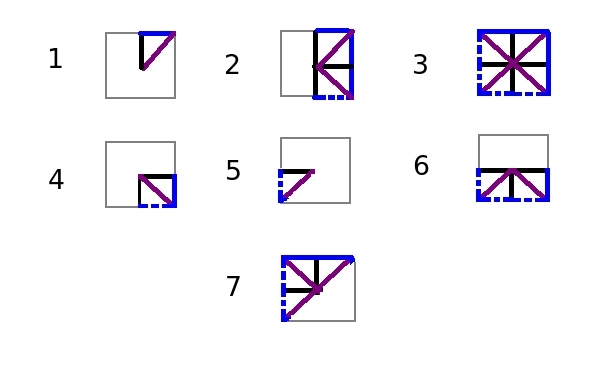}
\end{minipage}}
\caption{Schematic explanation of band folding for the lattices having square Brillouin zone. (a) primitive cell Brillouin zones contributing to the band structure of supercell of size $n$ (b) their involvement in supercell band structure according to their position and associated components of symmetry lines (c) IBZ boundariy lines filling Brillouin zone symmetry lines with one reflection and four rotation symmetries (d) primitive cell Brillouin zone with all seven possible arrangements of contributing components}
\label{fig_band_folding_square}
\end{figure}

The mechanism of band folding for the case of a lattice with square Brillouin zone may be explained in line similar to that of a lattice with hexagonal Brilluoin zones. As shown in Fig. \ref{fig_band_folding_square}(a), the square with solid bold (green) boundary line depicting the first Brillouin zone of the supercell and the portion of it between the arrowed (red) straight lines indicating the IBZ of the same. The zone contained within the arrowed (red) lines and the dashed grey line would indicate the IBZ of premitive cell with reference to the given supercell, having a particalar value of n, as depicted. The solid round (green) beads within this zone, including those on the boundaries, indicate, for any particalar value of n, the centres of those Brillouin zones of the supercell which have got at least some part of it housed within the IBZ of the primitive cell. For this case, the IBZ holds a reflective symmetry and four-fold rotational symmetry, and thereby fills the whole of the Brillouin zone. Hence, the IBZ boundary lines too, through such reflection and four-fold rotations represnt the whole set of symmetry lines in the Brillouin zone. This has been depicted in Fig. \ref{fig_band_folding_square}(c) showing first the Brillouin zone and the IBZ with boundaries marked, and subsequently the IBZ boundaries with reflection symmetry and the four-fold rotational symmetry, duly marked. The arrowed lines show the rotations with encircled counter 1 to 4. rotation 4 comes back to the starting point 0. The bands in the primitive cell IBZ,
for each of the lines in the supercell Brillouin zone, for all the supercell Brillouin zones positioned winthin the primitive cell IBZ, having centres marked with solid circular (green) beads, get folded
to the respective line in supercell IBZ. Fig. \ref{fig_band_folding_square}(c) is, like before,  a variation of Fig. \ref{fig_band_folding_square}(a) with the aforesaid
contributing symmetry lines marked corresponding to the position of primitive cell brillouin zone
in the supercell Brillouin zone. There could be upto twelve such contributions, due to four lines
each for the three lines in IBZ boundary (dashed lines are duplicates and hence discounted, due to
their overlapping with other supercell Brillouin zones and having been considered there) . There
are seven such configureations possible in which symmetry lines in the interior and boundary
supercell Brillouin zone contribute to the band folding, as shown by encircling their respective
position in Fig. \ref{fig_band_folding_square}(c). Fig. \ref{fig_band_folding_square}(d) shows all the seven possible combinations from the symmetry
lines in Brillouin zone and detailed in the Table \ref{table_square}

\begin{table}[!]
\centering
\caption{Possible arrangement of combination of the Brilluoin zone symmetry lines in their contribution to band fonding in supercell band structure of lattices with square Brillouin zone }
\label{table_square}
\begin{tabular}{lll}
\hline
Serial & Number of  components  & Components present\\ 
\hline \\
1 & 3 & a4, b2, c1 \\ 
\hline
2 & 8 & a4,a1,a2,b2,b3,b4,c1,c2 \\
\hline
3 & 12 & a4,a1,a2,a3,b1,b2,b3,b4,c1,c2,c3,c4 \\
\hline
4 & 4  & a1,a2,b4,c2 \\
\hline
5 & 2  & a3,c3 \\
\hline
6 & 6  & a1,a2,a3,b4,c2,c3 \\
\hline
7 & 7  & a4,a3,b1,b2,c1,c3,c4 \\ 
\hline
\end{tabular}
\end{table}

\begin{figure}[]
\subfloat[]{
\label{fig_nonsym_square:a}
\begin{minipage}{1.8in}
\includegraphics[width=1.4in]{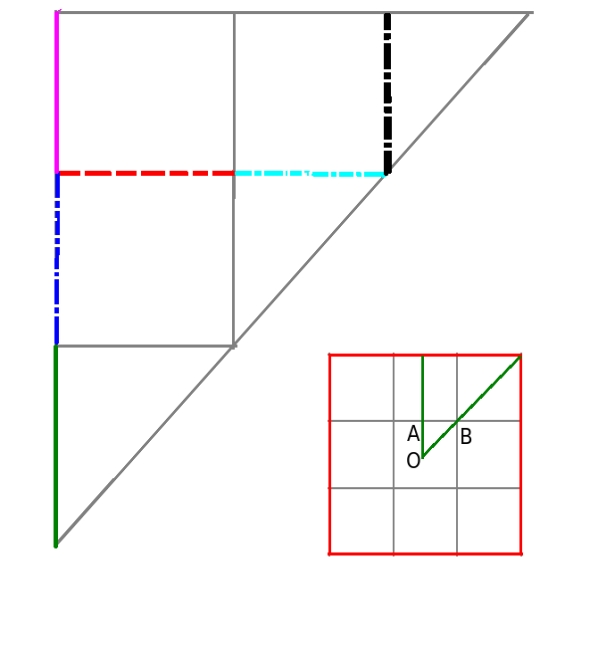}
\end{minipage}}
\subfloat[]{
\label{fig_nonsym_square:b}
\begin{minipage}{1.8in}
\includegraphics[width=1.4in]{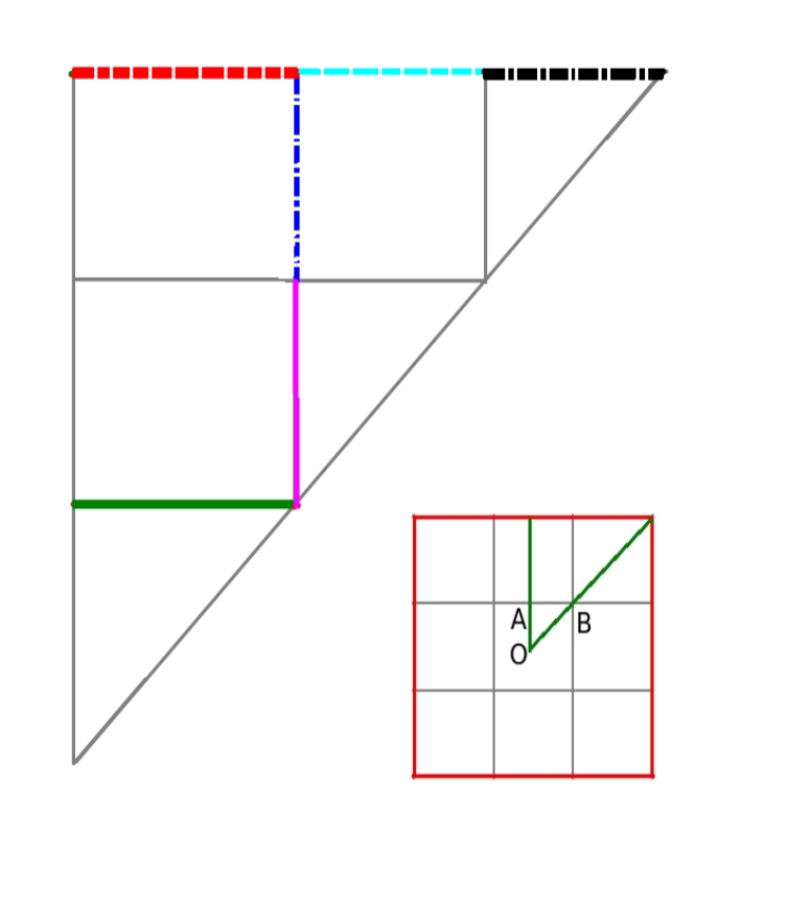}
\end{minipage}}
\subfloat[]{
\label{fig_nonsym_square:c}
\begin{minipage}{1.8in}
\includegraphics[width=1.4in]{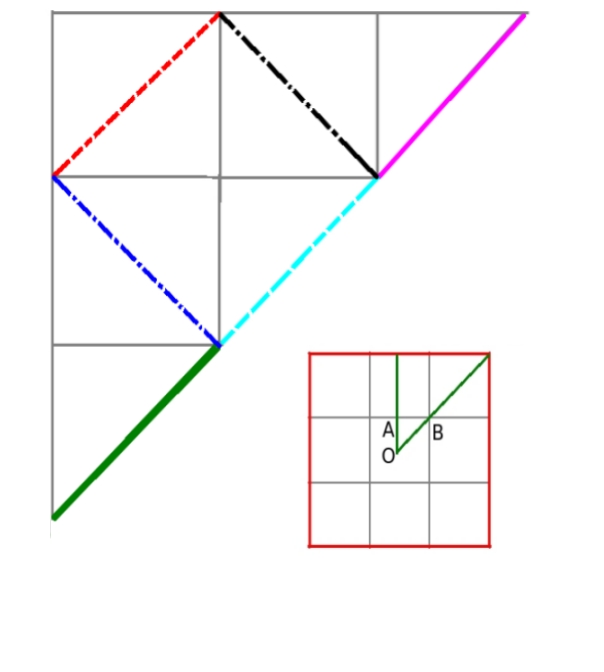}
\end{minipage}}
\par\medskip
\subfloat[]{
\label{fig_nonsym_square:d}
\begin{minipage}{1.4in}
\includegraphics[width=1.4in]{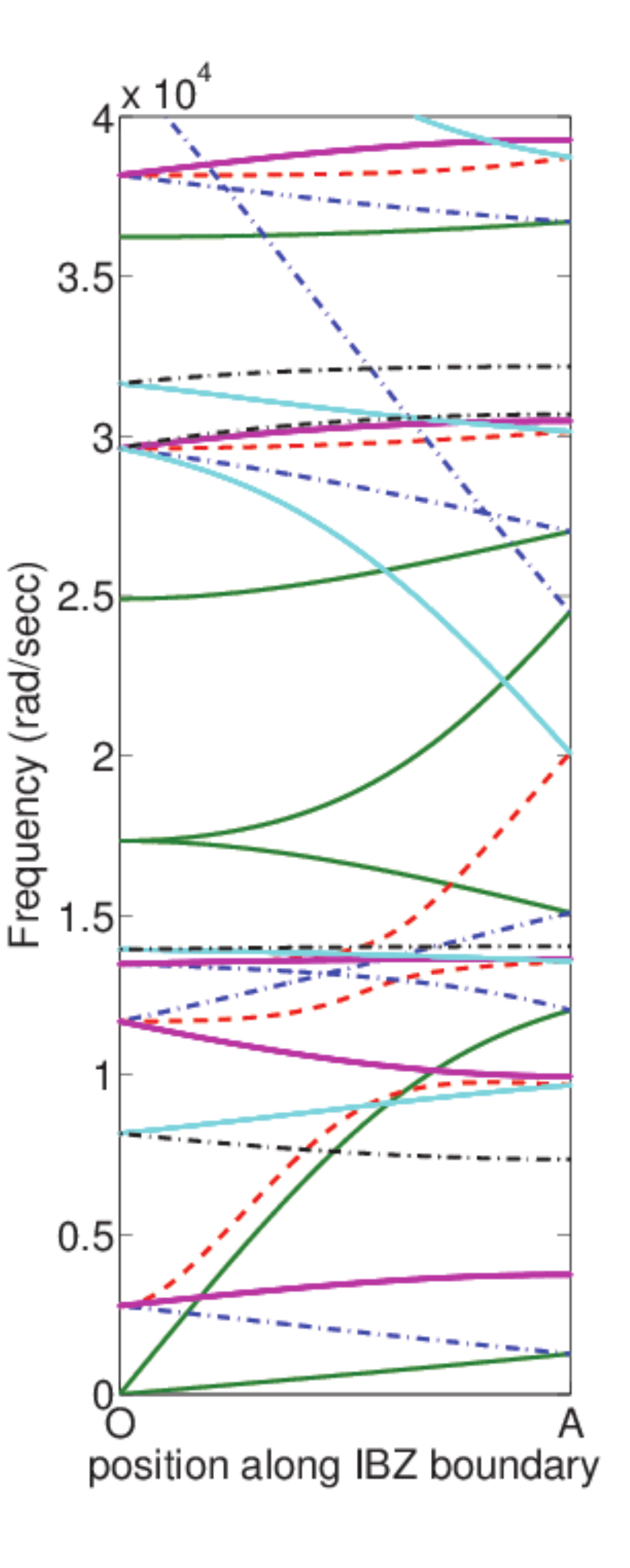}
\end{minipage}}
\subfloat[]{
\label{fig_nonsym_square:e}
\begin{minipage}{1.8in}
\hspace{0.4in}
\includegraphics[width=1.4in]{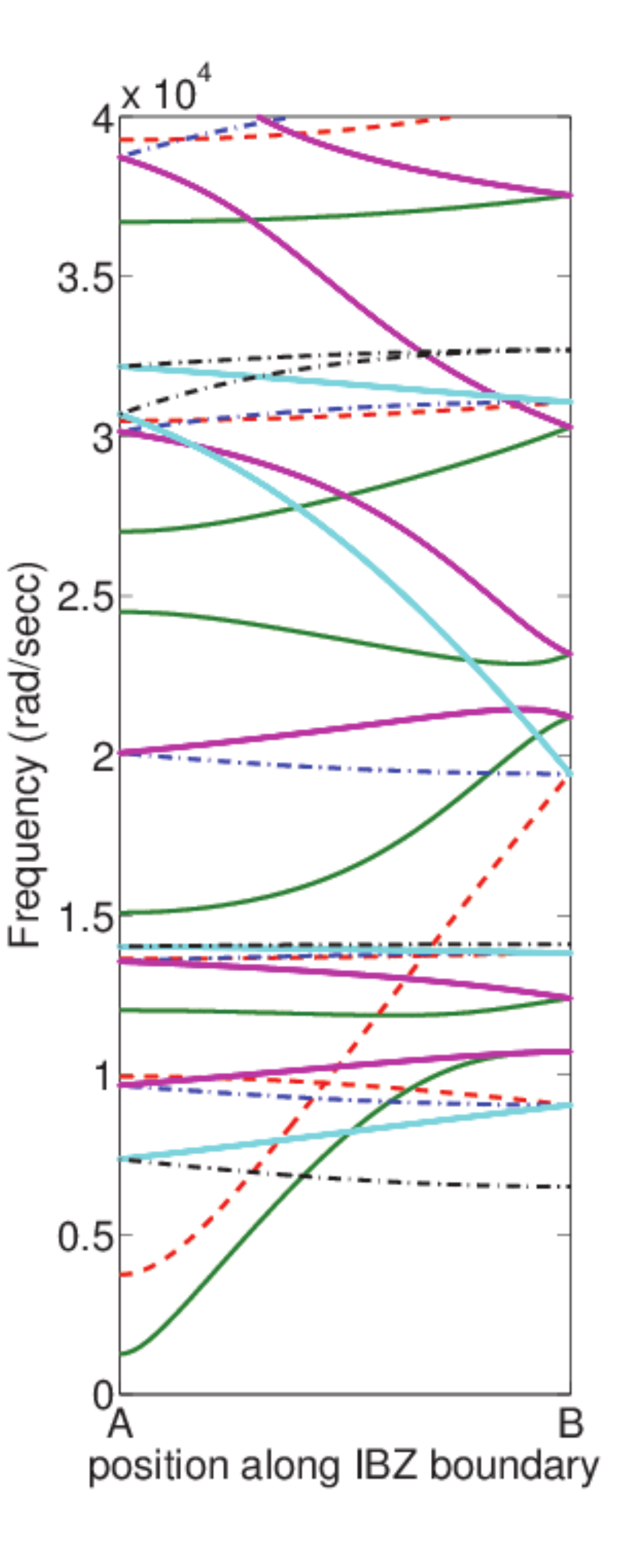}
\end{minipage}}
\subfloat[]{
\label{fig_nonsym_square:f}
\begin{minipage}{1.8in}
\hspace{0.4in}
\includegraphics[width=1.4in]{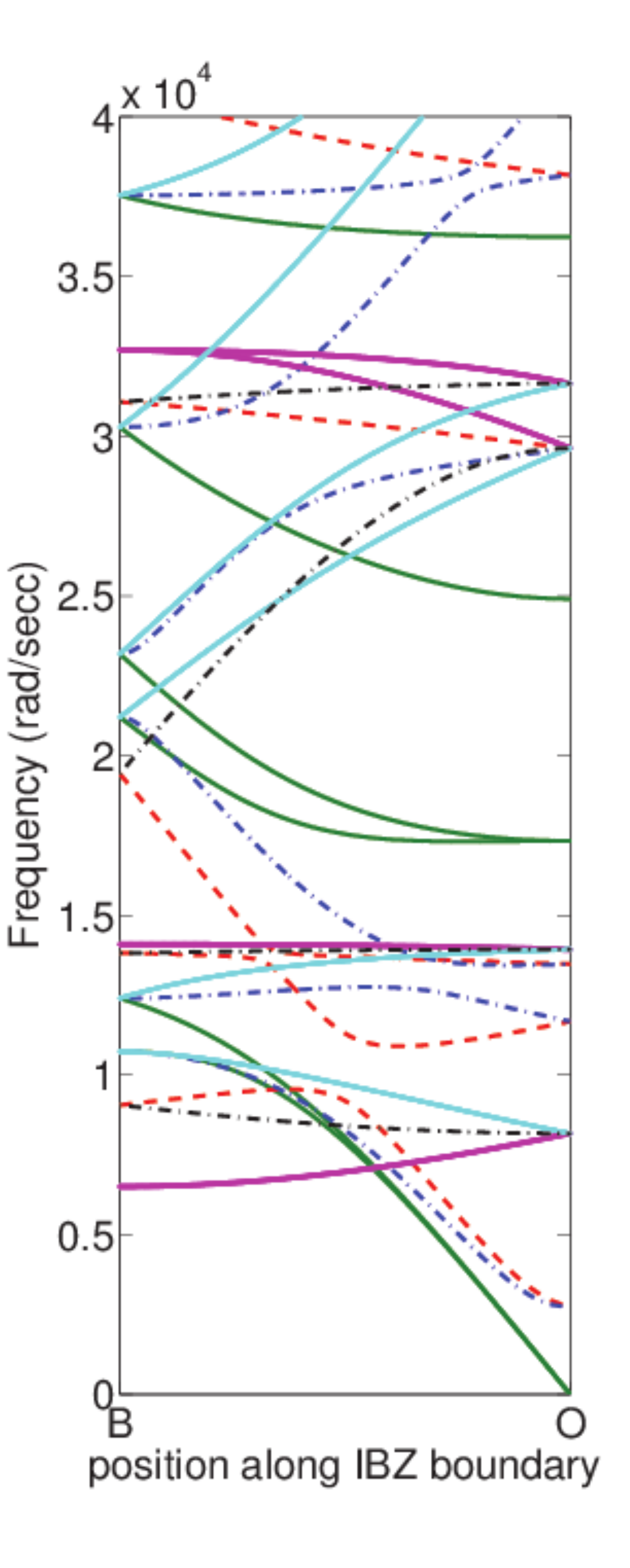}
\end{minipage}}
\caption{Construction of band structure of 3x3 supercell from that of primitive cell for a latiice with square Brillouin zone using equivalent symmetry and non-symmetry lines in the Brillouin zone. Explains origin of spurious modes. There are 6 such lines as  shown with different line style (and colors) on the IBZ at the top (a), (b), (c) and the corresponding lines in the band structure at the bottom (d),(e),(f) for the segments OA,AB and BO respectively}
\label{fig_nonsym_square}
\end{figure}

Regarding the validation of band folding for this case, folded bands in a 3x3 supercell has been generated from the symmetry and non-symmetry lines in the primitive cell band structure for the three zones in the IBZ boundary, as shown in Fig. \ref{fig_nonsym_square}. For this case too, there are six different lines of incidence, as shown in top subfigures Figs. \ref{fig_nonsym_square}(a), \ref{fig_nonsym_square}(b) and \ref{fig_nonsym_square}(c) for each of the three segments in IBZ,namely OA,AB and BO; those are marked with different line style and (and color) same line style (and colors) are used for the bands arising from each of this six lines of incidence as shown in the bottom subfigures Figs. \ref{fig_nonsym_square}(d), \ref{fig_nonsym_square}(e) and \ref{fig_nonsym_square}(f). It is noted that the three zones, when put together generate the 3x3 supercell band structure as shown in Fig. \ref{fig_square3}(a).

\subsection{Defects}  
\begin{figure}[]
\centering
\subfloat[]{
\label{fig_hex_bz:a}
\begin{minipage}{2.2in}
\includegraphics[width=2.2in]{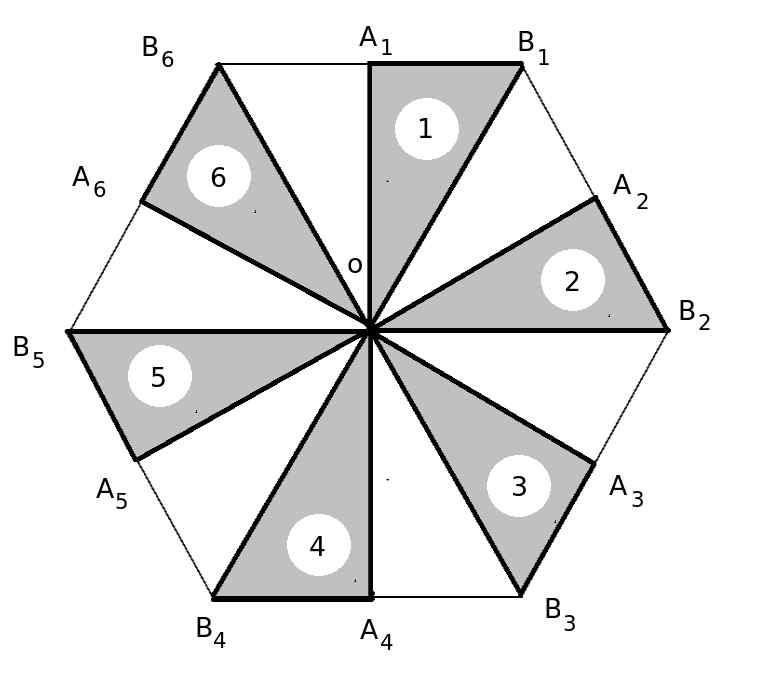}
\end{minipage}}
\subfloat[]{
\label{fig_hex_bz:b}
\begin{minipage}{2.2in}
\includegraphics[width=2.2in]{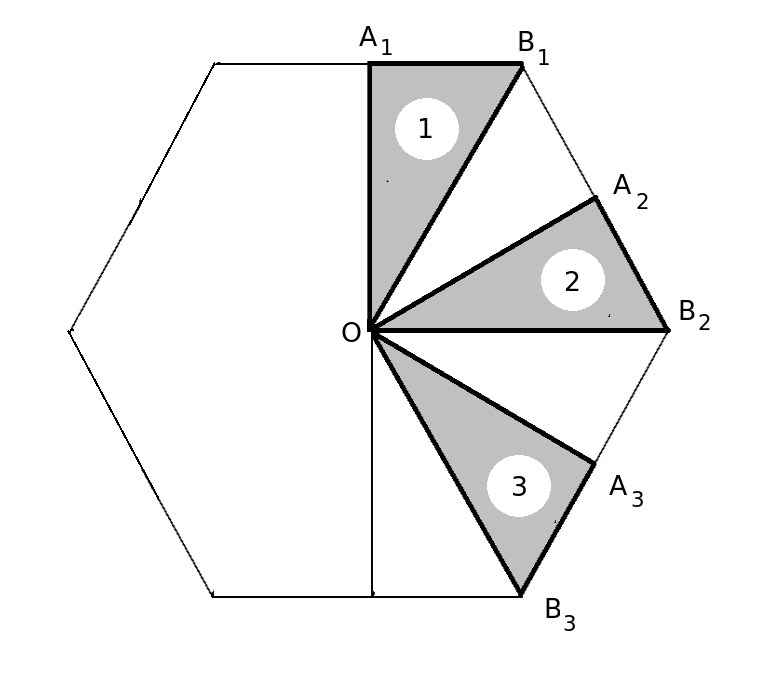}
\end{minipage}}
\caption{Line of depiction for the band structure along boundary of the (a) six traingles OA${_n}$B${_n}$- for an IBZ  having a regular hexagonal Brillouin zone and filling the whole of it, pertains to a lattice with arbitrary defect, arising out of complete lack of symmetry. (b) three triangle considering 180 rotational symmetry of the band structure} 
\label{fig_hex_bz}
\end{figure}

\begin{figure}[h]
\centering\includegraphics[width=1.5in]{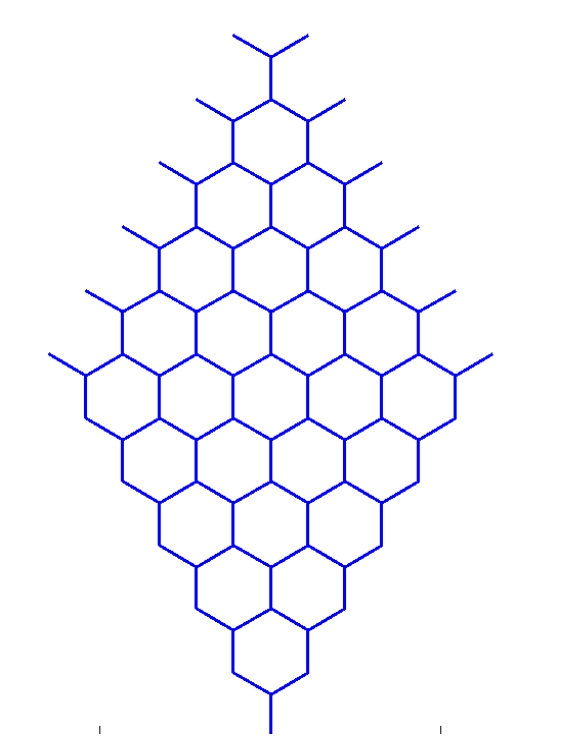}
\centering\includegraphics[width=1.5in]{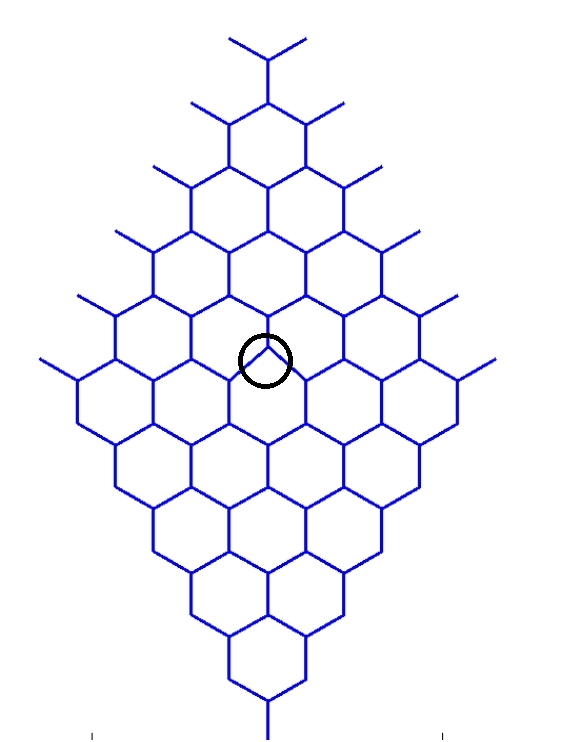}
\caption{Defect in a 6x6 honeycomb, on the left is a perfect lattice, on the right a single node dislocation defect with dislocation in positive y direction  (dislocations  marked with a circle) }
\label{fig_6x6uc_def1}
\end{figure}

For the case of lattice structures with defects, three different defect configurations have been taken, the density being 1 in a 6x6 supercell. Here, defect density indicates number of defects in a supercell. The first one, the supercell of which, with and without the defect has been shown in Fig. \ref{fig_6x6uc_def1} includes a single node dislocation defect on the middle node on the positive y direction, the defect magnitude being a deviation of 30\% of the element length.
 Toward generating the band structure for the said case, this should be noted that a lattice with defects does not retain  all the symmetries of the pure lattice, and in a general case, would be having none. So, the IBZ expands upto the whole of the Brillouin zone. This necessitates depiction of the whole of the Brillouin zone and may be accomplished by drawing the band strcuture along all the eighteen symmetry lines as shown in Fig. \ref{fig_band_folding_hex}(d) case 3 (and Table \ref{table_hex} serial 3), and may conveniently be presented along the boundaries of the six triangles as depicted in Fig.    \ref{fig_hex_bz}(a). However, it may be found that 180 degree rotation in k space preserves the band structure, i.e. one would obtain the same set of frequencies for the set $({\mu}_1,{\mu}_2)$ and $(-{\mu}_1,-{\mu}_2)$. That can be verified by considering a negative translation operator or pullback operator ${\bar{\bf{T}}}$ \cite{Farzbod1} and noticing that an alternative but equivalent equation is obtained as

\begin{equation}
\hat{\bf{K}}_r{\hat{\bf{q}}_r}={\hat{\bf{f}}_r} \hspace{0.5cm} where \hspace{0.5cm} \hat{\bf{K}}_r={{\bar{\bf{T}}}^{H}} \hat{\bf{K}}{{\bar{\bf{T}}}} \hspace{0.5cm} and \hspace{0.5cm}    {\hat{\bf{f}}_r}={\bar{\bf{T}}^{H}}{\hat{\bf{f}}} \hspace{0.5cm} and  \hspace{0.5cm}    {\hat{\bf{q}}}={\bar{\bf{T}}}{\hat{\bf{q}}_r}   
\end{equation}

\begin{figure}[]
\centering
\subfloat[]{
\label{fig_band_defect1:a}
\begin{minipage}{2.7in}
\includegraphics[width=2.8in]{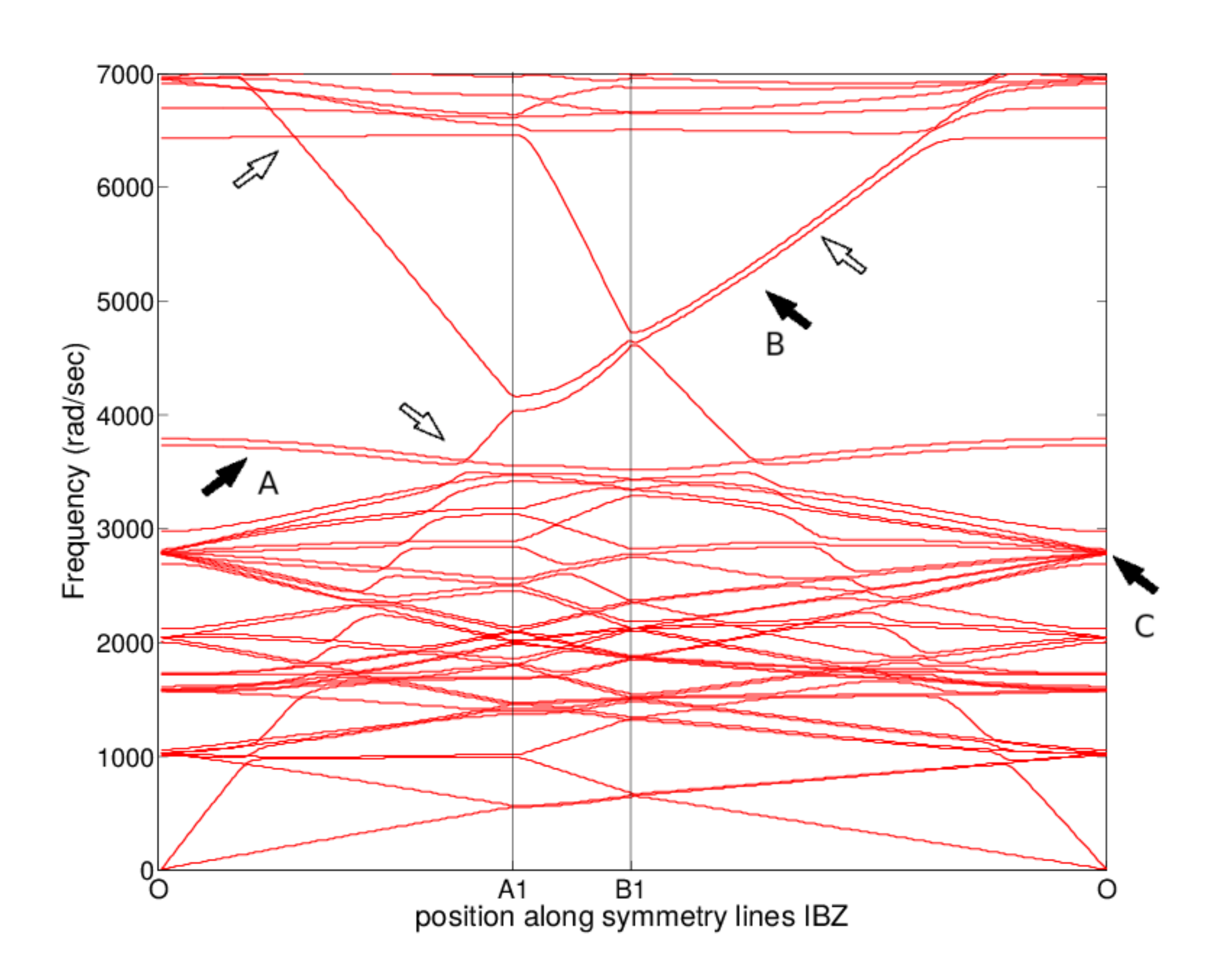}
\end{minipage}}
\subfloat[]{
\label{fig_band_defect1:b}
\begin{minipage}{2.7in}
\includegraphics[width=2.8in]{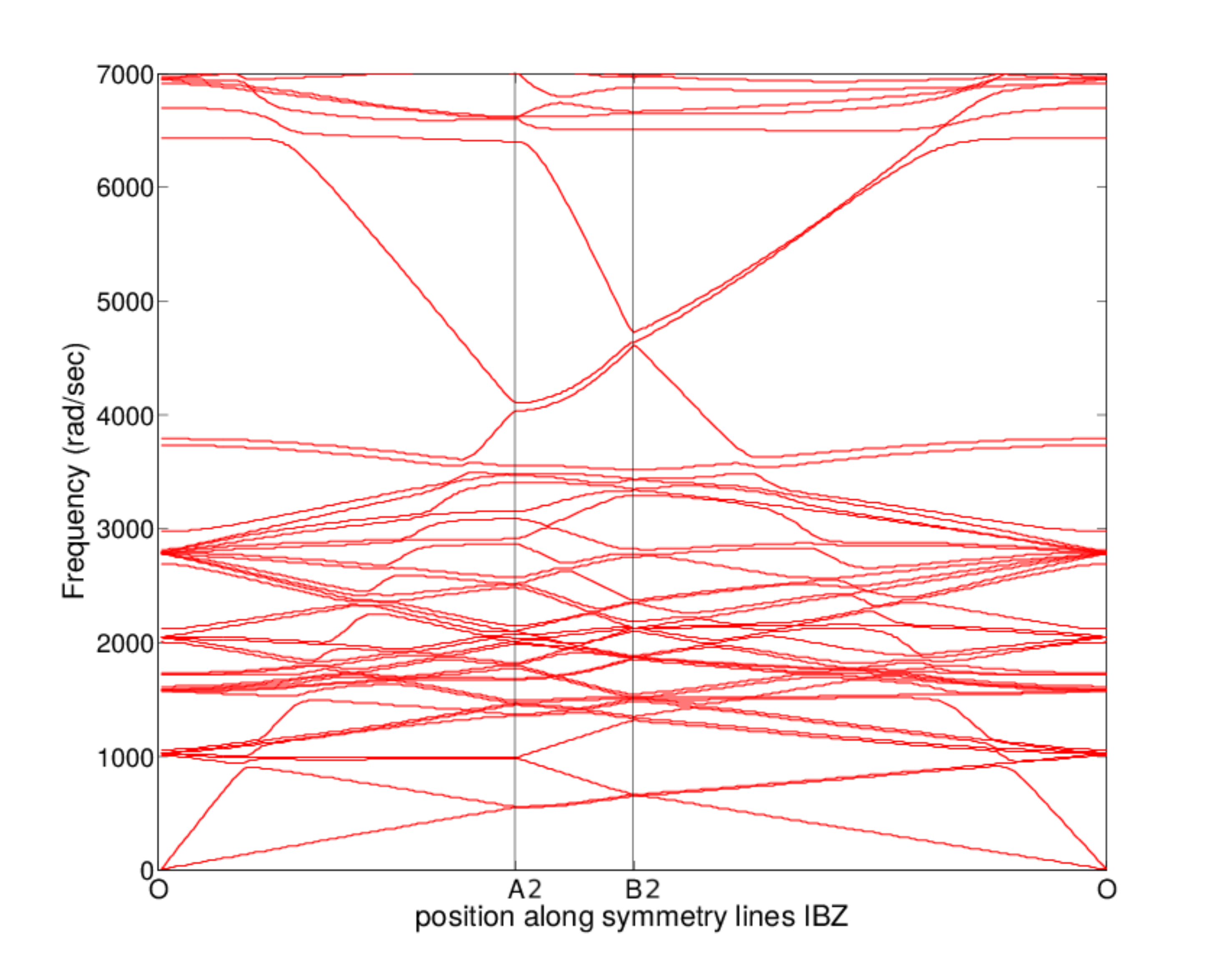}
\end{minipage}}
\par\medskip
\subfloat[]{
\label{fig_band_defect1:c}
\begin{minipage}{2.7in}
\includegraphics[width=2.9in]{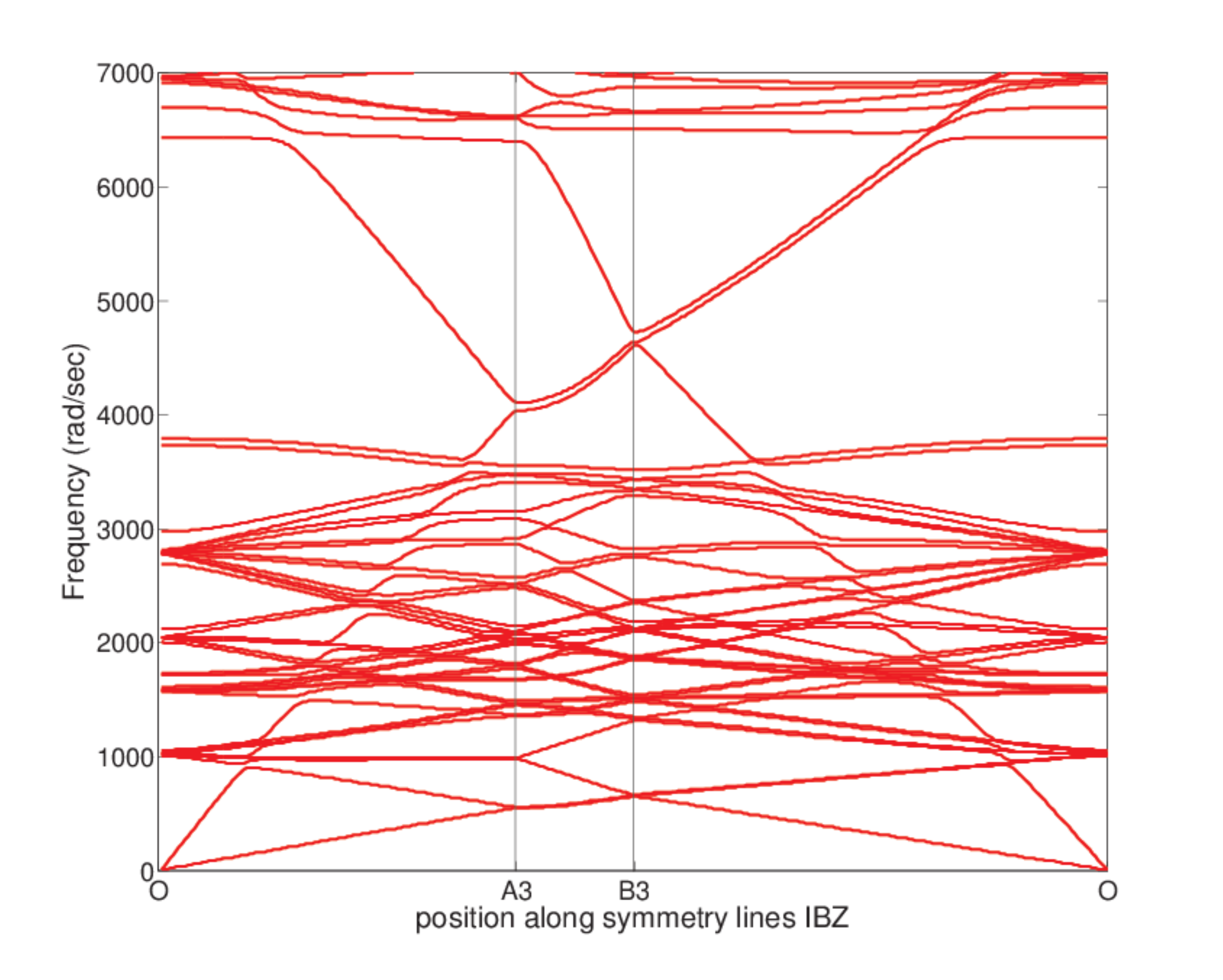}
\end{minipage}}
\subfloat[]{
\label{fig_band_defect1:d}
\begin{minipage}{2.6in}
\includegraphics[width=2.8in]{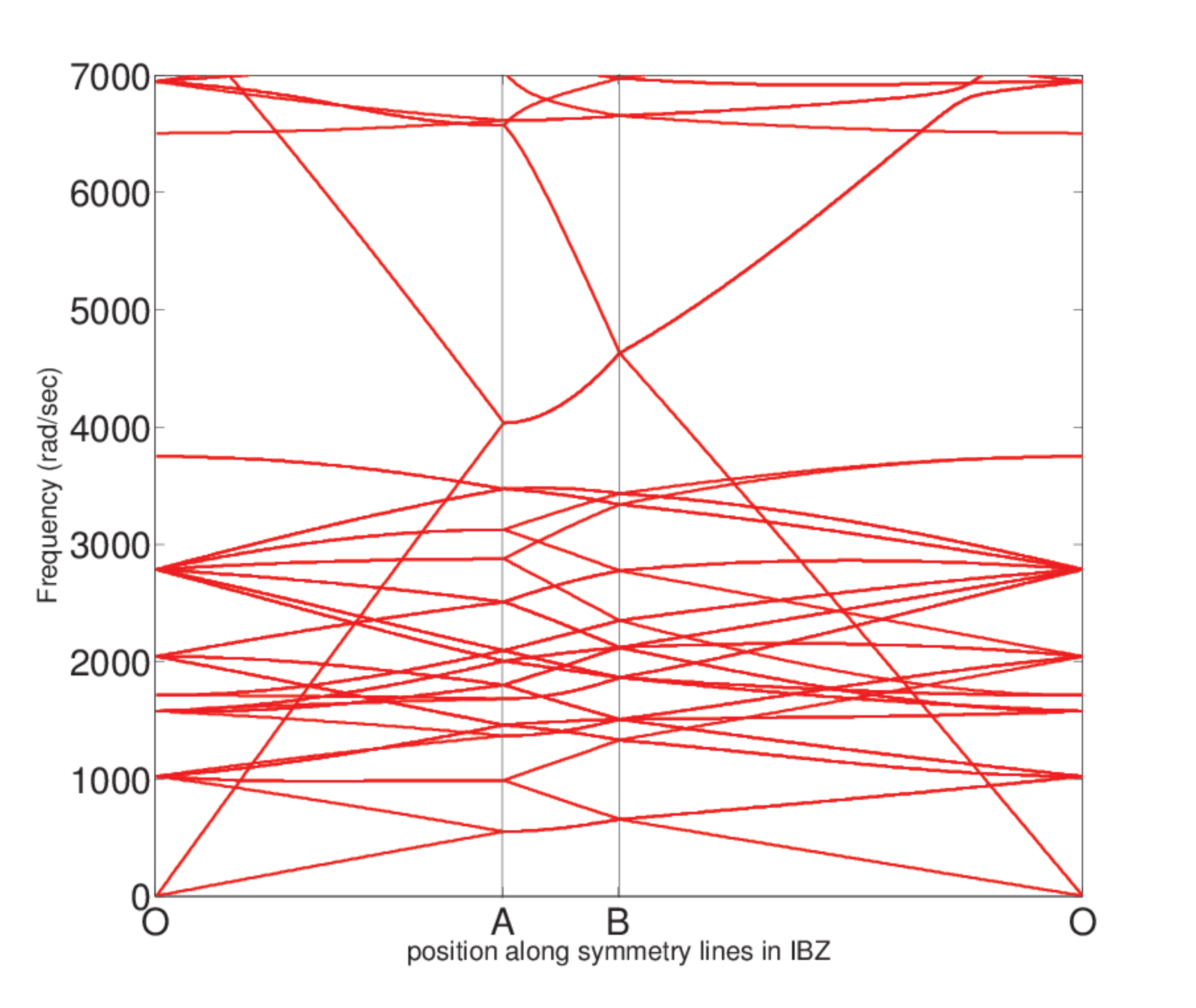}
\end{minipage}}
\caption{Band structure corresponding to the first defect case , as in Fig. \ref{fig_6x6uc_def1}, one dislocation defect in a 6x6 supercell of hexagonal honeycomb, drawn along the boundary of three traingles in IBZ as in Fig. \ref{fig_hex_bz}(b). i.e. depicted (a) along OA${_1}$B${_1}$, (b)  along OA${_2}$B${_2}$, (c) along OA${_3}$B${_3}$. (d) Band sturcture of the perfect 6x6 honeycomb lattice } 
\label{fig_band_defect1}
\end{figure}

But, the physical structure and the periodicity conditions having remained the same, the resulting eigenfrequencies also remain the same. On the other hand, the negative translation operator ${\bar{\bf{T}}}$, which is the complex conjugate of ${\bf{T}}$, refers to a wavenumber space position of $(-{\mu}_1,-{\mu}_2)$, thereby confirming the 180 degree rotaional invariance of the band structure. Hence, as shown in Fig. \ref{fig_hex_bz}(b), depiction along three triangles would suffice. 
  The complete band structure comprising all the three sets has been presented in Fig.\ref{fig_band_defect1}. The band structure corresponding to a perfect 6x6 supercell of regular hexagonal honeycomb is presented in  Fig. \ref{fig_band_defect1}(d) for comparison.
   
\begin{figure}[!h]
\centering\includegraphics[width=1.5in]{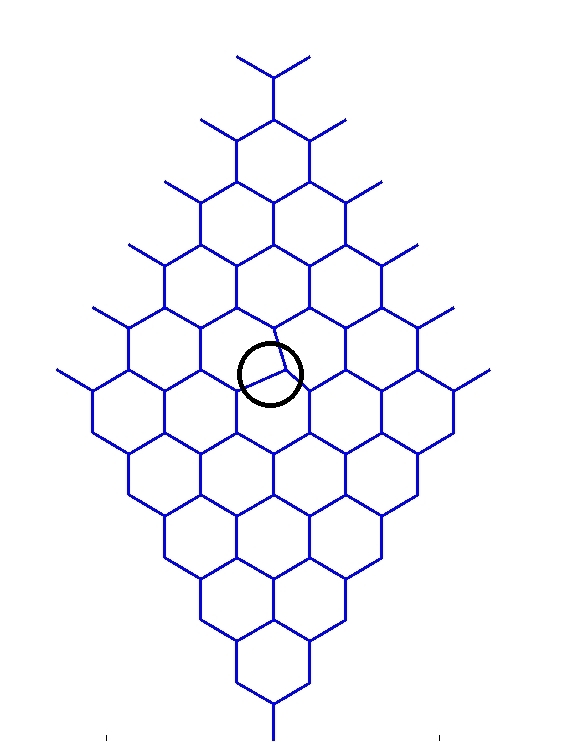}
\centering\includegraphics[width=1.5in]{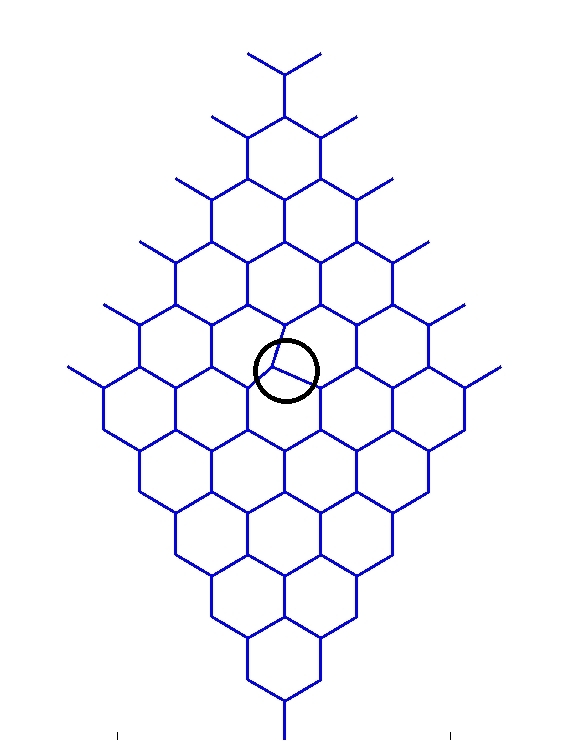}
\caption{Defect in a 6x6 honeycomb, two defect configuration have been presented having a single node dislocation in the middle, first one with dislocation in posive x direction second one with dislocation in negetive x direction (dislocations  marked with a circle)}
\label{fig_6x6uc_def2}
\end{figure}

\begin{figure}[!h]
\centering\includegraphics[width=4.0in]{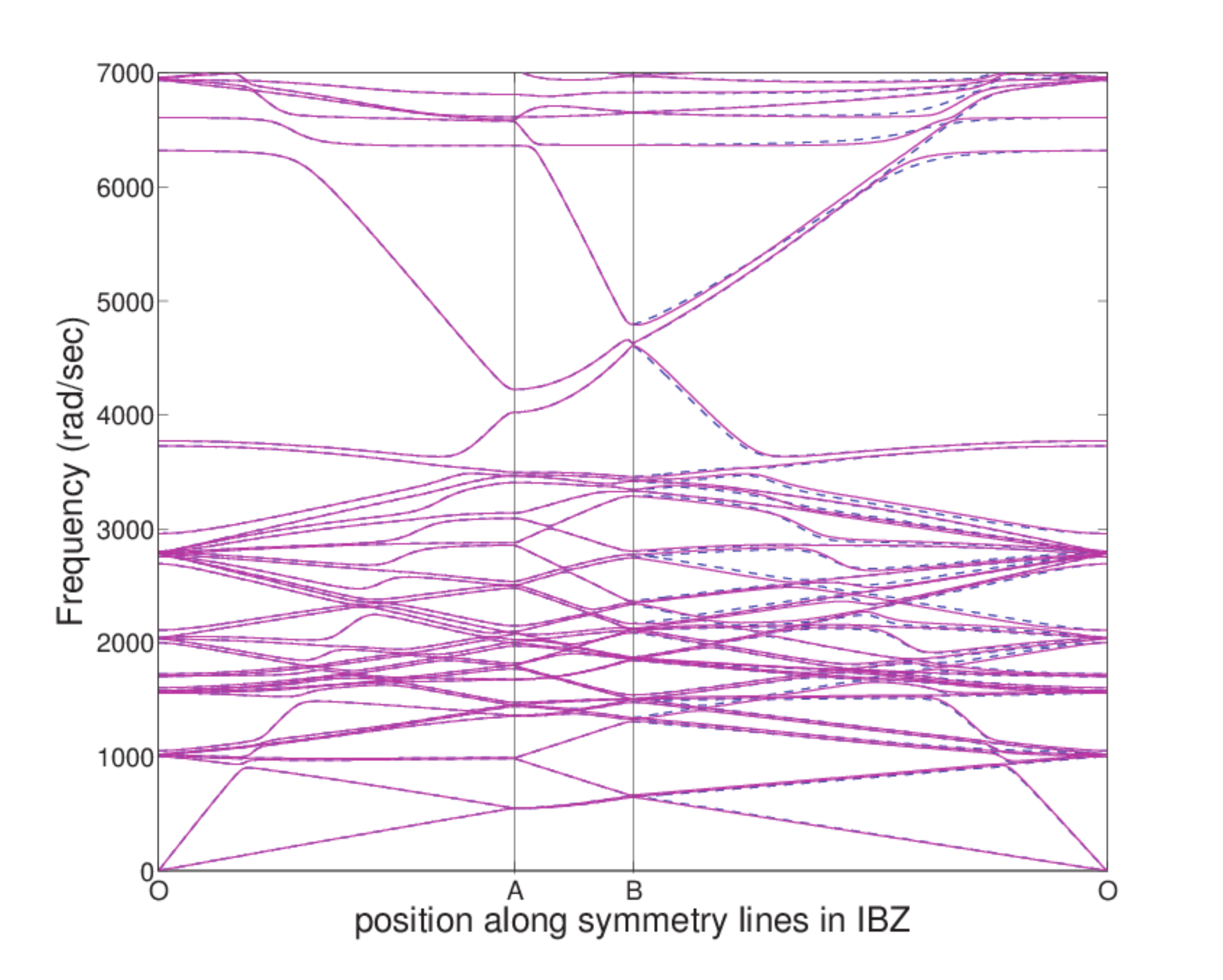}
\caption{Band structure of the hexagonal honeycomb using 6x6 supercell with a single dislocation defect each: one on the positive x, next on the negetive x , as in Fig. \ref{fig_6x6uc_def2}}
\label{fig_6x6_def1_def2}
\end{figure}

  Two other defect conditions, both for the 6x6 supercell, with a sigle node dislocation on each, have been considered. Respective unit cells, the first one having dislocation on the positive y and the second one having the dislocation on the negetive y (both defect magnitude being a deviation of 30\% of the element length.), have been shown in Fig. \ref{fig_6x6uc_def2}, and the corresponding band structures, along a single triangle in the supercell IBZ, have been generated- as shown in Fig. \ref{fig_6x6_def1_def2}

It can be seen from the band structure  of the lattices with defects that 

1. Defects induce breaking of degenaracy. The term degeneracy in paralllel with physics liturature, has been used here to indicate multiple bands getting fused together to manifest as a single band. A perfect lattice is characterised by such degeneracies- incidence of defects leads these being broken into multiple modes and manifest in the band structure through segeregation into multiple bands as lines running somewhat parallel to each other along the wavenumber line. In Fig. \ref{fig_band_defect1}(a) some of the degenaracy breaking points have been shown with solid arrows marked with A , B, C. 

2. The higher magnitude of defects show up as bands corresponding to these broken degeneracies getting further apart from each other.  

3. A few points of distinction within Figs. \ref{fig_band_defect1}(a), \ref{fig_band_defect1}(b) and \ref{fig_band_defect1}(c) are pointed out with hollow arrows. Though they are not markedly different, in view  of the low defect density ( 1 in 6x6 supercell) and low defect indensity, the diffences may still be recognised.

4. A particular observation can be made for the defects as in  Fig. \ref{fig_6x6uc_def2}, the band structure of which, corresponding to triangle OAB, are presented in Fig. \ref{fig_6x6_def1_def2} that on OA and AB line they are conincident and only differ on the zone BO. This is because, the k values along OA and AB result in the same effect for both these, cases arising out of symmetry with reference to the direction of wave propagtion.

 A rudimentary quantification can be made of the impact of defect on the band structure of the the lattice with defect compared to its defect free counterpart.
 This can be achieved by avaraging over the bands and the k points as
 
 \begin{equation}
 Average \hspace{0.04in} \% \hspace{0.04in}deviation= \frac{{\sum_{K \hspace{0.04in} values}} {\sum_{Bands}} \% \hspace{0.04in} deviation \hspace{0.04in} for \hspace{0.04in} the \hspace{0.04in} given \hspace{0.04in} band  \hspace{0.04in} and \hspace{0.04in} k  \hspace{0.04in} value   }{Total \hspace{0.04in}  k \hspace{0.04in} points* Number\hspace{0.04in}of \hspace{0.04in} bands}  
 \end{equation}

It may be seen that "Avarage \% deviation" is around 0.03$\%$  for the case of defects under consideration (i.e 1 defect in 6x6 supercell  with 30$\%$ defect deviation.)

As a general remarks with reference to all the band structures shown in ths paper,  it is worthy to mention that the rich fabric of wave phenomenon that takes place in the structure, which is the effect of propagation of wave as in an one dimensional frame member, does not involve shear mode of Timoshenko beam at the frequency range of depiction as the cut off occurs at a much higher frequency, presentable by ${\omega}_c=  \sqrt{G A K_1 /{\rho} I K_2}$ \cite{Doyle2}.

 \section{Conclusion}
 A spectral finite element based methodology has been demonstrated to generate phononic/acoustic band structure of reticulated periodic honeycomb lattice of varying geometry by treating constituent members as 1D waveguide. Bloch theory has been used to reduce the computation of the infinite spread of the structure to that of the repeatitive unit cell. Toward solving for the free vibration problem, aplication of spectral finite element and Bloch theroy result in an eigenvalue problem involving two independent wavenumbers and the implicit frequency. Wittrick Williams method, an iterative method for the solution of such nonlinear eigenvalue problem, has been used to efficiently solve this problem. Subsequently, band structures for a compound of cells known as supercells have been analysed and compared and reconciliated against the one for the elemental, primite unit cell. This provides a mechanism to treat defects in such a periodic structure. Defects are considered to be having certain periodic occurance and their implication on the band structure of the lattices are investigated for a few different cases and some general features are arrived at along with inferences about the behaviour. The proposed methodology, though involves significanty reduced computational cost, can still be very high for a high value of supercell $n$ and computations involving Monte Carlo or other simulation methodology.  The possibility of bucketing along k values and frequency values simultaneously, which fits very well with current framework, offers great opportunity of parallelization, and may be taken up as future work.

\bibliography{aps1}

\end{document}